\documentclass{emulateapj}  
\usepackage{natbib}
\usepackage{graphicx}
\usepackage{amsmath}
\usepackage{amssymb}
\usepackage{txfonts}
\usepackage{multirow}
\def\hii{\mbox{H\,{\sc ii}}}
\newcommand*{\textquotedouble}[1]{\textquotedblleft #1\textquotedblright}

\received{}
\revised{}
\accepted{}

\shorttitle{IRDC G333.73+0.37}
\shortauthors{Veena et al.}
\begin{document}

\title{Probing the massive star forming environment - a multiwavelength investigation of the filamentary IRDC G333.73+0.37}

\author{V. S. Veena$^1$ }
\author{S. Vig$^1$}
\author{B. Mookerjea$^2$}
\author{{\'A}. S{\'a}nchez-Monge$^3$}
\author{A. Tej$^1$}
\author{C. H. Ishwara-Chandra$^4$}

\affil{$^1$Indian Institute of Space Science and Technology, Thiruvananthapuram, 695 547, India}
\affil{$^2$Tata Institute of Fundamental Research, Mumbai, 600 005, India}
\affil{$^3$Physikalisches Institut, Universit{\"a}t zu K{\"o}ln, Z{\"u}lpicher Str. 77, 50937 K{\"o}ln, Germany}
\affil{$^4$National Centre for Radio Astrophysics (NCRA-TIFR), Pune, 411 007, India}

\begin{abstract}
We present a multiwavelength study of the filamentary infrared dark cloud (IRDC) G333.73+0.37. The region contains two distinct mid-infrared sources S1 and S2 connected by dark lanes of gas and dust. Cold dust emission from the IRDC is detected at seven wavelength bands and we have identified 10 high density clumps in the region. The physical properties of the clumps such as temperature: $14.3-22.3$~K and mass: $87-1530$~M$_\odot$ are determined by fitting a modified blackbody to the spectral energy distribution of each clump between 160~$\mu$m and 1.2~mm. The total mass of the IRDC is estimated to be $\sim4700$~M$_\odot$. The molecular line emission towards S1 reveals signatures of protostellar activity. Low frequency radio emission at 1300 and 610~MHz is detected towards S1 (shell-like) and S2 (compact morphology), confirming the presence of newly formed massive stars in the IRDC. Photometric analysis of near and mid-infrared point sources unveil the young stellar object population associated with the cloud. Fragmentation analysis indicates that the filament is supercritical. We observe a velocity gradient along the filament, that is likely to be associated with accretion flows within the filament rather than rotation. Based on various age estimates obtained for objects in different evolutionary stages, we attempt to set a limit to the current age of this cloud.

\end{abstract}


\section{Introduction}

The formation of a massive star is believed to proceed along an evolutionary path that is at variance with that of the lower mass counterparts, mainly due to the enhanced feedback mechanisms  expected from the former. While theoretical formulations have been proposed to explain their formation \citep{2007ARA&A..45..481Z}, observational studies of the early phases remain limited. The challenges are largely due to their rarity and short evolutionary timescales, in addition to obscuration produced by the associated molecular clouds. Possible clues to the earliest evolutionary phases of star formation can be found by examining dense clumps and cores nestled within Infrared Dark Clouds (IRDCs). 
IRDCs are believed to be the progenitors of massive stars and star clusters, that are characterized by dark extinction features seen against the bright-infrared Galactic background \citep{{2006ApJ...641..389R},{2009ApJS..181..360C}}. Initially detected by \textit{ISO} and then by \textit{MSX} \citep{{1996A&A...315L.165P},{1998ApJ...494L.199E}}, these massive clouds ($10^2-10^4~\textrm{M}_{\odot}$) vary widely in their morphology from elongated to compact structures. On larger scales, IRDCs are associated with filamentary structures and sizes are seen to range from a few to hundreds of parsecs \citep{{2010ApJ...719L.185J},{2014A&A...568A..73R}}. Several works have revealed the properties of these filaments such as the density structure, mass, stability, evolutionary stage, and kinematic properties \citep[e.g.,][]{{2010A&A...518L..95H},{2012A&A...540A.104M},{2013ApJ...764L..26B},{2015A&A...584A..67B},{2016MNRAS.463..146H}}. In addition, recent studies suggest that large filaments extending upto 100 pc and beyond are likely to be the `bones' of the Milky Way \citep{{2014ApJ...797...53G},{2015ApJ...815...23Z}}. On smaller scales, they fragment into dense clumps and cores \citep{{2010ApJ...721..222B},{2014A&A...570A..51S},{2016arXiv161108794Z}}. IRDCs are dense ($n>10^5~\textrm{cm}^{-3}$), cold ($\textrm{T}<20$~K), and bright at submillimeter wavelengths \citep{2000ApJ...543L.157C}. These extreme properties make them quintessential locales to forage for objects in the earliest phases in massive star formation. 

\par Comprehensive analyses at far-infrared and submillimeter wavelengths show that IRDCs possess significant substructures  within them that are undergoing a star-formation flurry. Molecular line emission from these clumps and cores often exhibit signatures of protostellar activity such as infall and outflow \citep{{2007ApJ...668..348B},{2016ApJS..225...21J}}, characterized by asymmetric line features. 
While some IRDCs harbour objects of different evolutionary stages such as  maser spots, starless cores, ultracompact \hii~regions and young stellar objects (YSOs), there are others that are devoid of any star formation activity \citep[][e.g.,]{{2016ApJ...819..139B},{2013A&A...553A.115B}}. The latter serve as good targets to study the earliest phases prior to collapse whereas those harbouring \hii~regions and YSOs can be used to decipher the conditions under which the infant massive stars evolve. Thus, IRDCs can be broadly categorised on the basis of the evolutionary stage of the cloud itself. However, such a study necessitates a detailed scrutiny of the star forming activity within clumps of these molecular clouds. In this work, we investigate the elongated IRDC G333.73+0.37 (hereafter, G333.73), using an assortment of markers to probe the diverse traits of the star forming activity in the cloud. Based on the results, we hope to be able to comment on the evolutionary stage of the cloud.
\begin{figure}
\hspace*{-0.7cm}
\centering
\includegraphics[scale=0.15]{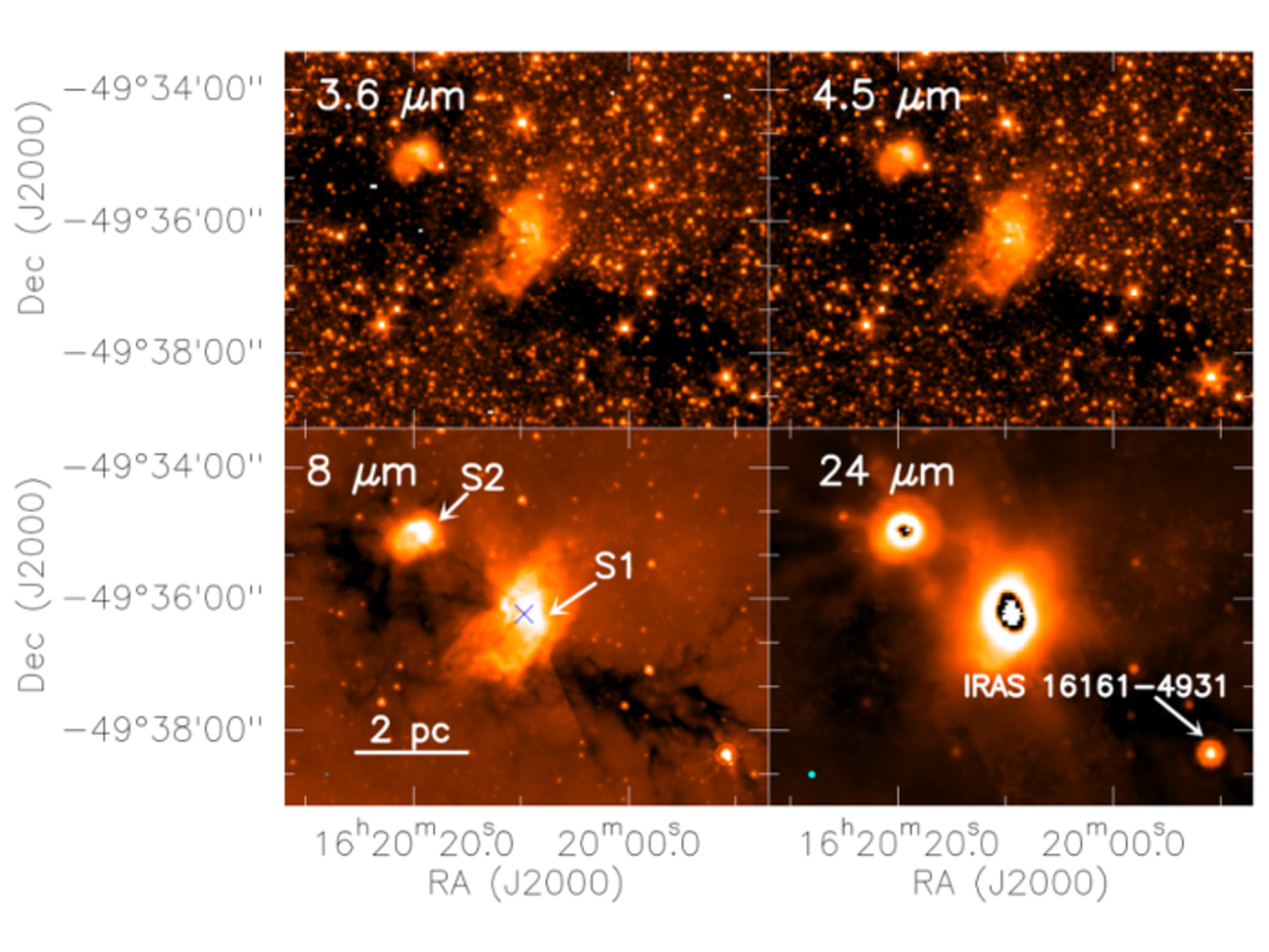}
\caption{Mid-infrared emission from the IRDC G333.73 at 4 wavelength bands from $\textit{Spitzer}$: 3.6, 4.5, 8.0 and 24.0~$\mu$m. The dark extinction filaments are clearly seen in the images. We have also marked the locations of two bright infrared sources S1 and S2 in the 8~$\mu$m image. The IRAS peak corresponding to S1 is indicated with a $\times$. The position of the IRAS source 16161-4931 is marked in the 24~$\mu$m image. The corresponding beam sizes (cyan) are shown at the bottom left corner of all panels.}
\label{mir}
\end{figure}
\par G333.73+0.37 is located at a distance of 2.6~kpc \citep{{2006A&A...447..221B},{2013A&A...550A..21S}}. Previous studies have reported signatures of massive star formation within this IRDC. \citet{2006A&A...447..221B} mapped the dust emission at 1.2~mm and identified 8 massive ($14-472$~M$_\odot$) cold dust clumps in this region. This IRDC is also associated with an infrared bubble (MWP1G333726+003642) identified by \citet{2012MNRAS.424.2442S}. High frequency radio continuum observations at 18 and 22.8~GHz by \citet{2013A&A...550A..21S} identified two sources in this region (beam size~$\sim$30$\arcsec$). These results are chiefly the outcomes of various surveys and hence provide limited information about the IRDC in its entirety. As our motivation is to examine the star forming potential across the entire IRDC, we use multiwavlength tracers including sensitive radio continuum observations at 1300 and 610~MHz to analyse the morphology and properties of ionised gas associated with newly formed massive stars. In addition, we have utilised the archival infrared and submillimeter data along with the molecular line emission of this region in various molecular species from the MALT90 and ThrUMMS surveys to probe the molecular cloud. Such a plethora of observational data enables a fair visualization of the physical properties, chemistry, kinematics as well as the evolutionary stage of G333.73.

\par The organization of the paper is as follows. The details of radio continuum observations and archival data are given in Section 2. Section 3 describes the results of our multiwavelength study while Section 4 elaborates on the analysis of the morphology of radio emission, fragmentation and evolution of cold dust clumps as well as the kinematics within the IRDC. We also attempt to estimate the age of this filamentary cloud based on objects in different evolutionary stages. Finally in Section 5, we present our conclusions. 

\begin{table}[hbt!]
\begin{center}
\footnotesize
\caption{Details of the radio continuum observations.}
\label{radio_tb}
\vspace{0.2cm}
\begin{tabular}{l c c}
\hline \hline 
    Frequency (MHz) & 610 & 1300\\
 \hline 
 Observation date & 2014 August 7 & 2014 August 31 \\
 On-source time (min) & 150 & 164 \\
 Bandwidth (MHz) & 32 & 32\\
 Primary Beam & $45'.8$ & $21'.8$\\
 Synthesized beam & $14''.5\times 5''.4$ & $5''.5\times2''.0$\\
 Position angle ($ ^\circ $) & 7.9 &5.3\\
 Noise ($\mu$Jy/beam) & 310 & 75\\
 \hline 
 \end{tabular}
\end{center}
 \end{table}
\section{Observations and Data Reduction}
 
\subsection{Radio Continuum Observations using GMRT}
The ionised gas emission from G333.73+0.37 is mapped using the Giant Metrewave Radio Telescope (GMRT), India \citep{1991CuSc...60...95S}. GMRT consists of 30 antennas each having a diameter of 45~m arranged in a Y-shaped configuration. Twelve antennas are distributed randomly in a central array within an area $\sim 1$~km$^2$ and the remaining 18 antennas are stretched out along three arms, each of length $\sim14$~km. The minimum and maximum baselines are 105~m and 25~km respectively that allows the simultaneous mapping of small and large scale structures. The radio continuum observations were carried out at two frequencies: 1300 and 610~MHz. The angular extent of the largest structure observable with GMRT at 1300~MHz is 7$\arcmin$ and 17$\arcmin$ at 610~MHz and our targets have sizes well within these limits. The radio source 3C286 was used as the primary flux calibrator while 1626-298 was used to calibrate the phases. The details of observations are listed in Table.~\ref{radio_tb}. 

\begin{figure}
\centering
\includegraphics[scale=0.25]{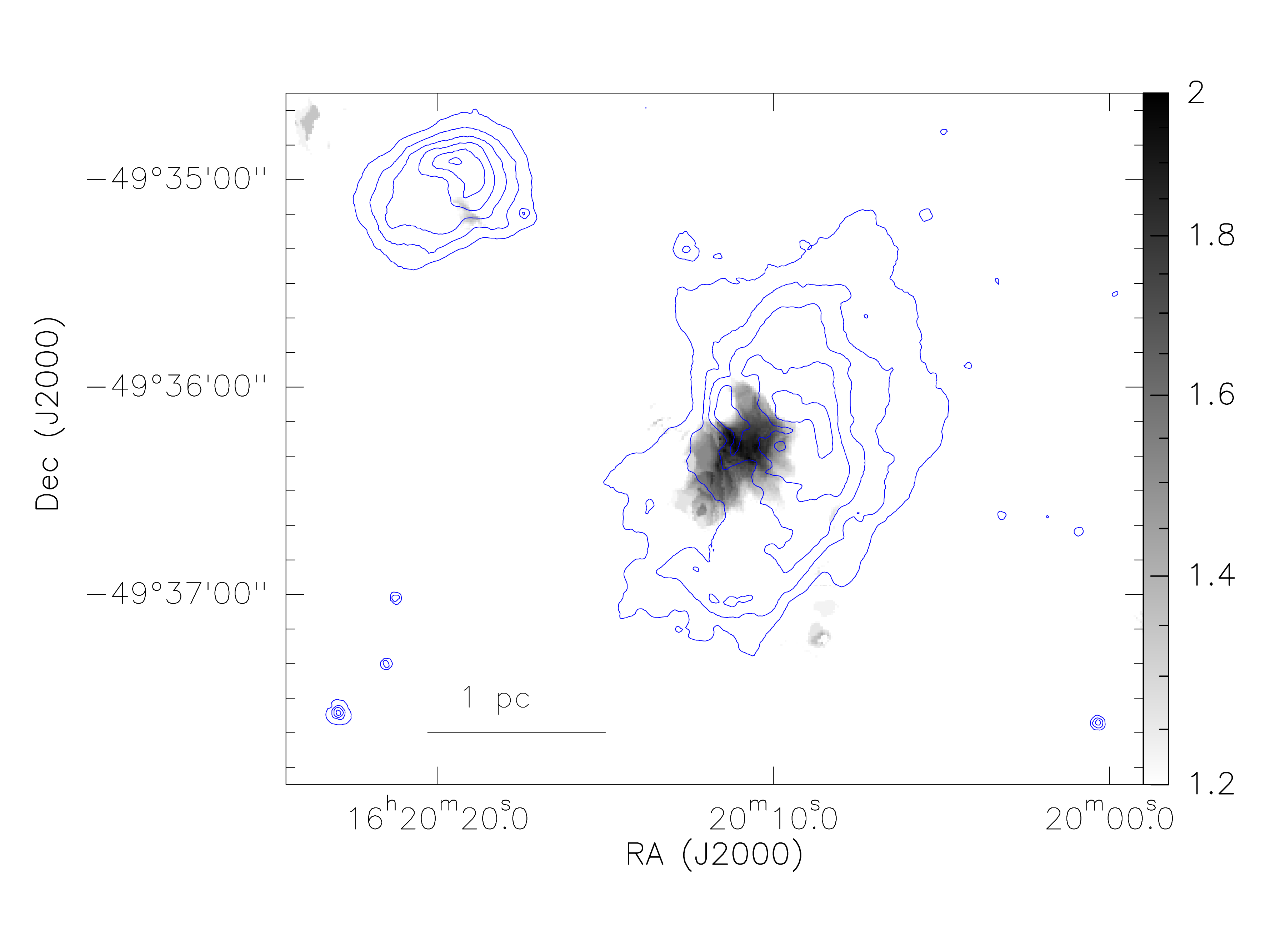}
\caption{Flux ratio map of $\textit{Spitzer}$ [4.5]/[3.6] overlaid with 8~$\mu$m warm dust contours. The contour levels are 100, 200, 400, 800, 1600 and 3200~MJy/Sr.}
\label{mirratio}
\end{figure}

\par We have carried out data reduction using the NRAO Astronomical Image Processing System ($\tt{AIPS}$). The tasks {\tt TVFLG} and {\tt UVFLG} were used to remove the visibilities affected by the  non-working antennas and radio frequency interference. The calibrated target data was cleaned and deconvolved using the task {\tt IMAGR} and we applied several iterations of self-calibration to minimize the amplitude and phase errors. In addition to creating a map using all the visibilities at 1300 MHz, we have constructed a lower resolution map at this frequency to examine the low brightness diffuse emission. This is achieved by limiting the UV range to 25~k$\lambda$. A system temperature correction to account for the Galactic plane emission, $\rm{(T_{gal}+T_{sys})/T_{sys}}$ has been used to scale the fluxes at each frequency, where $\rm{T_{sys}}$ is the system temperature corresponding to the flux calibrator located away from the Galactic plane. To estimate $\rm{T_{gal}}$, we have used the temperature map of \citet{1982A&AS...47....1H} at 408~MHz. Scaling factors are calculated by extrapolating $\rm{T_{gal}}$ to 610 and 1300~MHz by assuming a spectral index of $-2.6$ \citep{{1999A&AS..137....7R},{2011A&A...525A.138G}} which are then applied to the self-calibrated images. Finally, the flux-scaled maps were corrected for the primary beam using the task {\tt PBCOR}. 
 \begin{figure*}
\hspace*{-0.45cm}

\centering
\includegraphics[scale=0.38]{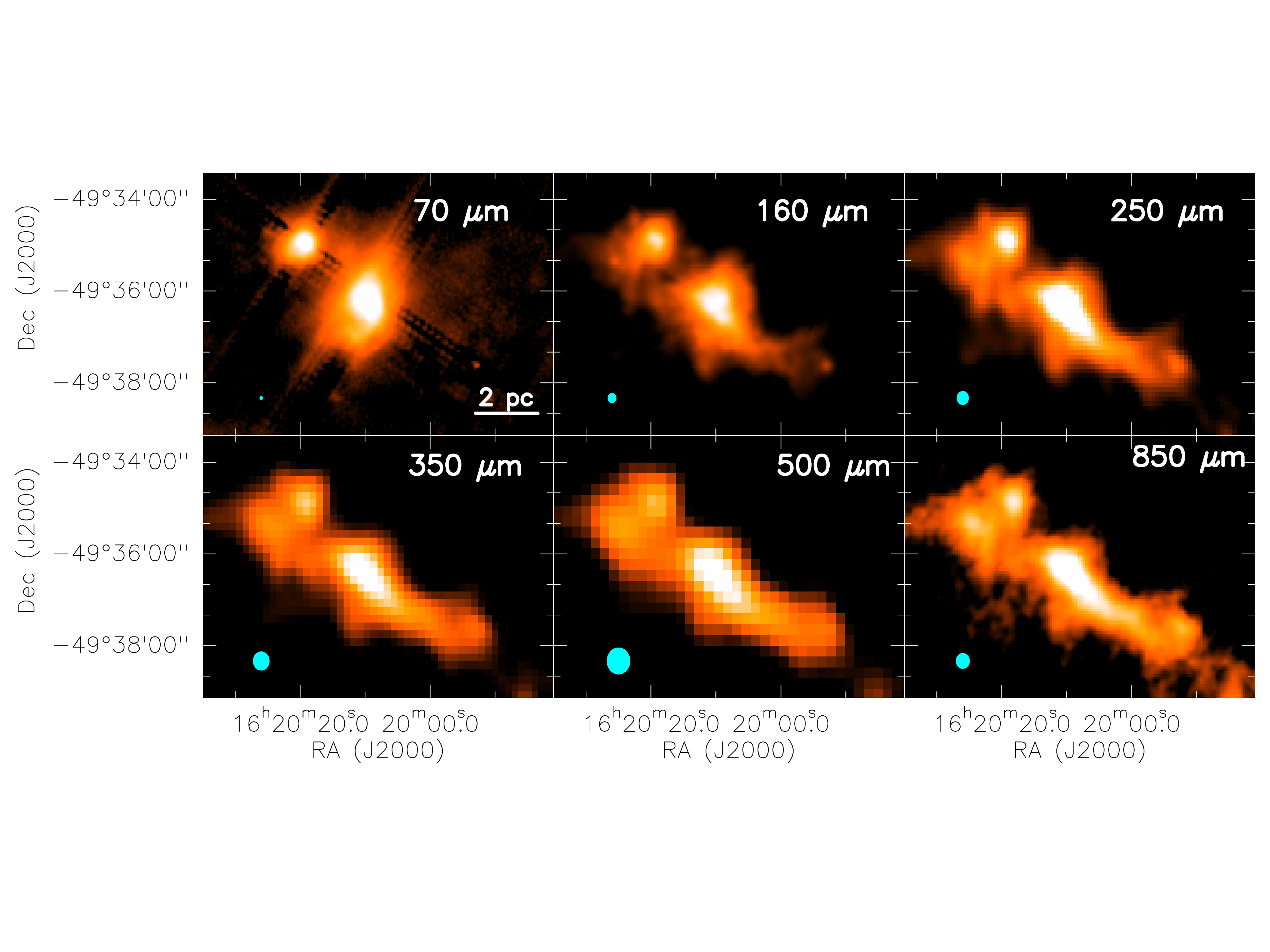}
\vspace*{-1cm}
\caption{Distribution of cold dust emission towards the G333.73 region at 6 wavelength bands: 70, 160, 250, 350 and 500~$\mu$m from $\textit{Herschel}$ and 850~$\mu$m from APEX+Planck. The corresponding beam sizes (cyan) are shown towards bottom left of individual panels.}
\label{submm}
\end{figure*}
\subsection{Archival Datasets}
Apart from radio observations, we have used archival data to investigate emission across the cloud at different wavebands. The properties of the warm dust associated with this region is investigated using mid-infrared \textit{Spitzer Space Telescope} data and cold dust emission is analysed using far-infrared and submillimeter maps from \textit{Herschel} Hi-GAL and APEX+Planck surveys. In addition, we have used the MALT90 and ThrUMMS spectral line surveys to examine the chemical properties and kinematics of the region. 

\subsubsection{\textit{Spitzer Space Telescope}}
We have used the mid-infrared maps of this region observed by \textit{Spitzer Space Telescope}, with a primary mirror of size 85~cm. 
The Infrared Array Camera (IRAC) is one of the three focal plane instruments that obtain simultaneous broadband images at 3.6, 4.5, 5.8 and 8.0~$\mu$m. The achieved resolutions are 1.7$\arcsec$, 1.7$\arcsec$, 1.9$\arcsec$ and 2.0$\arcsec$ at 3.6, 4.5, 5.8 and 8.0~$\mu$m respectively \citep{2004ApJS..154...10F}. We used the Level-2 Post-Basic Calibrated data (PBCD) images from the Galactic Legacy Infrared Mid-Plane Survey \citep[GLIMPSE;][]{2003PASP..115..953B} to study the nature of diffuse emission. In addition, we have also made use of the MIPS 24~$\mu$m image obtained as a part of the MIPSGAL survey \citep{2009PASP..121...76C}.

\subsubsection{\textit{Herschel} Hi-Gal Survey}
The cold dust emission from the molecular cloud is investigated using images from the \textit{Herschel Space Observatory}. The \textit{Herschel Space Observatory} is a 3.5~m telescope capable of observing in the far-infrared and submillimeter spectral range $55-671$~$\mu$m \citep{2010A&A...518L...1P}. The images are part of the \textit{Herschel} Hi-Gal Survey \citep{2010PASP..122..314M}. The instruments used in the survey are the Photodetector Array Camera and Spectrometer \citep[PACS;][]{2010A&A...518L...2P} and the Spectral and Photometric Imaging Receiver \citep[SPIRE;][]{2010A&A...518L...3G}. The Hi-Gal observations were carried out in parallel mode covering wavelengths $70 - 500~\mu$m. We used Level-2.5 PACS images at 70 and 160~$\mu$m and Level-3 SPIRE images at 250, 350 and 500~$\mu$m for our analysis.  The pixels sizes are 2$\arcsec$, 3$\arcsec$, 6$\arcsec$, 10$\arcsec$ and 14$\arcsec$ and the corresponding resolutions are 5$\arcsec$, 13$\arcsec$, 18.1$\arcsec$, 24.9$\arcsec$ and 36.4$\arcsec$ at 70, 160, 250, 350 and 500~$\mu$m respectively. We used the \textit{Herschel} Interactive Processing Environment (HIPE)\footnote{ HIPE is a joint development by the Herschel Science Ground Segment Consortium, consisting of ESA, the NASA Herschel Science Center, and the HIFI, PACS and SPIRE consortia.} to download and process the data.

\subsubsection{APEX+Planck data}
The Apex+Planck image is a combination of 870~$\mu$m data from the ATLASGAL survey \citep{2009A&A...504..415S} and 850~$\mu$m map from the Planck/HFI instrument. The data covers emission at larger angular scales, thereby revealing the structure of cold Galactic dust in greater detail \citep{2016A&A...585A.104C}. The pixel size and resolution achieved are 3.4$\arcsec$ and 21$\arcsec$, respectively.

\subsubsection{MALT90 Molecular Line Survey}
We have used Millimetre Astronomy Legacy Team 90~GHz Pilot Survey \citep{{2011ApJS..197...25F},{2013PASA...30...57J}} to understand the properties  of the associated molecular gas. This survey has mapped transitions of 16 molecular species near 90~GHz. The observations were carried out using the 8~GHz wide Mopra Spectrometer (MOPS). The data reduction was conducted by the MALT90 team using an automated reduction pipeline. The spatial and spectral resolutions are 72$\arcsec$ and 0.11~km\,s$^{-1}$ respectively. The data cubes available from the website are images of size $\sim$4$\arcmin$. The MALT90 data cube covers only a part of the cloud where there are signatures of active star formation. 

\subsubsection{ThrUMMS Molecular Line Survey}
In order to sample the molecular line emission from the entire IRDC filament, we have used $^{12}$CO and $^{13}$CO maps from The Three-mm Ultimate Mopra Milky Way Survey  \citep[ThrUMMS;][]{2015ApJ...812....6B}. The survey mapped $J$ = 1$\rightarrow$0 transition of $^{12}$CO, $^{13}$CO, C$^{18}$O, and CN lines near 112 GHz at a spectral resolution of 0.1~km\,s$^{-1}$  and spatial resolution of 66$\arcsec$. The data reduction was performed by ThruMMS team and the calibrated data is made available to the public through the website \footnote{http://alma-intweb.mtk.nao.ac.jp/~thrumms/}. In this work, we present only $^{12}$CO and $^{13}$CO molecular emission as C$^{18}$O and CN have not been detected due to relatively poor signal-to-noise ratio.

\let\cleardoublepage\clearpage

\section{Results}
We present our results in the following sequence. As IRDCs have been identified as dark structures against nebulous mid-infrared emission, we initiate our analysis with warm dust emission towards this region. Subsequently, we probe the properties of cold dust and gas in the cloud using far-infrared to millimetre wavelengths. The locations of star-forming flurry are realised using the distribution of ionised gas emission. Finally in this section, we examine the population of young stellar objects and their distribution across the cloud.

\subsection{Mid-infrared emission from warm dust}

 The mid-infrared maps of the filamentary IRDC G333.73 at 3.6, 4.5, 8.0 and 24~$\mu$m bands from $\textit{Spitzer Space Telescope}$ are shown in Fig.~\ref{mir}. Two prominent features visually discernible from the maps are the bright infrared objects that we designate S1 and S2. 
S1 is also catalogued as an infrared bubble \citep[MWP1G333726+003642;][]{2012MNRAS.424.2442S}. These sources appear to be connected by dark filamentary structures silhouetted against nebulous emission. We have not been able to deduce any previously reported information about S2 from our literature survey. We proceed with the assumption that both these regions belong to the same IRDC and the kinematic distance towards S2 is the same as that of S1, that is 2.6~kpc. The assumption receives support from the molecular line study towards this region which is discussed in a later section (Sect. 3.3). In the  24~$\mu$m image, S1 and S2 are bright and saturated towards the central regions. S1 is also identified as IRAS~16164--4929
indicated by a $\times$ symbol in the 8~$\mu$m map. In addition to S1 and S2, multiple point sources are also observed in the 24~$\mu$m map towards the IRDC elongation. A bright 24~$\mu$m source, associated with IRAS~16161--4931 is located towards the south-west of the IRDC filament. We discuss more about this source in Sect. 3.7.  

\begin{figure}
\hspace*{-1cm}
\centering
\includegraphics[scale=0.26]{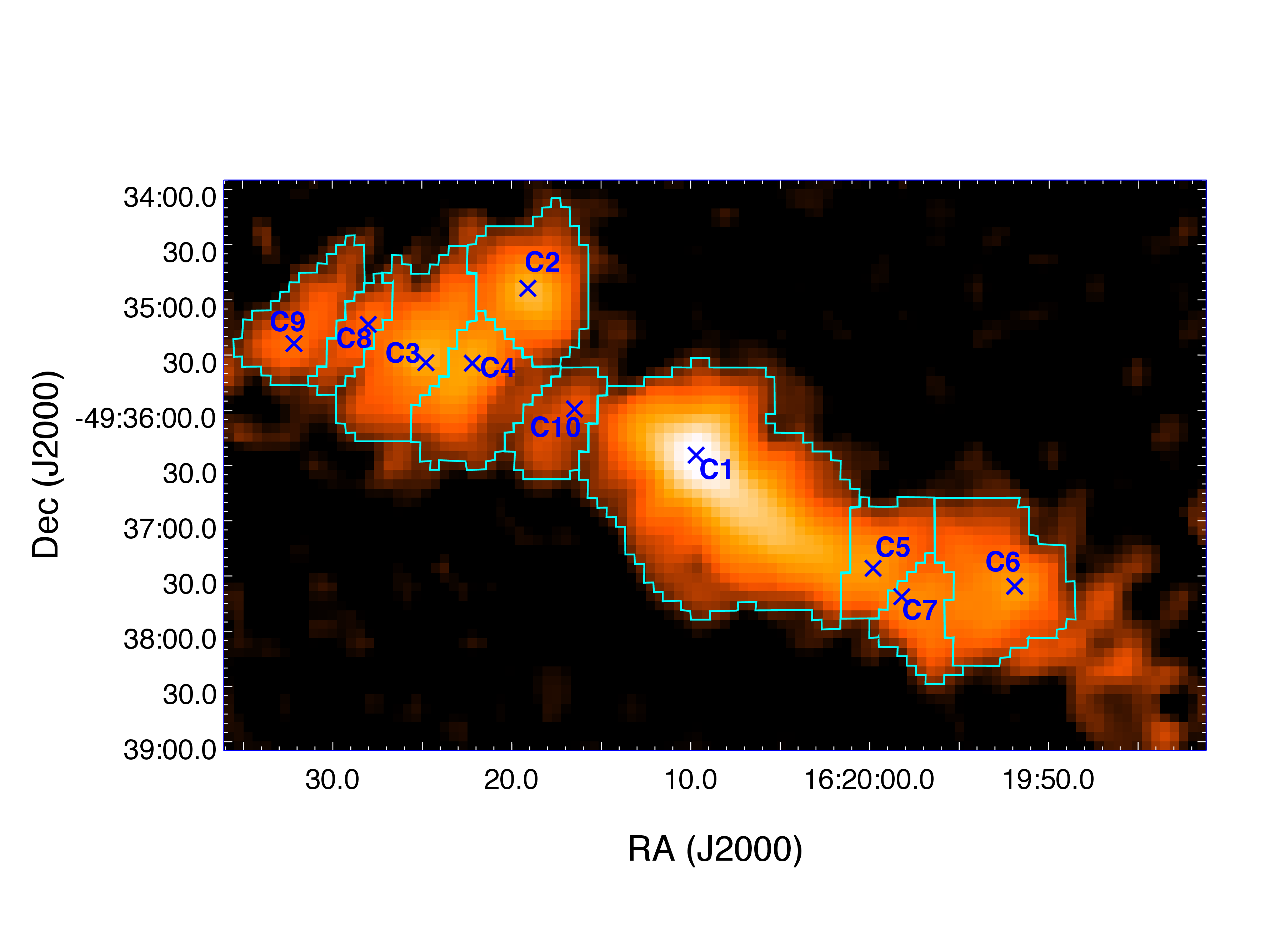}
\caption{Distribution of cold dust emission towards the G333.73 region at 1.2~mm from SEST-SIMBA. The apertures of 10 millimeter clumps identified using the FellWalker algorithm are marked in cyan and the corresponding peak positions are denoted by $\times$ points.}
\label{simba_cl}
\vspace*{0.3cm}
\end{figure}

\par The mid-infrared emission is mostly ascribed to small dust grains and could have contributions from (i) thermal emission from warm dust in the circumstellar envelope heated by direct stellar radiation, (ii) heating of dust due to Lyman-$\alpha$ photons resonantly scattering in the ionised region \citep{1991MNRAS.251..584H}, and (iii) emission due to excitation of polycyclic aromatic hydrocarbons (PAHs) by UV-photons in the photodissociation regions \citep[PDRs,][]{{2011A&A...535A.128B},{2016AJ....152..146N}}. The emission in the 4.5~$\mu$m band is believed to be dominated by molecular H$_2$ and CO emission, that traces the shocked molecular gas in active protostellar outflows \citep{{2004ApJS..154..352N},{2007MNRAS.374...29D}}. As the point response functions (PRFs) of 4.5 and 3.6~$\mu$m bands are similar, we have constructed a ratio map of [4.5~$\mu$m]/[3.6~$\mu$m] to study the signatures of outflow within the region. The [4.5]/[3.6] ratio map towards S1 is presented in Fig.~\ref{mirratio}.  It has been found that the [4.5]/[3.6] ratio is  $\sim$1.5 or larger for jets and outflows whereas it is lower for stellar sources \citep[$\ll$1.5;][]{{2010ApJ...720..155T},{2013ApJ...776...29L}}. In our map, we notice excess [4.5]/[3.6] ratio towards S1, that is located $\sim$20$\arcsec$ east of the millimeter peak.  If the large [4.5]/[3.6] does trace the distribution of shocked gas from the outflow, then it is possible that G333.73 harbours a protostellar outflow (or shocks/winds).

\begin{figure*}
\vspace*{-1cm}
\hspace*{-0.6cm}
\centering

\includegraphics[scale=0.45]{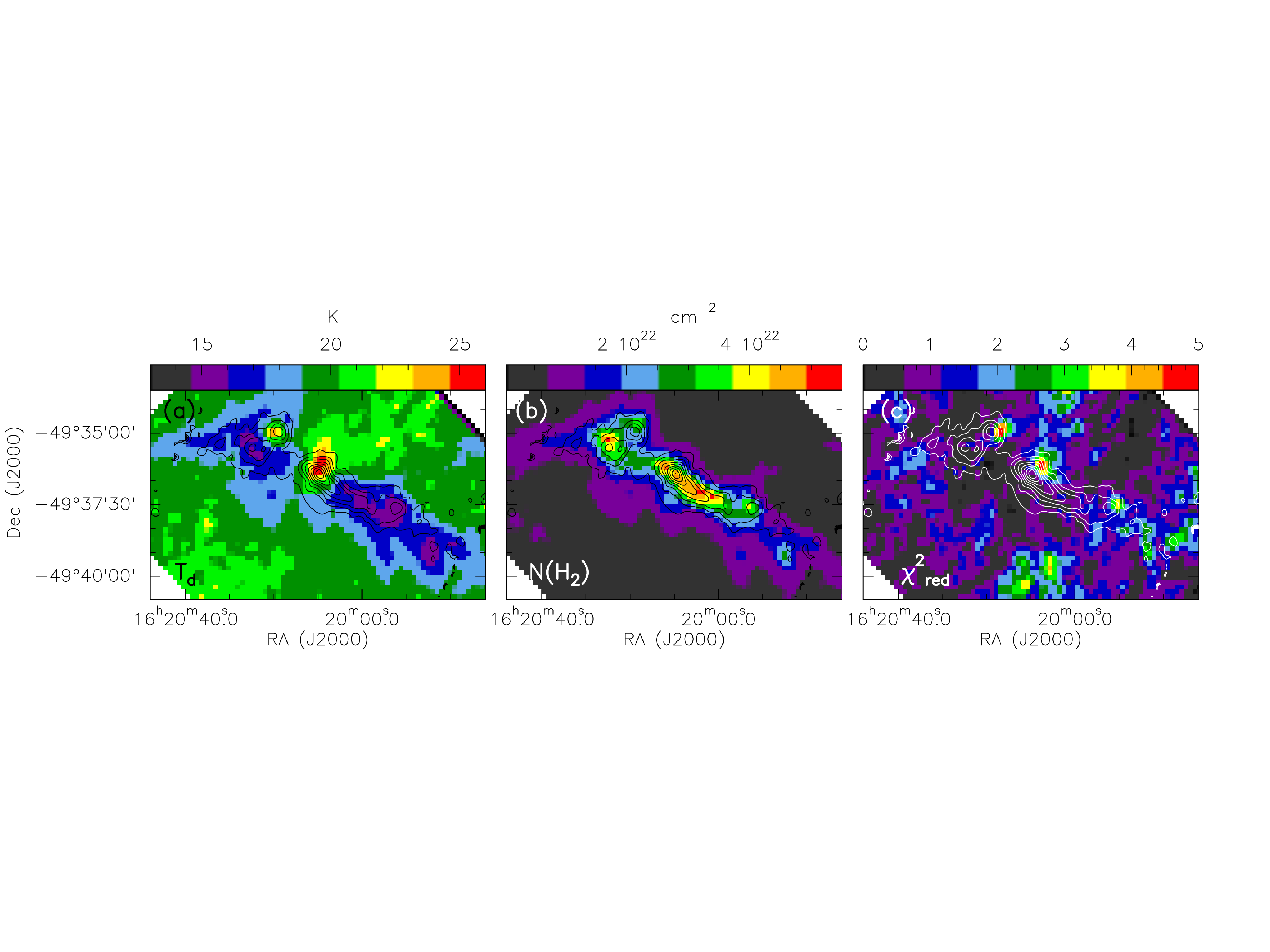}
\vspace*{-4cm}
\caption{(a) Dust temperature (T$_d$) (b) column density (N(H$_2$))  and (c) reduced chi-square ($\chi^2_\textrm{red}$) maps of G333.73 region obtained from the modified blackbody fit towards individual pixels overlaid with 1.2~mm cold dust contours. The contour levels are from 75 mJy/beam to 1200 mJy/beam in steps of 100~mJy/beam.}
\label{tempcd}
\end{figure*}

\tabcolsep=0.08cm
\begin{table*}
\footnotesize
\caption{Molecular clumps identified in this region.}
\begin{center}
\hspace*{-0.5cm}
\begin{tabular}{ c c c c c c c c c }\hline \hline
\setlength{\tabcolsep}{1pt}
Clump & $\rm{\alpha_{J2000}}$ & $\rm{\delta_{J2000}}$ &Area&Temperature &Column density& $\chi_{red}^2$&Mass &$\Sigma$\\
& $(^{h~m~s})$ &$(^{\degr~\arcmin~\arcsec})$&(pc$^{2}$) &(K)&(10$^{22}$ cm$^{-2}$)&&(M$_\odot$)&(g cm$^{-2}$)\\
\hline\\
C1 &16:20:09.693&$-49$:36:24.99&2.3 & $20.8\pm 1.7$ & $2.8\pm 0.6$ &2.4&1530&0.1\\
C2 &16:20:19.091&$-49$:34:53.54&0.6  & $22.3\pm 1.9$  &$1.9\pm 0.6$  &4.0&266&0.1\\
C3 &16:20:24.837&$-49$:35:34.14&0.6& $16.1\pm 3.2$  &$3.2\pm 0.9$  &4.8&456&0.2\\
C4 &16:20:22.226&$-49$:35:34.16&0.6& $16.4\pm 1.4$  &$2.9\pm 0.9$ &5.8&420&0.2\\
C5 &16:20:00.287&$-49$:37:25.91&0.4& $15.1\pm 1.1$ &$3.7\pm 0.9$  &4.2&350&0.2\\
C6 &16:19:51.925&$-49$:37:36.01&0.7 & $14.3\pm 1.2$ &$3.7\pm 0.1$  &7.2&612&0.2\\
C7 &16:19:58.196&$-49$:37:41.13&0.3& $15.6\pm 1.4$ &$2.5\pm 0.8$  &6.1&180&0.1\\
C8 &16:20:27.990&$-49$:35:13.52&0.2& $15.9\pm 1.5$&$1.8\pm 0.6$&7.1&87&0.1\\
C9 &16:20:32.144&$-49$:35:23.76&0.5& $16.4\pm 1.4$&$3.8\pm 1.1$&5.4&451&0.2\\
C10&16:20:16.481&$-49$:35:59.42&0.3& $15.1\pm 0.7$&$4.9\pm 0.9$&1.5&353&0.3\\
\hline
\end{tabular}
\label{cltb}
\end{center}
\end{table*}

\subsection{Properties of cold dust emission}

The cold dust emission towards the IRDC is examined using far-infrared and submillimeter maps at seven wavelength bands (70, 160, 250, 350, 500, 850~$\mu$m and 1.2~mm). The wavelength dependent variation of emission towards this region is apparent from the $\textit{Herschel}$ and APEX+Planck maps ($70-850~\mu$m), presented in Fig.~\ref{submm}. The cold dust emission maps exhibit a clumpy structure of the IRDC that spans a region $9\arcmin.5\times2\arcmin.0$, which corresponds to 7.2~pc~$\times$~1.5~pc. The 70~$\mu$m map is morphologically similar to that of the 24~$\mu$m warm dust emission. Unlike the longer wavelength emission maps, the resemblance of the 70~$\mu$m emission to the warm dust emission at 24~$\mu$m can be attributed to the fact that apart from the thermal emission due to cold dust, the emission at 70~$\mu$m could also have contribution from very small dust grains \citep[VSGs,][]{2013A&A...554A..42R}. The regions S1 and S2 appear to be connected by cold dust filaments as perceived from the longer wavelength emission maps. 

\par A visual inspection of the 1.2~mm map shows that there are cold dust peaks towards this region in addition to the eight dust clumps identified by \citet{2006A&A...447..221B}. We have used the FellWalker algorithm \citep{2015A&C....10...22B} to identify clumps in this region. The FellWalker algorithm uncouples peaks based on local gradients, assigning each pixel to the peak that the local gradient point towards. We used a detection threshold of 5$\sigma$ for identification of peaks and all the pixels outside the 5$\sigma$ contour are considered as noisy. We also set the parameter MinPix as 5 which excluded all clumps with pixels less than 5. Using this algorithm, we detected 10 clumps in G333.73. The peak positions of the clumps are shown in Fig.~\ref{simba_cl}. These clumps are labeled as C1, C2...C10 in order of their decreasing peak brightness. Overplotted on the image are the apertures corresponding to the area covered by each clump using the FellWalker algorithm. 

In order to characterise the individual clumps that can be regarded as sites of local star formation, we have constructed their spectral energy distributions (SEDs). This is achieved by integrating the flux densities within the clump apertures for the wavelengths: $70$~$\mu$m to 1.2~mm.  An average sky background, estimated from a nearby field that is $\sim3.5'$ away (centred at $\alpha_{J2000} =16^h$ 19$^m$51.36$^s$,$\delta_{J2000}=-49^\circ$~34$\arcmin$~53.5$\arcsec$), and devoid of bright diffuse emission is appropriately subtracted to account for the zero offsets at each wavelength. We fitted the flux densities ($F_{\nu}$) of the clumps using a modified blackbody function of the form \citep{{1987ApJ...316..258G},{1990MNRAS.244..458W}}:

\begin{equation}
F_{\nu}=\Omega B_{\nu}(T_\textrm{d})(1-e^{-\tau_\nu})
\label{bbody}
\end{equation}

\noindent where

\begin{equation}
\tau_\nu=\mu\  m_\textrm{H} \, \kappa_{\nu} \, N(\textrm{H}_2)
\label{tau}
\end{equation}

\noindent Here, $\Omega$ is the solid angle subtended by the clump, $B_{\nu}(T_\textrm{d})$ is the blackbody function at dust temperature $T_\textrm{d}$, $\mu$ is the mean weight of molecular gas taken to be 2.86 assuming that the gas is 70$\%$ molecular hydrogen by mass \citep{2010A&A...518L..92W}, $m_\textrm{H}$ is the mass of hydrogen atom, $\kappa_\nu$ is the dust opacity and $N(\textrm{H}_2)$ is the molecular hydrogen column density. The dust opacity is estimated using the expression \citep{2010A&A...518L..92W}, 

\begin{equation}
\kappa_{\nu}=0.1\left(\frac{\nu}{1000\,\rm{GHz}}\right)^{\beta} 
\label{opacity}
\end{equation}

\begin{figure*}
\hspace*{-0.6cm}
\centering
\includegraphics[scale=0.8]{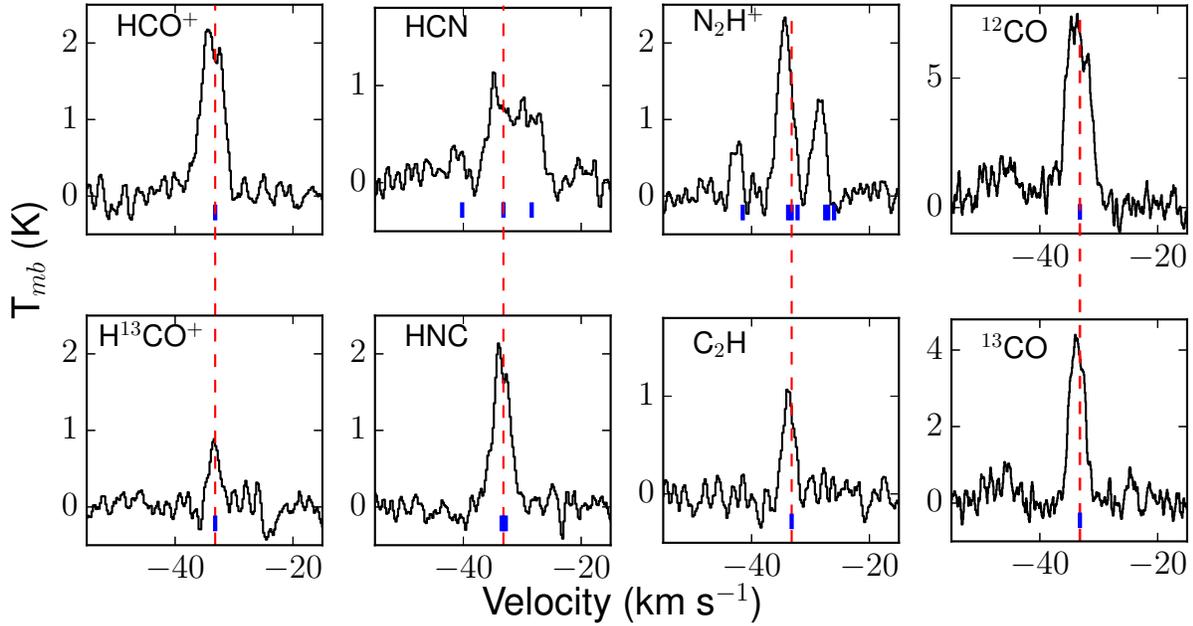}
\caption{The spectra of 8 molecular lines towards the HCO$^+$ peak position smoothed to a velocity smoothing of 0.4~km\,s$^{-1}$. The LSR velocity of the cloud is estimated from H$^{13}$CO$^+$ line as $-33.2$~km\,s$^{-1}$, indicated by a red dashed line in all the six panels. The hyperfine components of each species are marked as blue lines in individual panels. The velocities of hyperfine components are estimated assuming an LSR velocity of $-33.2$~km\,s$^{-1}$.}
\label{molspec}
\end{figure*}

\noindent where $\nu$ is the frequency and $\beta$ is the dust emissivity index. We have assumed $\beta$ = 2 in our analysis \citep{{2012A&A...542A..10A},{2013A&A...554A..42R}}. The best fits were obtained using non-linear least squares Marquardt-Levenberg algorithm, considering $T_d$ and $N(\textrm{H}_2)$ as free parameters. We have assumed a flux density uncertainty of 15$\%$ in all bands \citep{{2006A&A...447..221B},{2009A&A...504..415S},{2013A&A...551A..98L}}. We find that the fits that include the 70~$\mu$m show larger $\chi_{red}^2$ (upto a factor of 3) as well as larger errors in the parameters  (upto 60\%) when compared to fits carried out by excluding the 70~$\mu$m flux densities. This is evident from the fits to the SEDs displayed in Appendix~A. It is evident that the  70~$\mu$m point exhibits excess emission. Such excess has been observed in other star forming clouds  and has been attributed to the contribution from transiently heated very small grains \citep[e.g.,][]{{2009ApJ...696..676S},{2010ApJ...724L..44C},{2013A&A...554A..42R}} and its inclusion could overestimate the dust temperature. We proceed with the parameters of fits that exclude the 70~$\mu$m emission as this characterises the cold dust in the IRDC. The values of the derived parameters for all the clumps are listed in Table~\ref{cltb}. We also note that the ground-based SEST-SIMBA observations failed to pick up large scale diffuse emission at low flux levels owing to poor sensitivity. Clump 10, being the faintest of all the clumps, has relatively lower flux at 1.2~mm compared to the other bands (Fig.~\ref{clsed}). We have, therefore, excluded this 1.2~mm data point from the SED fit in order to get more robust estimate of parameters for this clump.

\par The temperature in the clumps lie in the range: $14.3 - 22.3$~K whereas the column density values lie between $1.8-4.9\times$10$^{22}$~cm$^{-2}$. Clump C2 exhibits highest dust temperature whereas the clump C10 possesses the highest column density. Note that these estimates represent average values over the entire clump. We have also used the column densities of the clumps to estimate their masses ($M_c$), using the following expression:

\begin{equation}
M_c = N(\textrm{H}_2)\,\mu\,m_\textrm{H}\,A
\end{equation}

\noindent Here $A$ represents the physical area of the clump. The clump masses lie in the range $87-1530$~M$_\odot$. The total cloud mass is estimated to be $\sim$4700~M$_\odot$. This is nearly $\sim5$ times larger than the 992~M$_\odot$ obtained by \citet{2006A&A...447..221B}. This difference could be attributed to the following: (i) our estimate of cloud mass is based on the modified blackbody fits using 6 far-infrared wavelength bands unlike the latter obtained from only the 1.2~mm map, and (ii)  \citet{2006A&A...447..221B} used a value of dust opacity $\kappa_\nu\sim1$~cm$^2$g$^{-1}$ at 1.2~mm, whereas we have used a different form of dust opacity law whose value depends on $\beta$. Note that $\beta=2$ leads to  $\kappa_\nu\sim0.6$~cm$^2$g$^{-1}$ at 1.2~mm. 

\par We have also estimated the surface density ($\Sigma$) of the individual clumps, defined as $M_c/A$, a parameter that can be used to probe massive star formation in the clumps. According to \citet{2008Natur.451.1082K}, clouds with a minimum surface density of $\Sigma \sim$1~g~cm$^{-2}$ would be able to form massive stars by suppressing fragmentation. The surface density values for clumps in G333.73 are listed in Table~\ref{cltb}. The values of $\Sigma$ for the ten clumps lie in the range $0.1-0.3$~g~cm$^{-2}$. The maximum $\Sigma$ is observed towards clump C10 that also displays the largest column density among clumps. Observational studies towards a large sample of massive star forming cores, such as those by \cite{2010A&A...517A..66L}, \cite{2010A&A...520A.102M} and \cite{2013A&A...556A..16G} have shown that massive star forming cores possess lower surface densities of the order of $\sim$0.2~g~cm$^{-2}$. According to this latter gauge, six of our clumps have the potential to form massive stars. This is substantiated by our assertion that the surface density value of a clump represents a sort of average, and  the actual surface density could be higher near the peak emission or dense core considering that the size of clumps are large ($>0.6$~pc).

\subsubsection{Maps of Column density and Dust Temperature}

\par In addition to the clump SEDs, we have constructed the line-of-sight averaged molecular hydrogen column density and dust temperature maps of this region with an intention to understand the small scale variations across the IRDC in addition to comparing this with molecular line emission maps. The maps are created by carrying out a pixel-to-pixel greybody fit in the selected wavelength regime (160~$\mu$m-1.2~mm) using the equations discussed earlier. If we consider all the wavelengths, the  resolution of the map is limited by emission at the wavelength that has the lowest resolution, i.e. 36.4$\arcsec$ at 500~$\mu$m. As the longer wavelength data is well sampled, we prefer to construct higher resolution maps. To achieve this, we excluded the 500~$\mu$m image from the analysis. The remaining maps at 160, 250, 350, 850~$\mu$m and 1.2~mm are  convolved and regridded to the resolution (25$\arcsec$) and pixel size (10$\arcsec$) of the 350~$\mu$m image. As the sensitivity of the 1.2~mm map is lower,  we are unable to sample the diffuse emission extending beyond the high density regions. For these pixels, the values of $\chi^2_\textrm{red}$ are larger. To obtain better fits, the pixels with $\chi^2_\textrm{red}>$2  due to noisy 1.2~mm emission, were fitted anew by excluding the 1.2~mm data point from the SED fit.

\begin{figure}
\centering
\includegraphics[scale=0.25]{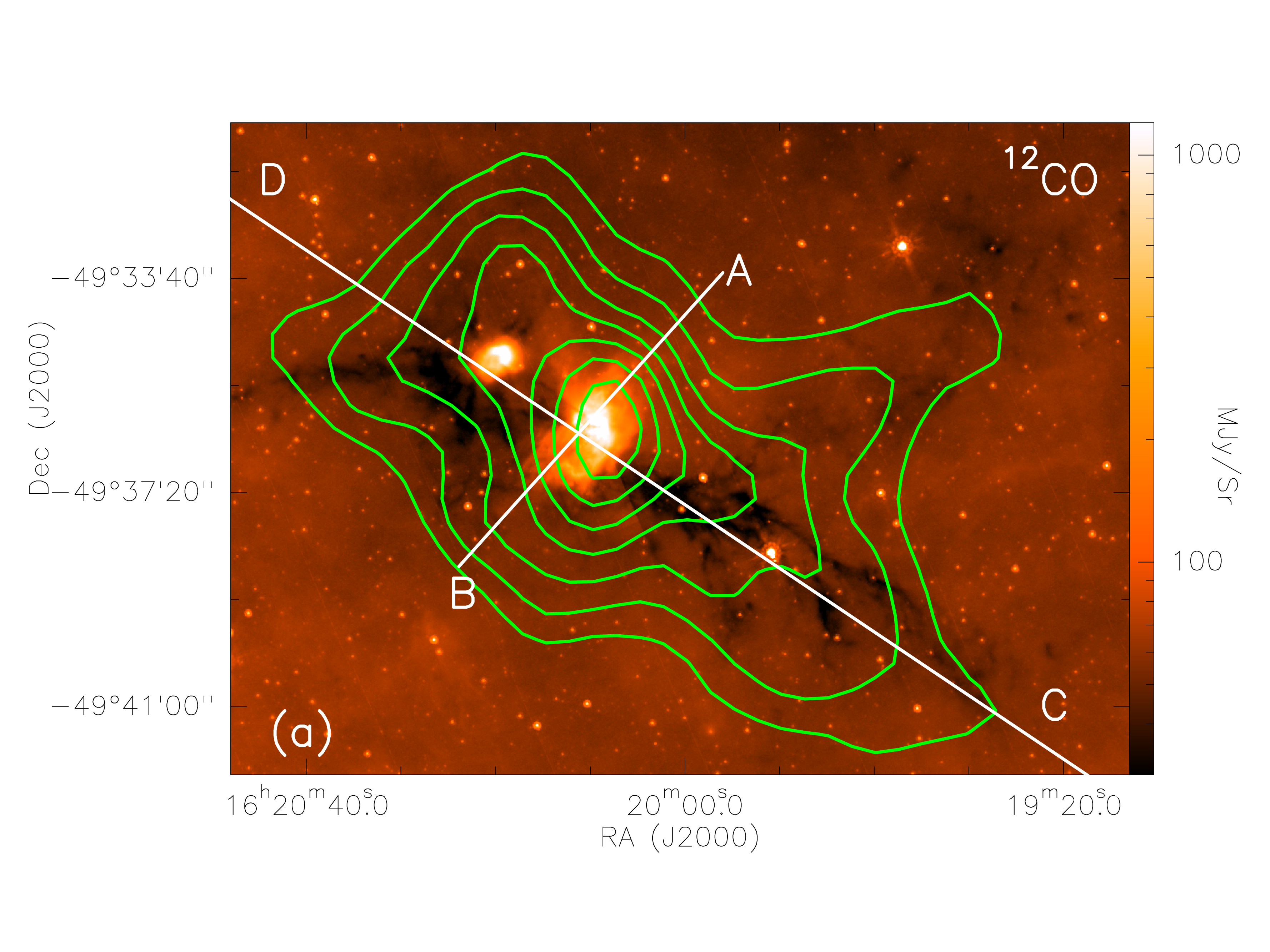}\vspace*{-0.7cm}\quad \includegraphics[scale=0.25]{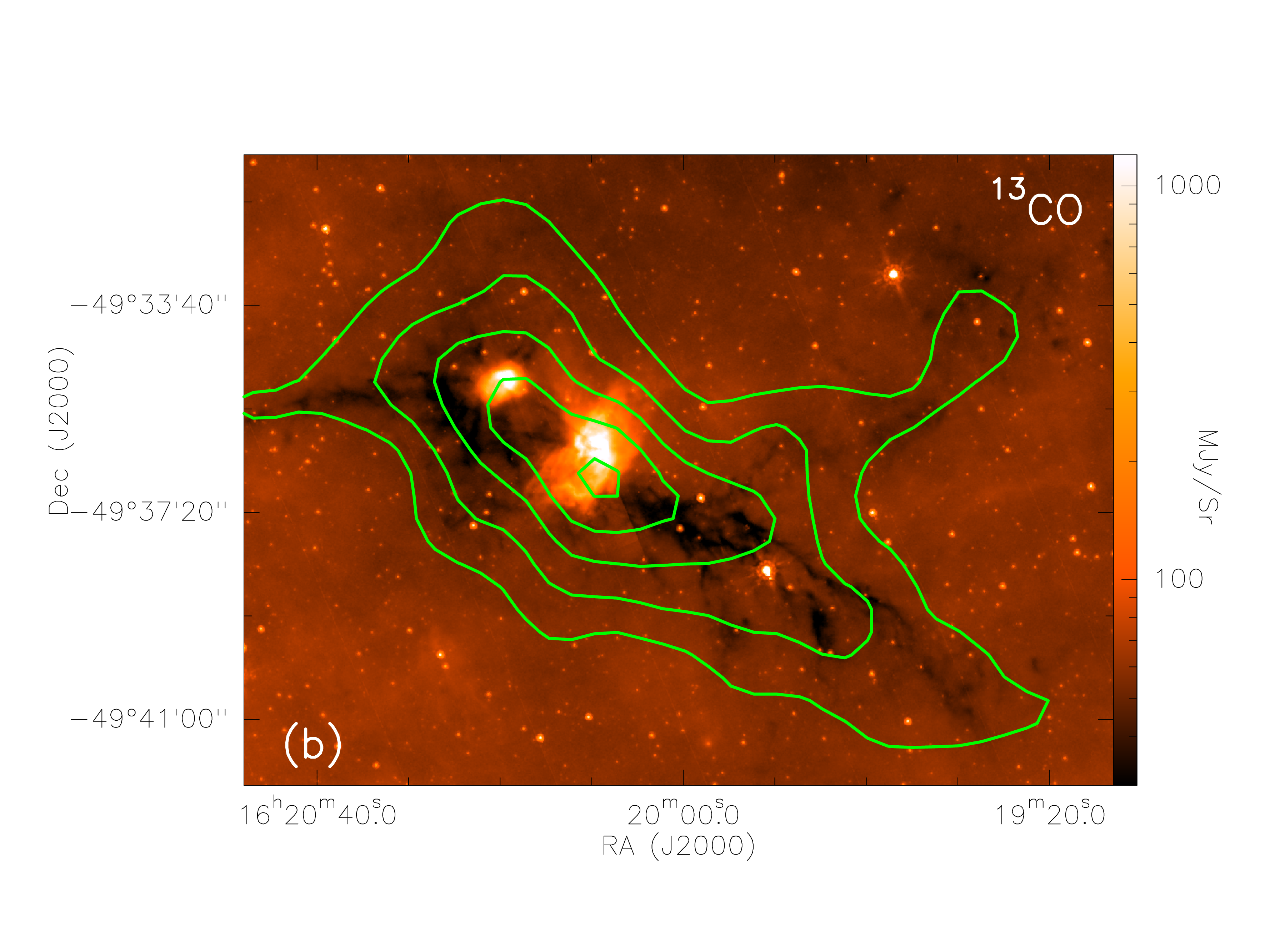}
\caption{$\textit{Spitzer}$ 8~$\mu$m map overlaid with (a)$^{12}$CO and (b) $^{13}$CO contours. Contour levels for (a) are 13 to 35~K\,km\,s$^{-1}$ in steps of 3.1~K\,km\,s$^{-1}$ and for (b), 4.5 to 15~K\,km\,s$^{-1}$ in steps of 2.1~K\,km\,s$^{-1}$. Lines AB and CD are the lines along which PV diagrams are constructed.}
\label{comaps}
\end{figure}
 
\par The dust temperature, column density and reduced chi-square ($\chi^2_\textrm{red}$) maps are presented in Fig.~\ref{tempcd}. For further analysis, we have considered pixels within the 5$\sigma$ contour of 1.2~mm map. 
The peak column density is $5.5\times10^{22}$~cm$^{-2}$ whereas the mean column density is $2.4\times 10^{22}$~cm$^{-2}$. The column density distribution is clumpy in nature exhibiting multiple peaks. The temperature within the IRDC ranges from $14.8-25.4$~K with a mean value of 18~K. The temperature map is peaked towards the location of S1. The temperature map also reveals an additional peak that matches with the location of S2. These temperature peaks can be understood based on the morphology of 160~$\mu$m emission. The 160~$\mu$m emission is the shortest wavelength used in the SED construction and traces the warmest dust emission components. Hence, pixels with significant emission at 160~$\mu$m is weighted by the correspodning flux density leading to a higher dust temperature that signifies higher levels of star formation activity here. The low values of dust temperature are observed towards the dark filaments in the 8~$\mu$m map.

\subsection{Molecular line emission from G333.73}
The kinematics and chemistry of IRDCs can be investigated using molecular line emission. For the IRDC G333.73, we use molecular line data from the MALT90 pilot survey that covers a region of size $\sim4'$ centred on S1 in clump C1. Six molecular species have been detected towards this region: HCO$^+$, H$^{13}$CO$^+$, HCN, HNC, N$_2$H$^+$ and C$_2$H. The spectra of these molecules at the location of emission peak of HCO$^+$ is shown in Fig.~\ref{molspec}. The LSR velocity of the region (hence IRDC) is estimated using a single transition H$^{13}$CO$^+$ assuming the line to be optically thin. We have fitted a single Gaussian profile to the spectrum and determined the  LSR velocity as $-33.2$~km\,s$^{-1}$. This is consistent with the LSR velocity of $-33.0$~km\,s$^{-1}$ estimated from the CS(2-1) line \citep{1996A&AS..115...81B}. The hyperfine components of HCN and N$_2$H$^+$ molecules are clearly discerned in the velocity profiles. HCN has 3 hyperfine components which are well separated (+5 and $-7$~km\,s$^{-1}$ respectively). N$_2$H$^+$ has 7 hyperfine components. The profiles of HCO$^+$ and HNC lines exhibit a blue asymmetric feature characterized by self-absorption dips in the lines, with relatively strong blue peaks with respect to red peaks. We explore the likely origin of the asymmetric profile in the next section.

\par As the MALT90 survey has limited coverage, we are unable to sample the molecular gas kinematics of the entire IRDC filament. We, therefore, utilise the $^{12}$CO and $^{13}$CO data from ThrUMMS survey for this purpose. The CO spectra towards the peak position of HCO$^+$ emission are shown in Fig.~\ref{molspec} (last column). These spectra exhibit blue asymmetric profiles similar to those of HCO$^+$ and HNC. The distribution of CO emission with respect to warm dust emission is shown in Fig.~\ref{comaps}. From the maps, it is evident that the CO emission extends well beyond the apparently dark filamentary structure. This is in accordance with expectations as the $^{12}$CO and $^{13}$CO lines also sample the diffuse envelope, being low density tracers.

\subsubsection{Blue asymmetry of HCO$^+$ and HNC profiles}
The HCO$^+$ line is optically thick based on the expected ratio of line intensties of HCO$^+$ and H$^{13}$CO$^+$. Similarly, we proceed with the supposition that HNC is optically thick. Although both display a double peaked structure, HCO$^+$ is a single transition line whereas HNC has three hyperfine components  within 0.5~km\,s$^{-1}$, marked in Fig~\ref{molspec}. These lines are considered as good infall and outflow tracers. An examination of the HCO$^+$ and HNC velocity profiles show that they exhibit significant blue asymmetry in their profiles indicative of infall in this region \citep{{2012A&A...540A.104M},{2016ApJS..225...21J}}. Blue asymmetry could also arise from rotation and outflow \citep[e.g.,][]{2004MNRAS.352.1365R}. The velocity of the absorption dip agrees well with that of the LSR velocity estimated from the H$^{13}$CO$^+$ line. The velocities of the blue shifted peaks of the HCO$^+$ and HNC lines relative to the LSR velocity are $-0.9$~km\,s$^{-1}$ and $-0.8$~km\,s$^{-1}$, respectively. Similarly, the velocities of the red shifted peaks with respect to the LSR velocity are $1.0$~km\,s$^{-1}$ and $0.8$~km\,s$^{-1}$, respectively. These values indicate that the red and blue peaks are quite symmetric with respect to the LSR velocity. 
We next scrutinise the intensities, and to quantify the blue-skewed profile, we have used the asymmetry parameter $\delta$V. This is defined as the difference between the peak velocities of optically thick line, $V_\textrm{thick}$ (of HCO$^+$/HNC in our case), and optically thin line, $V_\textrm{thin}$ (of H$^{13}$CO$^+$), divided by the FWHM of the optically thin line represented as $\Delta V$ \citep{2013RAA....13...28Y}:
$$\delta\mathrm{V} =\frac{V_\textrm{thick}-V_\textrm{thin}}{\Delta V_\textrm{thin}}$$ 
Using V$_\textrm{thin}$ as $-33.2$~km\,s$^{-1}$ and $\Delta \textrm{V}_\textrm{thin}$ as 2.2~km\,s$^{-1}$ from the Gaussian fit to the H$^{13}$CO$^+$ profile, we obtain $\delta V$ as $-0.4$ for both the lines. This is characterised as a blue profile according to the criterion of  \citet{1997ApJ...489..719M}, who use $\delta V < -0.25$ to assign a profile as blue. 
\begin{figure}
\centering
\hspace*{-0.6cm}
\includegraphics[scale=0.55]{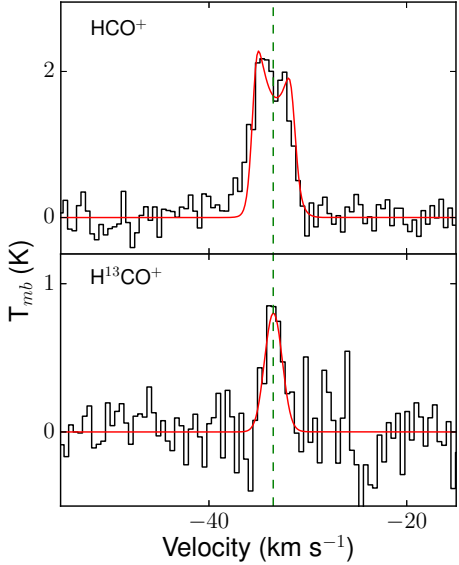}
\caption{Spectrum towards the peak of HCO$^+$ and H$^{13}$CO$^+$ emission. The red line represents the best-fit model using the two-layer infall process for optically thick HCO$^+$ line and its optically thin isotopologue H$^{13}$CO$^+$ line. The dashed line (green) indicates the LSR velocity of $-33.2$~km\,s$^{-1}$.}
\vspace*{0.3cm}
\label{hcopcassis}
\end{figure}

\begin{table}
\scriptsize
\caption{Best fit parameters for the two layer infall model.}
\begin{center}
\setlength{\tabcolsep}{1.8pt}
\hspace*{-0.2cm}
\begin{tabular}{l c c } \hline \hline \\
Parameter     &Component 1&Component 2 \\
\hline\\
Column density ($10^{14}$~cm$^{-2}$)&$16.1\pm2.7$&$0.2\pm0.03$\\
T$_\textrm{ex}$ (K)&$31.1\pm0.5$&$7.1\pm0.5$\\
FWHM (km\,s$^{-1}$)&$1.9\pm0.1$&$2.8\pm0.2$\\
Size ($\arcsec$) &$23.7\pm1.4$&$31.1\pm4.1$ \\
V$_\textrm{LSR}$ (km\,s$^{-1}$)&$-33.4\pm0.1$&$-33.0\pm0.1$\\
\hline\\
\end{tabular}
\label{cassis}
\end{center}
\end{table}


\subsubsection{HCO$^+$ line profile analysis using LTE modelling}

\begin{figure}
\centering
\includegraphics[scale=0.32]{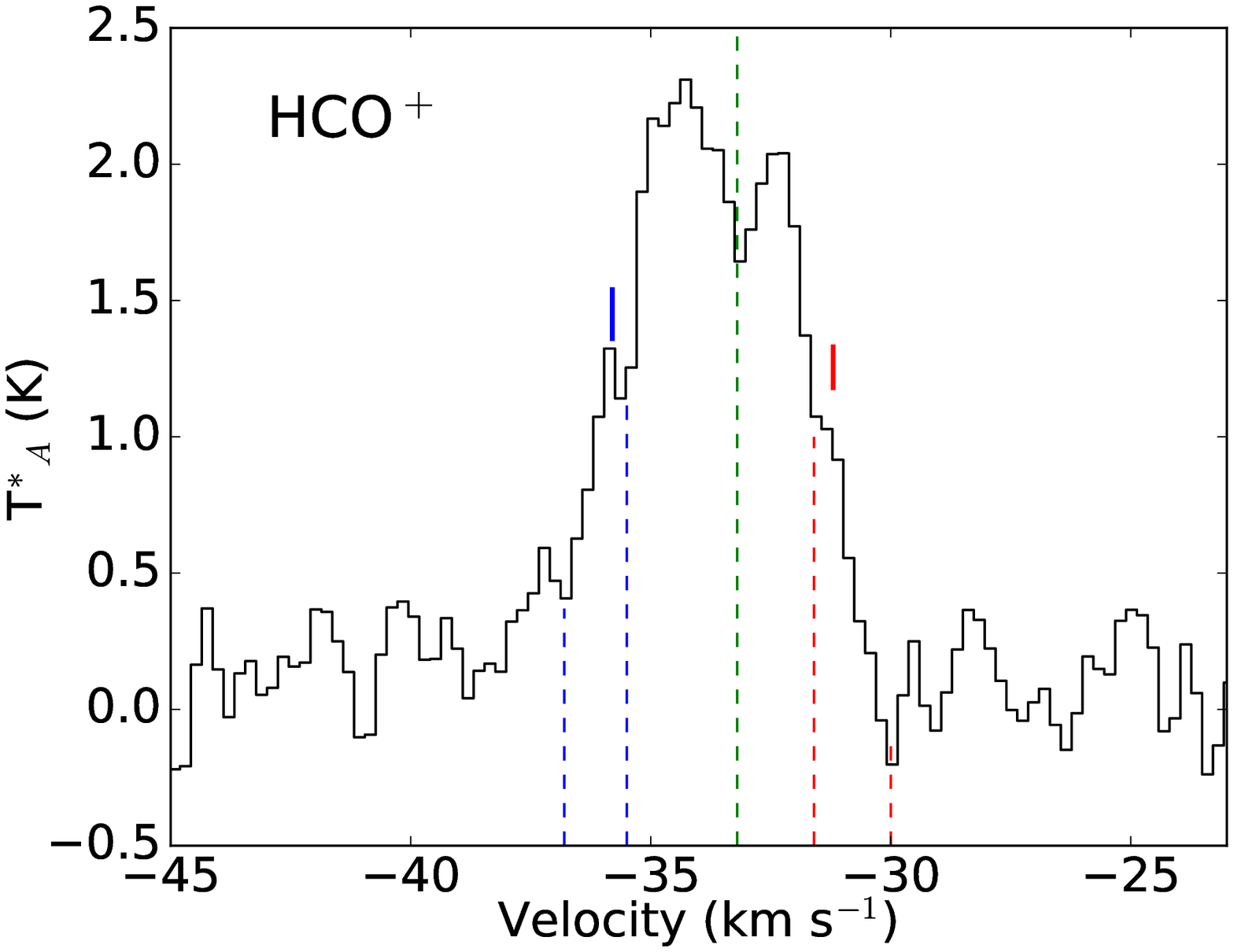}\quad \includegraphics[scale=0.32]{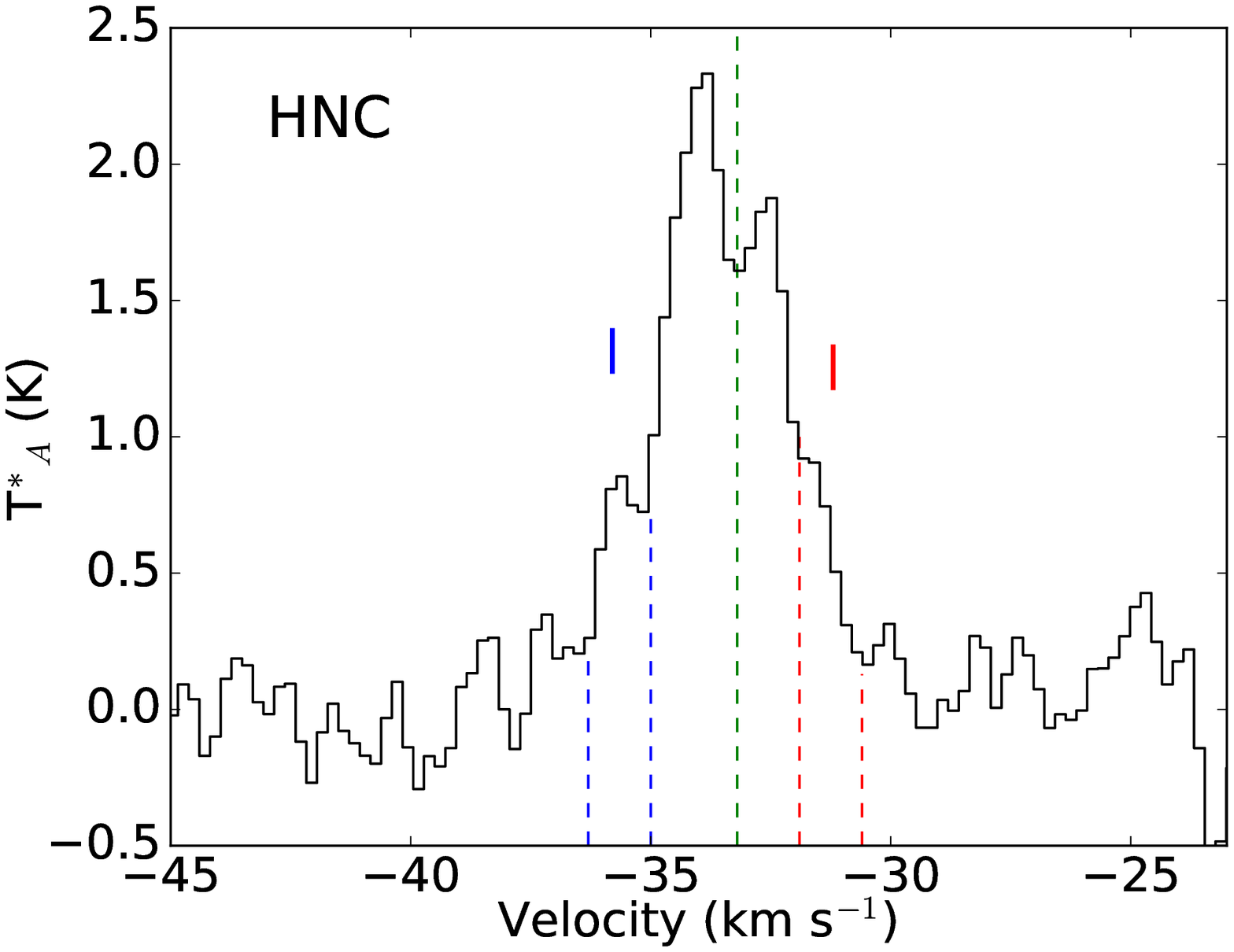}
\caption{Spectrum of HCO$^+$ and HNC molecules towards peak emission. Location of blue and red wings seen towards both the spectra are indicated with solid lines.}
\label{outflowpeak}
\end{figure}

To study the likely mechanisms responsible for the observed blue asymmetry in the HCO$^+$ line, we carried out a two-component LTE modelling in CASSIS software \citep{2011IAUS..280P.120C}. For the modelling, we considered the HCO$^+$ line as well as its isotopologue, H$^{13}$CO$^+$. The observed self-absorbed profile of HCO$^+$ line can be explained if we use a two-layer model where there is a warm emitting component and a cold absorbing component. For a better signal-to-noise ratio, we have integrated the emission within 20$\arcsec$ of the HCO$^+$ peak for both HCO$^+$ and H$^{13}$CO$^+$ lines. The best fit to the profiles are obtained by varying the parameters such as linewidth, V$_\textrm{LSR}$, excitation temperature, column density and source size. The [HCO$^+$]/[H$^{13}$CO$^+$] abundance ratio is assumed as 50 \citep{2006MNRAS.367..553P}. The fitted spectrum is shown in Fig.~\ref{hcopcassis} and the results of the radiative analysis are presented in Table~\ref{cassis}. The characteristics of the two components: a warm component with an excitation temperature of 31.1~K and column density of $1.6\times10^{15}$~cm$^{-2}$, and a cold, absorbing component with lower excitation temperature (7.1~K) and column density ($2.0\times10^{13}$~cm$^{-2}$). The velocity of the cold component is red shifted by 0.4~km\,s$^{-1}$ with respect to the warm component. This could be construed as cold molecular gas in the outer envelope receding towards the inner warmer regions and interpreted as protostellar infall. The overall blue asymmetric profile fits well using LTE modelling although we see some additional red and blue components that cannot be explained through the infall scenario alone (see Fig.~\ref{outflowpeak}). These additional peaks require multiple components suggesting the presence of small scale outflows in the region. Observations with better spatial resolution and sensitivity are essential to enhance our understanding of the profiles.

\subsubsection{Mass infall rate}

\par Considering that the blue asymmetry of HCO$^+$ and HNC lines suggest protostellar infall, the mass infall rate ($\dot{\textrm{M}}_\textrm{inf}$) of the circumstellar envelope can be estimated using the expression: $\dot{\textrm{M}}_\textrm{inf}$ = 4$\pi$\,R$^2\,$V$_\textrm{inf}\,\rho$ \citep{2010A&A...517A..66L}, where V$_\textrm{inf}$ = $V_\textrm{thin}-V_\textrm{thick}$ = $\textrm{V}_{\textrm{H}^{13}\textrm{CO}^+}-\textrm{V}_{\textrm{HCO}^+}$ is an estimate of the infall velocity, $\rho$ = M/(4/3\,$\pi$\,R$^3$) is the average clump volume density and R is the radius of the clump, calculated using the dust continuum emission. We estimate V$_\textrm{inf}$ as 0.9~km\,s$^{-1}$ and M$\sim$1530~M$_\odot$ and R$\sim$0.9~pc for clump C1 and we obtain $\dot{\textrm{M}}_\textrm{inf}$ as $4.7\times10^{-3}$~M$_\odot$\,yr$^{-1}$. This is in congruence with that estimated towards other infall candidates. For example \citet{2010A&A...517A..66L} obtained infall rates ranging from $10^{-3}-10^{-1}$~M$_\odot$\,yr$^{-1}$ for a sample of high mass star forming clumps. 
\citet{2015MNRAS.450.1926H} inferred median mass infall rates of $\mathrm{7-8\times10^{-3}}$~M$_\odot$\,yr$^{-1}$ for pre-stellar, proto-stellar and ultracompact \hii~region stages from their sample of massive star forming regions. They concluded that the infall rate is independent of the evolutionary stage. 

\begin{figure*}
\hspace*{-0.6cm}
\centering
\includegraphics[scale=0.4]{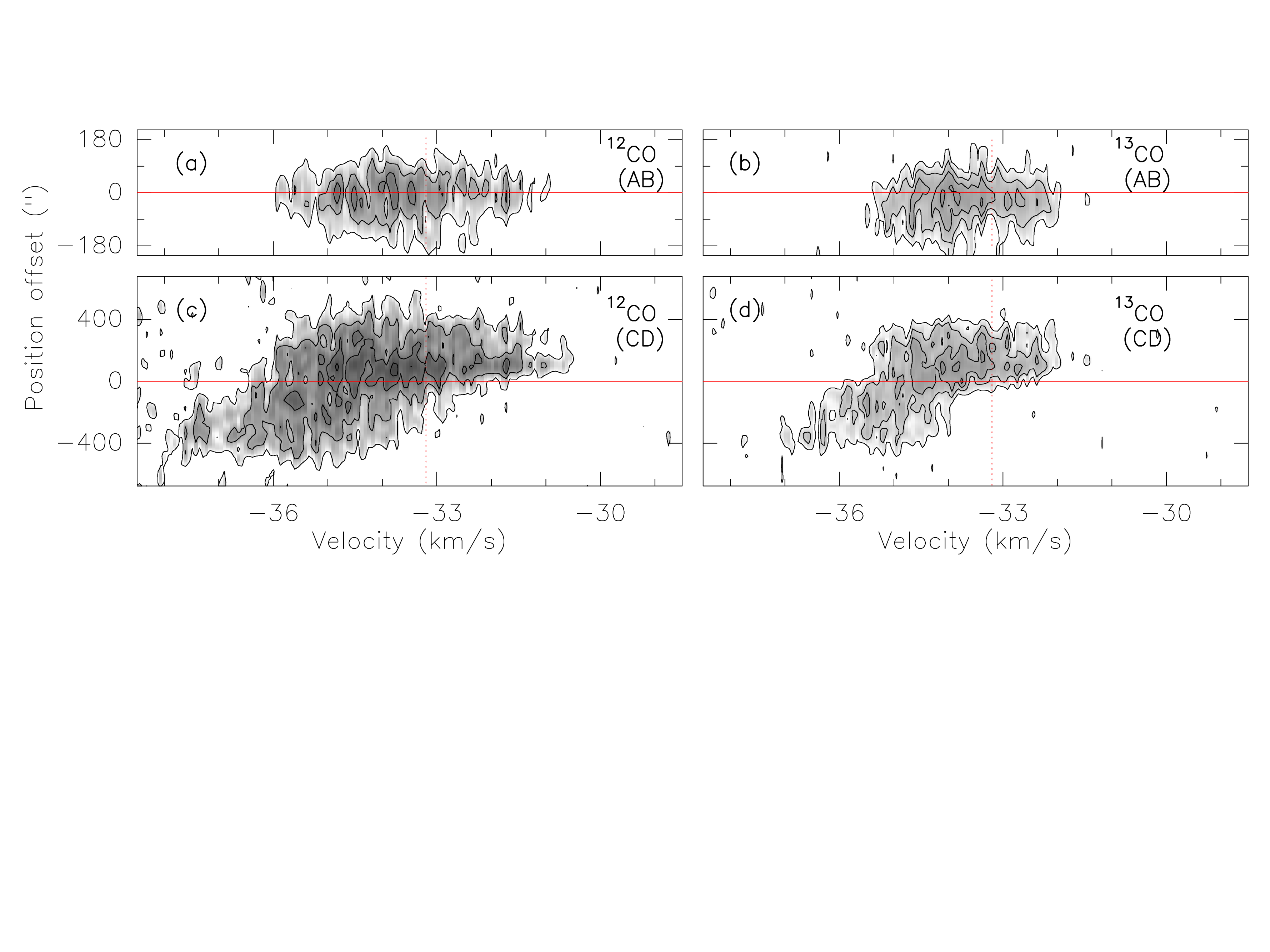}
\vspace*{-4cm}
\caption{Position-velocity diagrams of $^{12}$CO and $^{13}$CO molecules. (a) and (b) are along the cut AB shown in Fig.~\ref{comaps}(a) and (c) and (d) along CD shown in Fig.~\ref{comaps}(a). Solid line marks the location of zero offset and dashed line denotes the LSR velocity.}
\label{PV}
\end{figure*}

\begin{figure}
\hspace*{-0.2cm}
\centering
\includegraphics[scale=0.25]{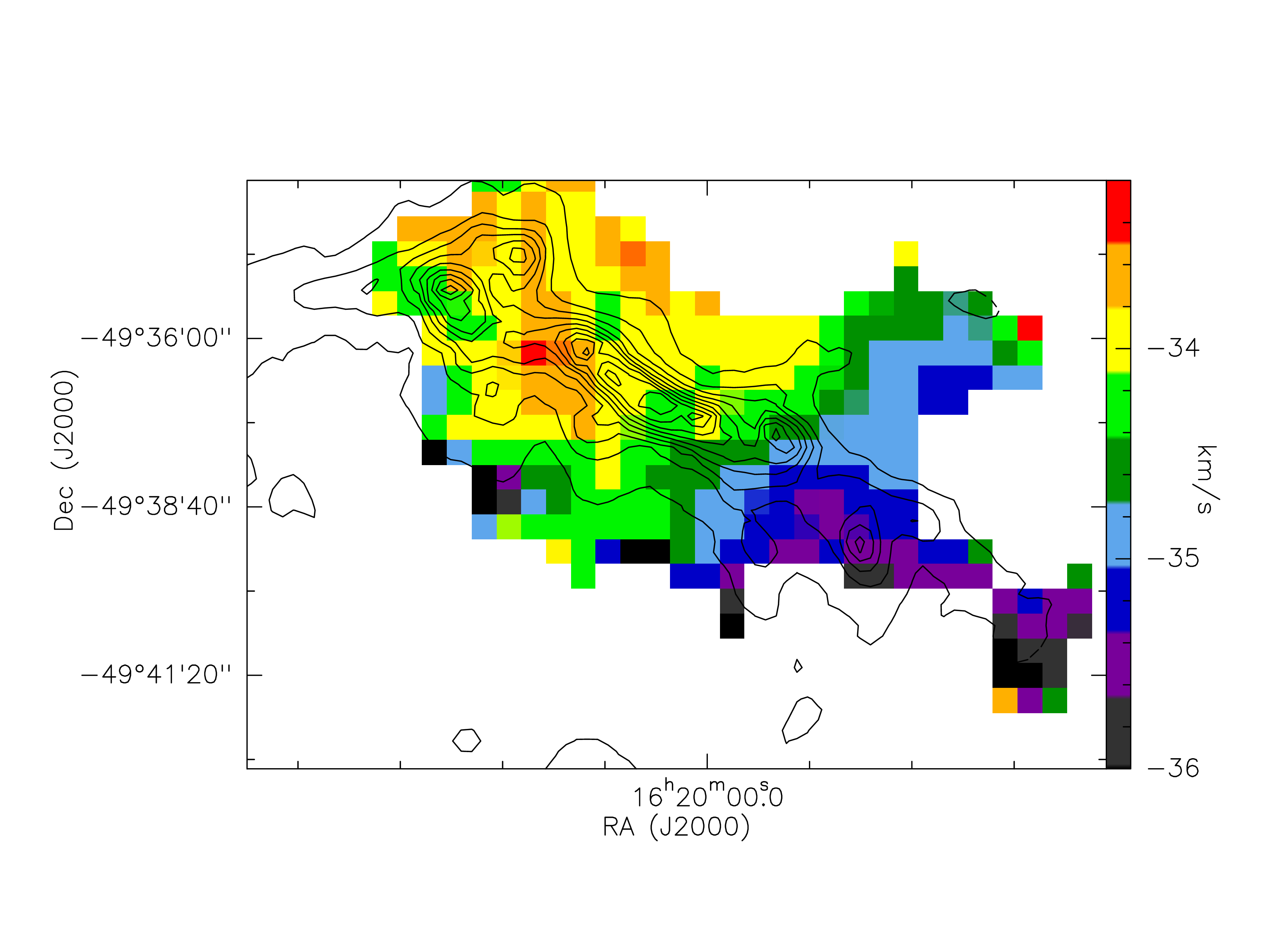}
\caption{Velocity (first moment) map of $^{13}$CO line overlaid with column density contours. The contour levels are from $1\times10^{22}$~cm$^{-2}$ to $6\times10^{22}$~cm$^{-2}$ in steps of $7.3\times10^{21}$~cm$^{-2}$.}
\label{13covel}
\vspace*{0.3cm}
\end{figure}

\subsubsection{Velocity structure of the cloud}

The position-velocity (PV) diagram serves as useful tool to understand the large scale kinematics of a region. The PV diagrams of $^{12}$CO and $^{13}$CO are constructed along two directions: (i) AB, perpendicular to the long-axis of the cloud (P.A=47.3$^\circ$) and (ii) CD, that is parallel to the IRDC long axis (P.A=134.2$^\circ$). These directions are shown in Fig.~\ref{comaps}(a). The PV plots are presented in Fig.~\ref{PV}. The zero offset in the PV diagrams corresponds to the position $\alpha_{J2000} =16^h$ 20$^m$10.7$^s$ and $\delta_{J2000}=-49^\circ$~36$\arcmin$~18.8$\arcsec$. Along AB towards the centre position, the blue and red components are clearly visible in both species, with the blue component brighter than red, suggesting infall. Along CD, we observe a velocity gradient from C to D (i.e. south-west to north-east). The overall velocity gradient is approximately 5~km\,s$^{-1}$ in magnitude, spanning a region of 10$\arcmin$ from west to east i.e. 0.7~km\,s$^{-1}$\,pc$^{-1}$. We also detect few additional substructures in velocity, evident from the $^{13}$CO velocity map shown in Fig.~\ref{13covel}. Velocity gradients of this nature have been observed in other star forming regions. For example, \citet{2017A&A...606A.133S} find a velocity gradient of 0.2~km\,s$^{-1}$\,pc$^{-1}$ in the IRDC filament G035.39-00.33 of length 6~pc. From their study towards a sample of 54 filaments in the northern Galactic plane, \citet{2016ApJS..226....9W} estimate a mean velocity gradient of 0.4~km\,s$^{-1}$\,pc$^{-1}$ towards the filaments. The systematic velocity gradient observed in the IRDC studied here could hint at the rotation and/or accretion flows along the cloud. This is explored in detail in Sect.4.4. 

\subsection{Intensity distribution of molecular gas}
\begin{figure*}
\hspace*{-0.6cm}
\centering
\includegraphics[scale=0.45]{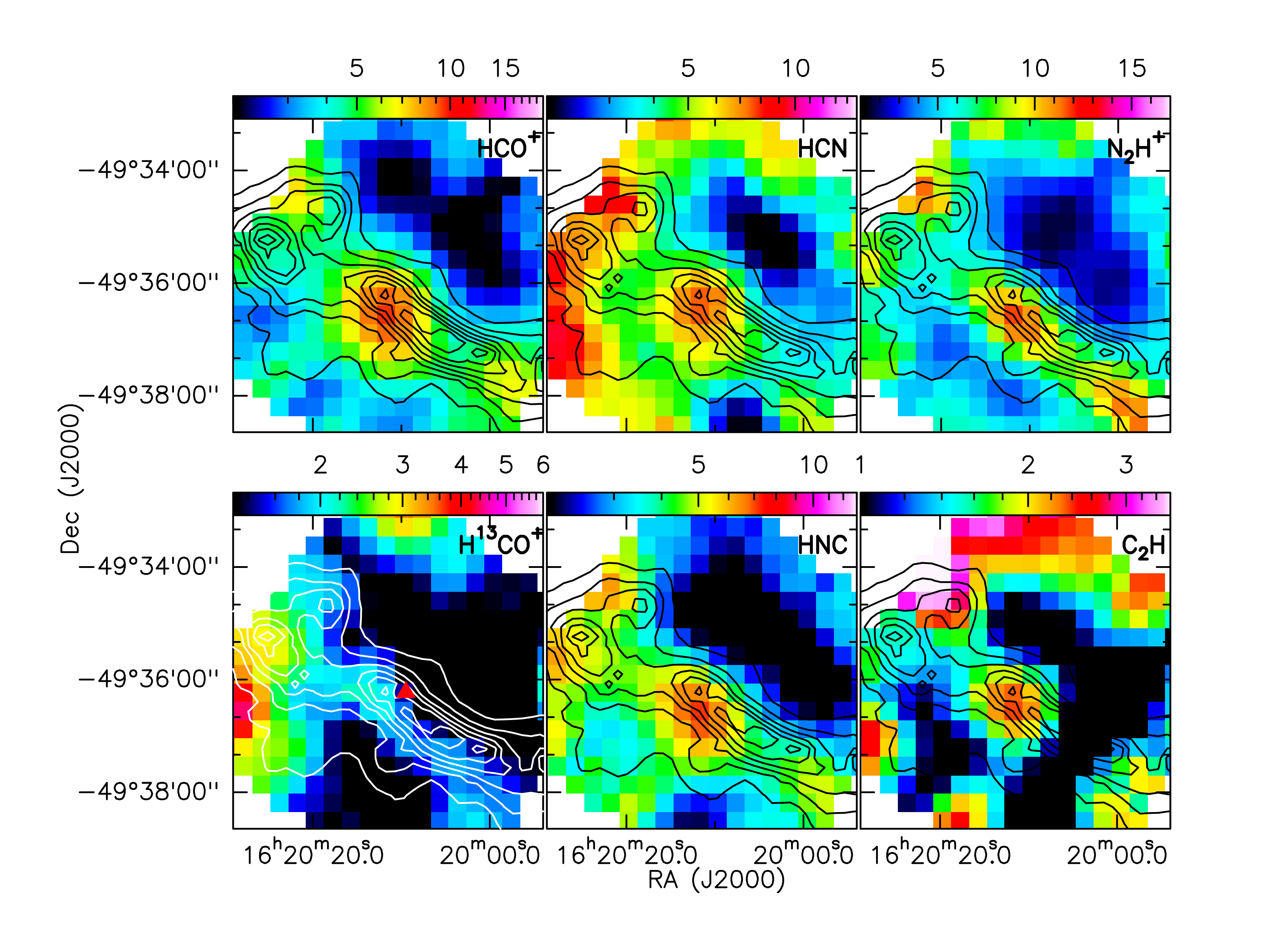}
\caption{Integrated intensity maps of 6 molecular lines (labeled in the top right corners) detected towards G333.73 overlaid with column density contours. The contour levels are from $1\times10^{22}$~cm$^{-2}$ to $6\times10^{22}$~cm$^{-2}$ in steps of $7.3\times10^{21}$~cm$^{-2}$.  We show the integrated intensity maps of HCO$^+$ (V=[$-38.5,-29.5$]~km\,s$^{-1}$, H$^{13}$CO$^+$ (V=[$-35.5,-30.7$]~km\,s$^{-1}$, HCN (V=[$-37.7,-23.9$]~km\,s$^{-1}$, HNC (V=[$-38.5,-29.0$]~km\,s$^{-1}$, N$_2$H$^+$ (V=[$-44.5,-25.2$]~km\,s$^{-1}$ and C$_2$H (V=[$-36.3,-31.5$]~km\,s$^{-1}$. The colour intensity scale is in units of K\,km\,s$^{-1}$. Triangle in panel 4 (H$^{13}$CO$^+$) corresponds to IRAS peak position.}
\label{mom0}
\end{figure*}

In this section, we examine the morphology of the molecular line emission associated with the IRDC. The distribution of CO is apparent from Fig.~\ref{comaps} while Fig.~\ref{mom0} shows the integrated intensity (zeroth moment)  maps of the six molecular species from MALT90 survey. The peak of the molecular line emission appears shifted towards the south of the column density peak (estimated from dust continuum emission) by $\sim15\arcsec$ (within Clump C1). This could be attributed to resolution effects as the beam size of the molecular gas emission is nearly three times larger than that of dust continuum emission. Besides, the role of optical depth effects cannot be ruled out.
The detailed properties of individual species are discussed below.

\subsubsection{$^{12}$CO and $^{13}$CO (Carbon monoxide)}
CO is the most easily observed molecular line in the interstellar medium and is present even in fairly tenuous gas \citep{2001ApJ...547..792D}. Generally, the $^{12}$CO line is optically thick and $^{13}$CO being relatively optically thin, can trace higher density gas ($n_{\textrm{H}2}>10^3$~cm$^{-3}$) in molecular clouds. In the present case, $^{13}$CO is also optically thick as evident from Fig.~\ref{molspec}. The distribution of $^{12}$CO and $^{13}$CO emission in G333.73 is shown in Fig.~\ref{comaps} and displays elongated morphology consistent with cold dust emission and column density maps. The aspect ratio of $^{13}$CO ($\sim$5) is more than twice that of $^{12}$CO ($\sim$2). Moreover, the $^{12}$CO emission towards S2 is extended compared to  $^{13}$CO. This could be due to the fact that $^{12}$CO is a low density gas tracer compared to the latter and hence traces the extended envelope surrounding the dense gas. We also see an extension towards the north-west in both the CO maps that overlaps with an extinction filament in the warm dust emission.

\subsubsection{HCO$^+$ and H$^{13}$CO$^+$ (Formylium)}
The HCO$^+$ ion has been used to investigate the infall and outfall motions \citep[e.g.,][]{{2001A&A...376..271C},{2005A&A...442..949F},{2011ApJ...729..124C}} and hence, the HCO$^+~J=1-0$ is believed to be a good tracer of kinematics in star forming regions \citep[e.g.,][]{{2009MNRAS.392..170S},{2013A&A...549A...5R}}.  
However, this transition could be optically thick as a result of contributions from various mechanisms and gas motions within the clumps. Consequently, higher transitions of HCN, HNC and HCO$^+$ have been suggested as more favourable infall tracers \citep{2014MNRAS.444..874C}.

\par HCO$^+$ is detected close to the peak of the cold dust emission and the distribution is nearly spherical (see Fig.~\ref{mom0}). Weak HCO$^+$ emission is detected towards the other millimeter peaks of the IRDC (C2, C7) within the sampled region. H$^{13}$CO$^+$ is a high density tracer and is generally assumed to be optically thin. The distribution of H$^{13}$CO$^+$ is morphologically different as compared to the HCO$^+$ emission and we discern that the intensity is relatively weak towards the peak location of other molecular species such as HCO$^+$. This is a region where most of the ionised gas emission is distributed. The lower intensity of H$^{13}$CO$^+$ emission towards S1 could be attributed to the destruction of this species by UV radiation and high density electrons as the abundance of H$^{13}$CO$^+$ is a factor of 50 lower than that of HCO$^+$ \citep[e.g.,][]{{2009A&A...498..771G},{2017MNRAS.465.4219V}}.

\subsubsection{HCN (hydrogen cyanide) and HNC (hydrogen isocyanide)}

The HCN molecule and its metastable geometrical isomer HNC  are typically employed as dense gas tracers in analysing the chemistry of star forming regions \citep[e.g.,][]{{2014A&A...562A...3M},{2013MNRAS.431...27L}}. In particular, HNC molecule is considered as a good tracer of infall motion \citep{2013ApJ...766..115K}. In the IRDC G333.73, we detect both HCN and HNC molecules with optically thick profiles. The morphologies of HCN and HNC emission are similar to that of HCO$^+$ emission. The hyperfine components of HCN are visible in the spectrum, but displays heavy self-absorption. HNC exhibits a strong blue asymmetry similar to the HCO$^+$ line. Similar to HCO$^+$ integrated intensity map, additional peaks (associated with clumps) are also seen towards north-east and south-west directions.

\subsubsection{N$_2$H$^+$ (diazenylium) and C$_2$H (ethynyl)}
 
 The N$_2$H$^+$ ion is regarded as a good tracer of dense gas as it resists freeze-out on dust grains compared to the carbon bearing species \citep{{1997ApJ...486..316B},{1997MNRAS.291..455C}}. Thus, it is favoured in the studies of cold molecular clumps and cores where other species such as CO and CS are depleted. The distribution of N$_2$H$^+$ emission in G333.73 is similar to the HCO$^+$, HCN and HNC molecules, but the shape of the clump is elongated (similar to continuum emission from dust) unlike the other species that show a spherical distribution. 

The species C$_2$H is believed to form through photodissociation of acetylene molecule and is acknowledged as a good tracer of PDRs \citep{2012A&A...543A..27G}. A recent study by \citet{2008ApJ...675L..33B} has shown that C$_2$H is observed in all stages of high mass evolution from infrared dark clouds to massive protostellar objects to ultracompact \hii~regions. The distribution of C$_2$H is spherical in morphology and similar to other molecules such as HCO$^+$, HCN and HNC towards the peak emission region. However, unlike the other species, the emission is not extended in the direction of filament but appears rather confined to the clump C1. The lack of C$_2$H emission towards the immediate south-west of peak emission, where the extinction is high, is noticeable. Evidence of secondary peaks are observed towards S2 and towards the south-west of the IRDC. As the location of peak emission matches with that of other high density tracers, we infer that the C$_2$H emission close to the continuum peak is possibly originating from the molecular cloud itself rather than from the PDR.

\subsection{Ionised gas emission}
\begin{figure*}
\hspace*{-0.7cm}
\centering
\includegraphics[scale=0.17]{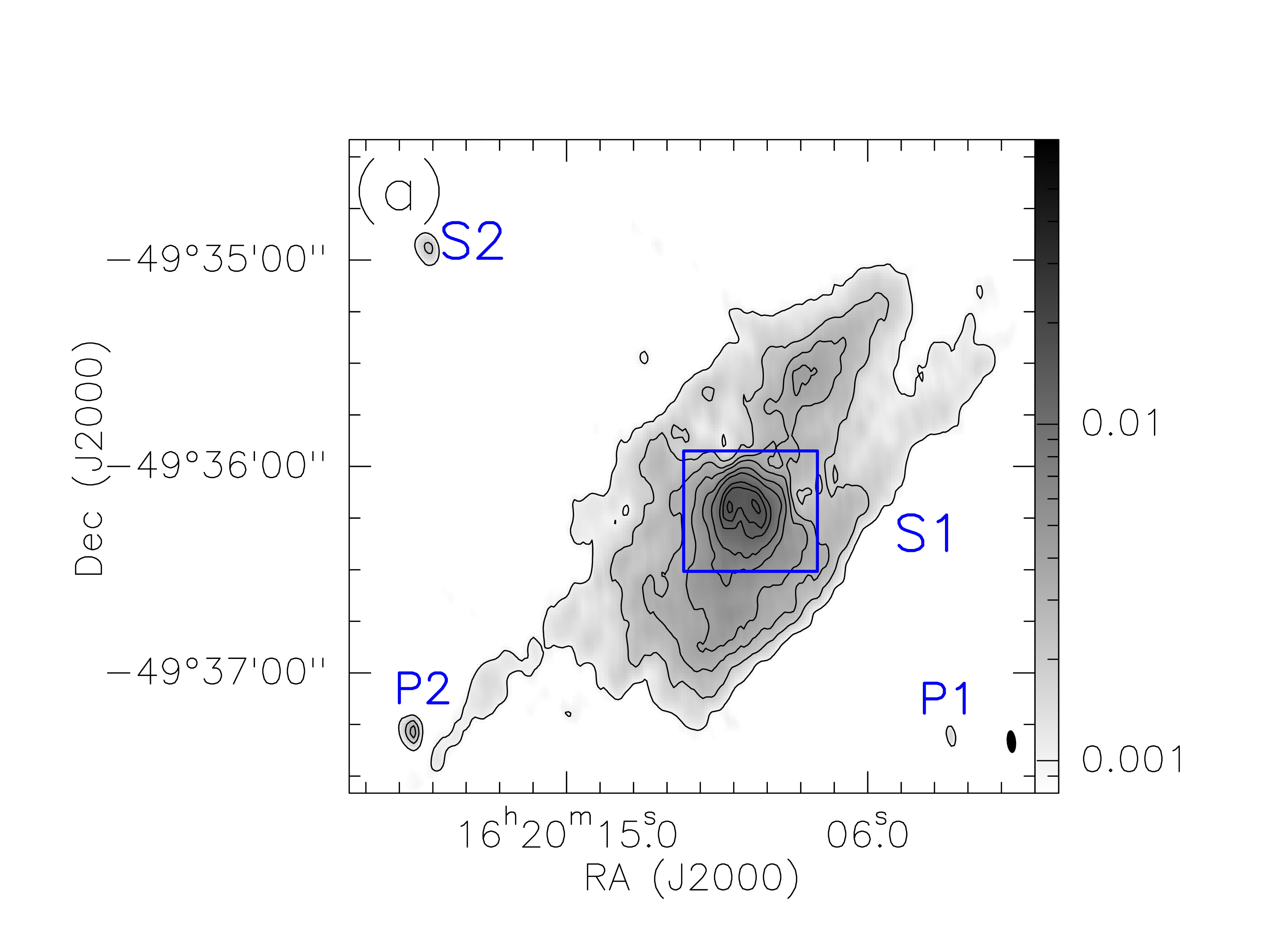} \hspace*{-0.35cm} \quad \includegraphics[scale=0.17]{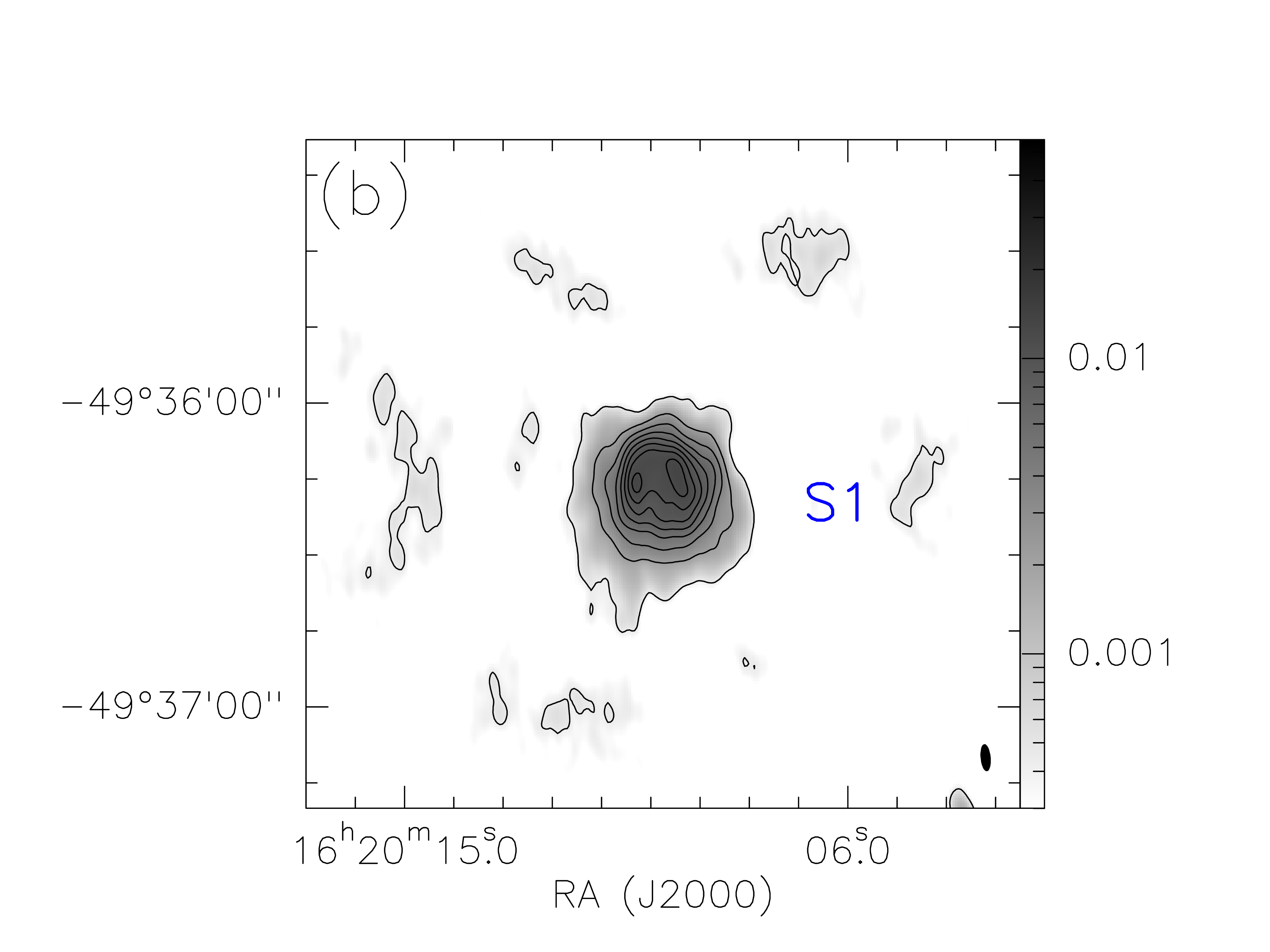} \hspace*{-0.35cm} \quad \includegraphics[scale=0.17]{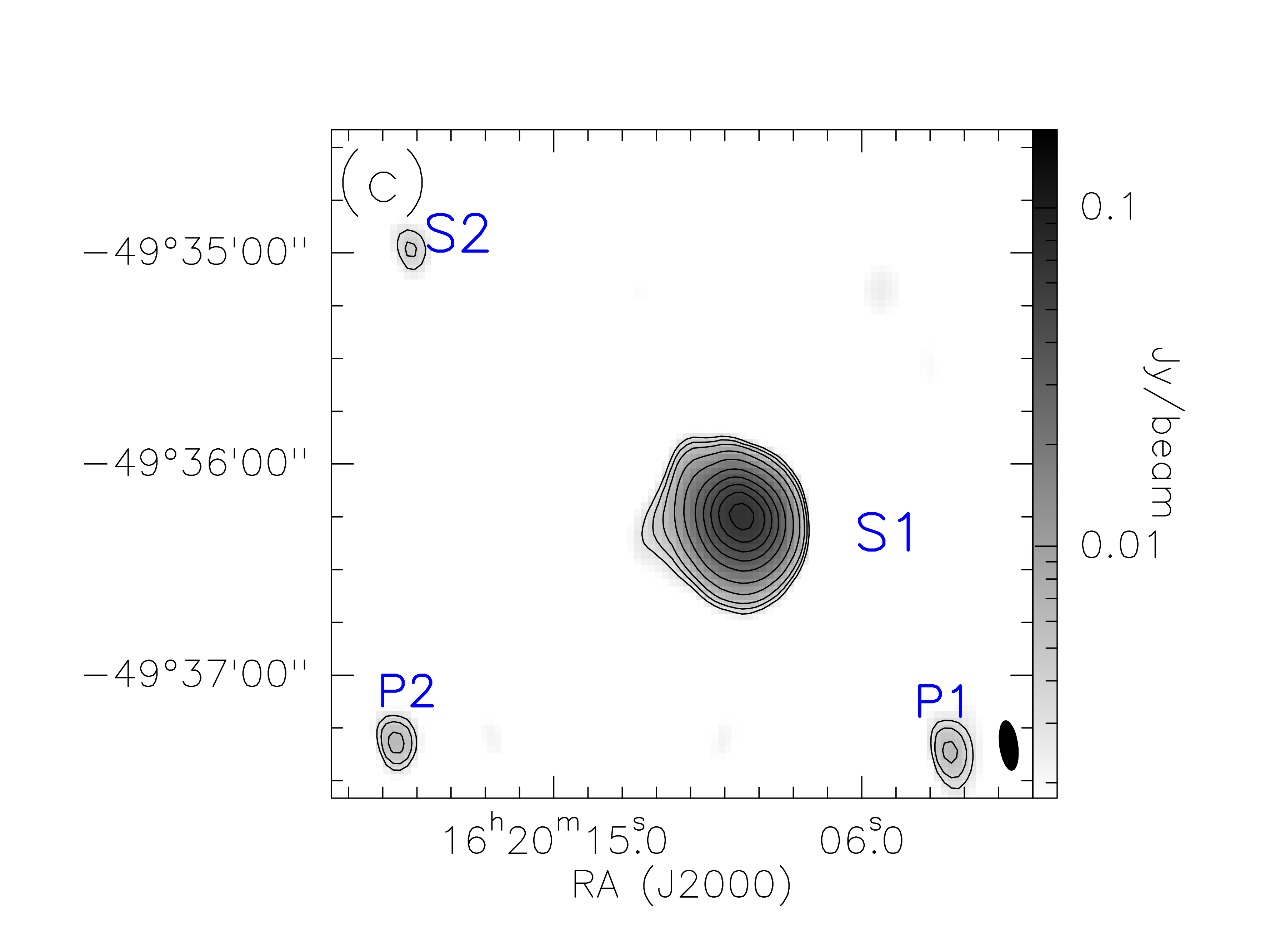} 
\caption{(a) Low resolution radio continuum map of G333.73+0.37 at 1300~MHz. The contour levels are at 1.4, 2.1, 3.1, 4.1, 6.1, 8.1, 10.1, 12.1, 14.1 and 16.1~mJy/beam with beam size of 6.6$\arcsec\times$2.7$\arcsec$. The compact shell structure is enclosed within the rectangle. (b) High resolution map of the rectangular region shown in (a). The contour levels are from 2~mJy/beam to 19~mJy/beam in steps of 2.2~mJy/beam. The corresponding beam size is 5.5$\arcsec\times$2.0$\arcsec$. (c) Radio continuum map of G333.73+0.37 at 610~MHz. The contour levels are from 3.5~mJy/beam to 98~mJy/beam in steps of 10~mJy/beam. The corresponding beam size is 14.5$\arcsec\times$5.4$\arcsec$. The beams are represented as filled ellipses towards bottom right of all panels.}
\label{1300}
\end{figure*}

The radio continuum emission from G333.73+0.37 at 1300 and 610~MHz are shown in Fig.~\ref{1300}. The ionised gas emission at 1300~MHz towards S1 reveals a shell-like structure surrounded by a low surface brightness diffuse envelope as seen in Fig.~\ref{1300}(a). The shell structure is more evident in the high resolution map, displayed in Fig.~\ref{1300}(b) with two peaks separated by lower flux density towards the center that gives the appearance of a cleft ring. The angular diameter of the shell-structure is $\sim$30.5$\arcsec$ that corresponds to 0.4~pc at a distance of 2.6~kpc. The radio emission at 610~MHz shown in Fig.~\ref{1300}(c) shows a more compact structure and traces of the shell are not evident. 
While resolution effects could play a role, the data quality is poor compared to the higher frequency image as the diffuse structure is not visible either. There, however, exists a possibility that optical depth effects could hamper our viewing of the shell structure. We also detect radio emission towards the source S2 where the emission is highly compact. This suggests that S2 is relatively young and or excited by a lower mass zero age main sequence (ZAMS) star. Additionally, two point-like sources, designated P1 and P2, are detected towards the south-west and south-east of S1, marked in Fig.~\ref{1300}(a). These are not positionally coincident with any infrared emission/source.

We have computed the spectral indices ($\alpha_{610-1300}$) of the compact sources by integrating flux densities at 610 and 1300~MHz. For S2, we estimate the spectral index as $-0.28$. For P1 and P2, we obtain steeper spectral indices of $-2.0$ and $-0.4$, respectively. These values are indicative of non-thermal contribution to the radio emission \citep{1999ApJ...527..154K} as thermal emission from \hii~regions typically falls in the range $-0.1\leq\alpha\leq2$ \citep{1975A&A....39..217O}. The latter two are likely to be background sources of extragalactic origin and we exclude them from further analysis in this work.

We estimate the radio properties of both the \hii~regions, S1 and S2, to learn about the source(s) of excitation as well as the physical conditions such as ionised gas densities within these regions. The emission measure (EM), electron density ($n_e$) and the Lyman continuum photon rate ($\rm {\dot N_{Lyc}}$) under the assumptions of optically thin emission and negligible absorption by dust  are given by the following relations \citep{2016A&A...588A.143S}.

\begin{equation}
 \left( \rm{\frac{EM}{pc\,\,cm^{-6}}}\right) = 3.217\times10^7 \left( \frac{S_\nu}{\rm Jy}\right) \left( \frac{T_e}{\rm K}\right)^{0.35} \left( \frac{\nu}{\rm GHz}\right)^{0.1} \left( \frac{\theta_{src}}{\rm arcsec}\right)^{-2}
 \label{EM}
\end{equation}

\begin{equation}
\begin{split}
\left( {\frac{n_e}{\rm cm^{-3}}}\right) = 2.576\times10^6 \left( \frac{S_\nu}{\rm Jy}\right) ^{0.5}\left( \frac{T_e}{\rm K}\right)^{0.175} \left( \frac{\nu}{\rm GHz}\right)^{0.05} \left( \frac{\theta_{src}}{\rm arcsec}\right)^{-1.5} \\ 
\left( \frac{d}{\rm pc}\right)^{-0.5}
\end{split}
\label{ne}
\end{equation}

\begin{equation}
\left( {\frac{\rm{N}_{Lyc}}{\rm s^{-1}}}\right) = 4.771\times10^{42} \left( \frac{S_\nu}{\rm Jy}\right)\left( \frac{T_e}{\rm K}\right)^{-0.45} \left( \frac{\nu}{\rm GHz}\right)^{0.1} \left( \frac{d}{\rm pc}\right)^{2}
\label{nly}
\end{equation}

\noindent where $S_{\nu}$ is the flux density at frequency $\nu$, $T_e$ is the electron temperature, $\theta_{src}$ is the angular source size, and $d$ is the distance to the source. In order to estimate the electron temperature in this region, we apply the electron temperature gradient curve across the Galactic disk \citep{{1978A&A....70..719C},{2006ApJ...653.1226Q}} and obtain a value of  6800~K for a Galactocentric distance of 6.3~kpc. To determine the properties of S2, we use the same kinematic distance of 2.6~kpc as S1, since the LSR velocity of molecular gas close to S2 is similar to what we measured towards S1. The radio properties of S1 and S2 determined from the above equations are listed in Table~\ref{rad_pro}. Assuming that the HII regions are excited by a single ZAMS star, S1 is ionised by a late O or early B type star while S2 is powered by an early B star. The electron density towards S2 is nearly factor of two larger than that towards S1. Equipped with the knowledge of the Lyman continuum flux as well as electron density, we estimate the radius of the Str\"{o}mgren sphere, defined as the radius at which the rate of ionization equals that of recombination under the assumption that the \hii~region is expanding in a homogeneous and spherically symmetric medium. The radius of the Str\"{o}mgren sphere, $R_s$ is given by the expression

\begin{equation}
R_s = \left(\frac{3\,N_{Lyc}}{4\,\pi\,n_0^2\,\alpha_B}\right)^{1/3}
\end{equation}

\noindent Here $\alpha_B$ is the radiative recombination coefficient assumed to be $2.6\times10^{-13}~\textrm{cm}^3\,\textrm{s}^{-1}$ \citep{osterbrock1989astrophysics}. $n_0$ represents the mean number density of atomic hydrogen which is estimated from the column density map using the expression $n_0$ = 3$N(\textrm{H}_2)$/2R where R is the radius of the clump.$n_0$ is $1.6\times10^4$~cm$^{-3}$ and $2.3\times10^4$~cm$^{-3}$ for clumps C1 and C2 that corresponds to S1 and S2 respectively. From the above expression, we found $R_s$ to be 0.03 and 0.02~pc for S1 and S2, respectively. If we compare this with the observed radii of S1 and S2, we find that the observed radii are an order of magnitude higher compared to $R_s$ values determined. This signifies that the \hii~regions have expanded beyond the Str\"{o}mgren spheres and are in the second expansion phase, where pressure disturbances from within  the \hii~region are able to cross the ionization front and create an expanding shock. We can estimate the dynamical age, t$_{dyn}$ of these HII regions based on a simple model of expanding photoionised nebula, in a homogeneous medium using the size of radio emission \citep{1980pim..book.....D}. The expression for t$_{dyn}$ is given by

\begin{equation}
t_{dyn} = \left[\frac{4\,R_s}{7\,c_i}\right] \left[\left(\frac{R}{R_s}\right)^{7/4}-1\right]
\end{equation}   

\noindent where $R$ represents the radius of the spherical HII region and $c_i$ is the isothermal sound speed in the ionised gas, assumed to be 10~km\,s$^{-1}$ for typical \hii~regions \citep{2005fost.book.....S}. $R$ is the radius of the source.  The estimated dynamical ages for S1 and S2 are found to be 0.2 and 0.01~Myr, respectively. This hints at the youth of S2 relative to S1. It is to be noted that the dynamical age has been calculated assuming a medium which is homogeneous and spherically symmetric. This is unlikely to represent the factual situation. Hence $t_{dyn}$ should be considered as representative at best.

\begin{table}
\footnotesize
\caption{Properties of the sources S1 and S2 from radio continuum data}
\begin{center}
\setlength{\tabcolsep}{3pt}
\begin{tabular}{l c c c c c c c} \hline \hline
 &  \\
Source    &Diameter  &EM &$n_e$&N$_\textrm{Lyc}$ & Spectral type&$R_s$& $t_{dyn}$  \\
&(pc)&(pc\,cm$^{-6}$) & (cm$^{-3}$) &(10$^{46}$~s$^{-1}$)&&(pc)&(Myr)\\
\hline\\
S1 &0.38&$2.4\times10^5$  &734&22.9&O9.5 - B0&0.03&0.2 \\
\hline\\
S2 &0.15&$2.9\times10^5$ &1313&4.2&B0 - B0.5&0.02&0.01\\
\hline\\ 
\end{tabular}
\label{rad_pro}
\end{center}
\end{table}

\subsection{Young stellar objects associated with G333.73}
\par Color excess at infrared wavelengths has been extensively used to identify the young objects and to broadly categorise them according to evolutionary stages. Recent studies in nearby star forming regions have shown that the $\textit{Spitzer}$ IRAC color-color diagrams are particularly useful in identifying the young stellar population in these regions as the IRAC bands are highly sensitive to the emission from the circumstellar disks and envelopes \citep[e.g.,][]{{2004ApJS..154..363A},{2004ApJS..154..367M},{2005ApJ...629..881H}}. Examining different models and combining them with the observations, they have found that the different classes of YSOs such as Class I (central source+disk+envelope) and Class II (central source+disk) objects occupy distinct regions in the color-color diagram. In addition to this, the color-color diagrams that combine the IRAC and MIPS 24~$\mu$m data are often used to identify highly embedded stars and sources with significant inner holes \citep[e.g.,][]{{2006ApJ...643..965R},{2006AJ....131.1574L}}.
\begin{figure}
\hspace*{-0.5cm}
\centering
\includegraphics[scale=0.39]{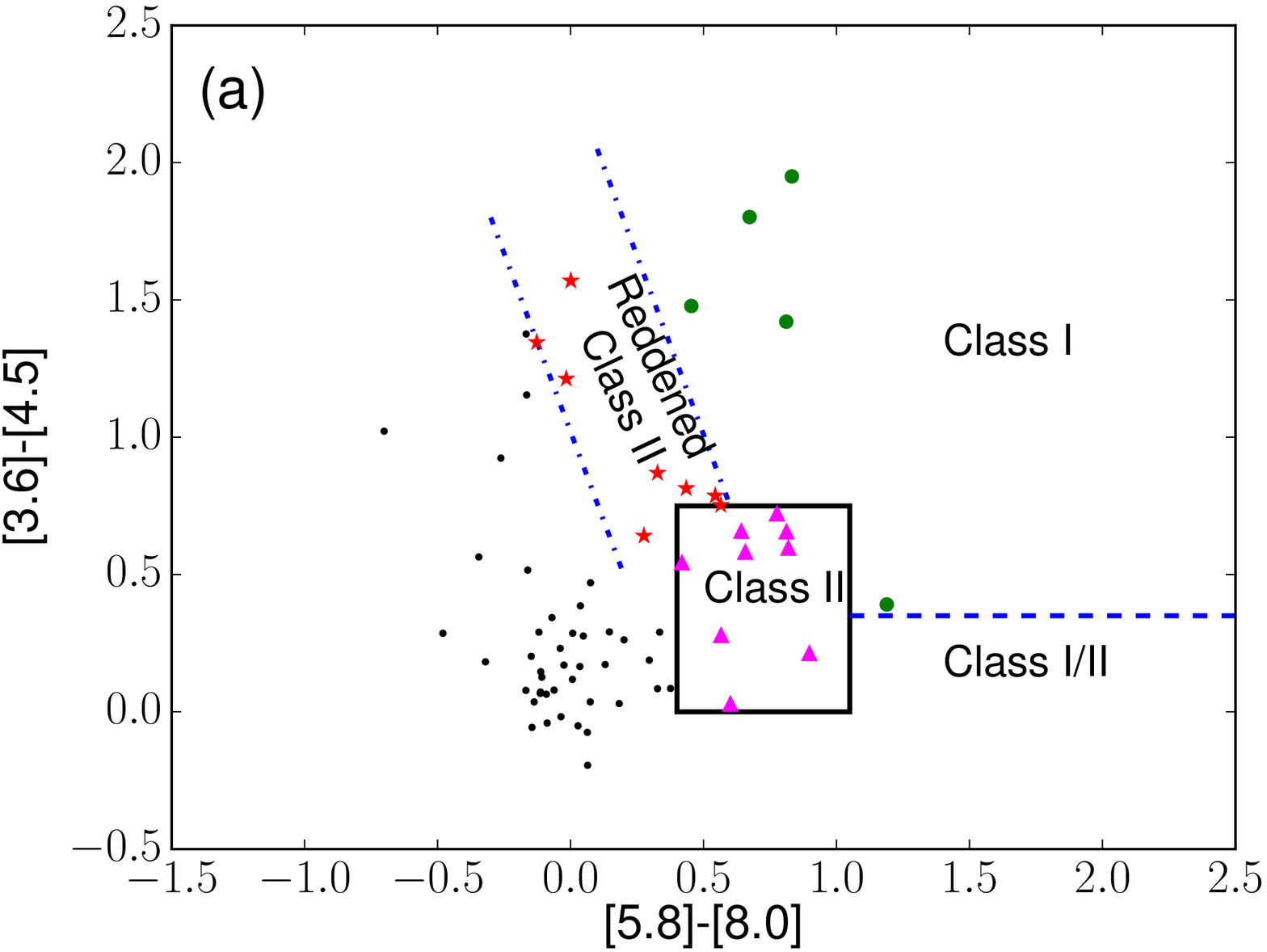} \quad \hspace*{-0.5cm}\includegraphics[scale=0.39]{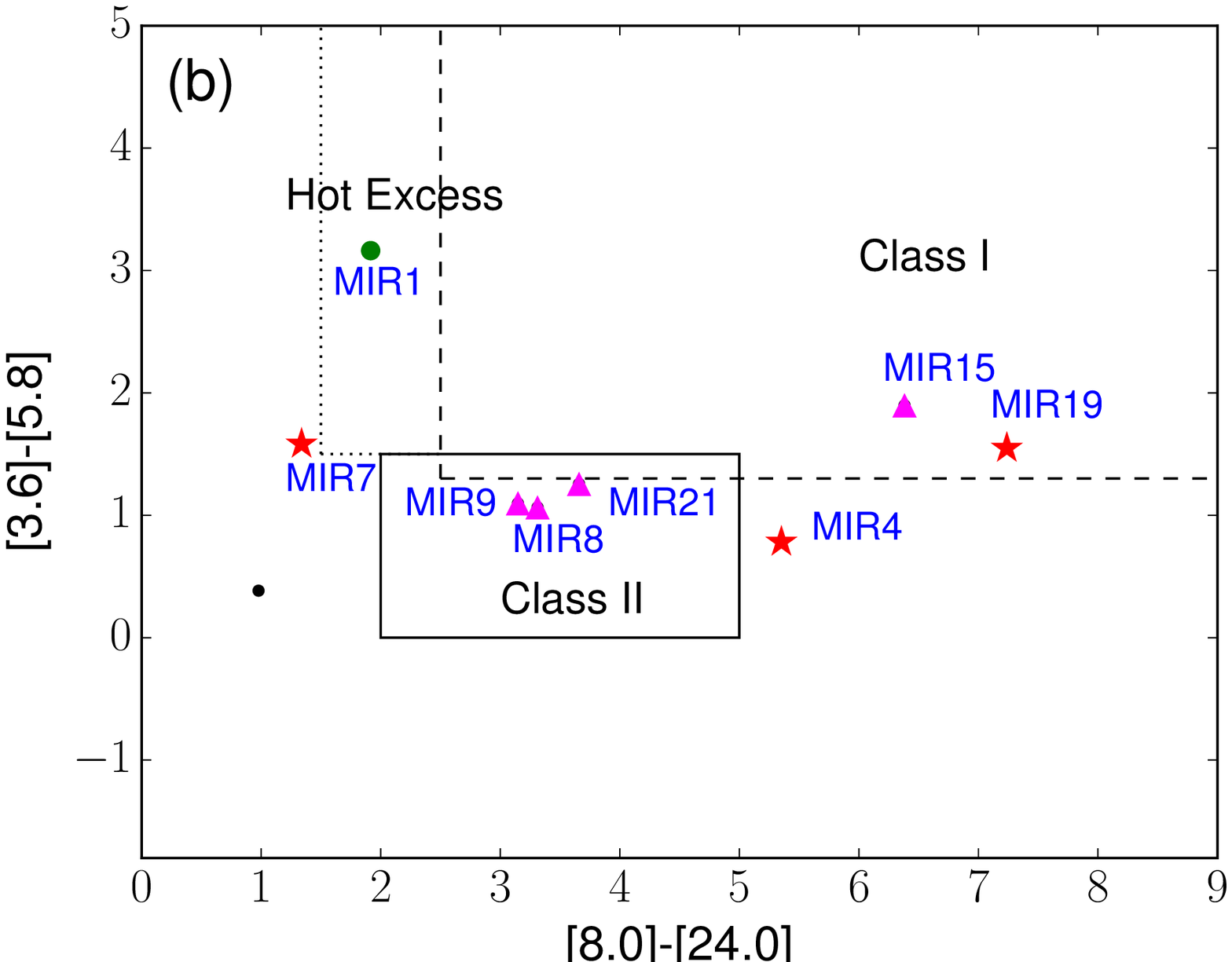} 
\caption{(a) $\textit{Spitzer}$-IRAC color-color diagram using the colors [3.6]-[4.5] versus [5.8]-[8.0]. Solid rectangle indicates the position occupied by Class II objects while parallel dotted lines represent the region occupied by reddened Class II objects. The top-right region of the color-color diagram is occupied by Class I sources and the region to the right of Class II sources is occupied by Class I/II objects. Filled circles (green) represent Class I YSO candidates and triangles (magenta) denote Class II objects. Reddened Class II objects are represented with star symbol (red). Other sources are marked as filled circles (black). (b) Color-color diagram of [3.6]-[5.8] versus [8.0]-[24.0] using mid-infrared magnitudes from $\textit{Spitzer}$-IRAC and MIPS. Solid line box indicates the region occupied by Class II objects whereas region marked by dashed line represents the region occupied by Class I objects. Dotted line indicates the location of \textquotedouble{hot excess} sources (see text).}
\label{mircc}
\end{figure}

\begin{figure}
\hspace*{-1.5cm}
\centering
\includegraphics[scale=0.52]{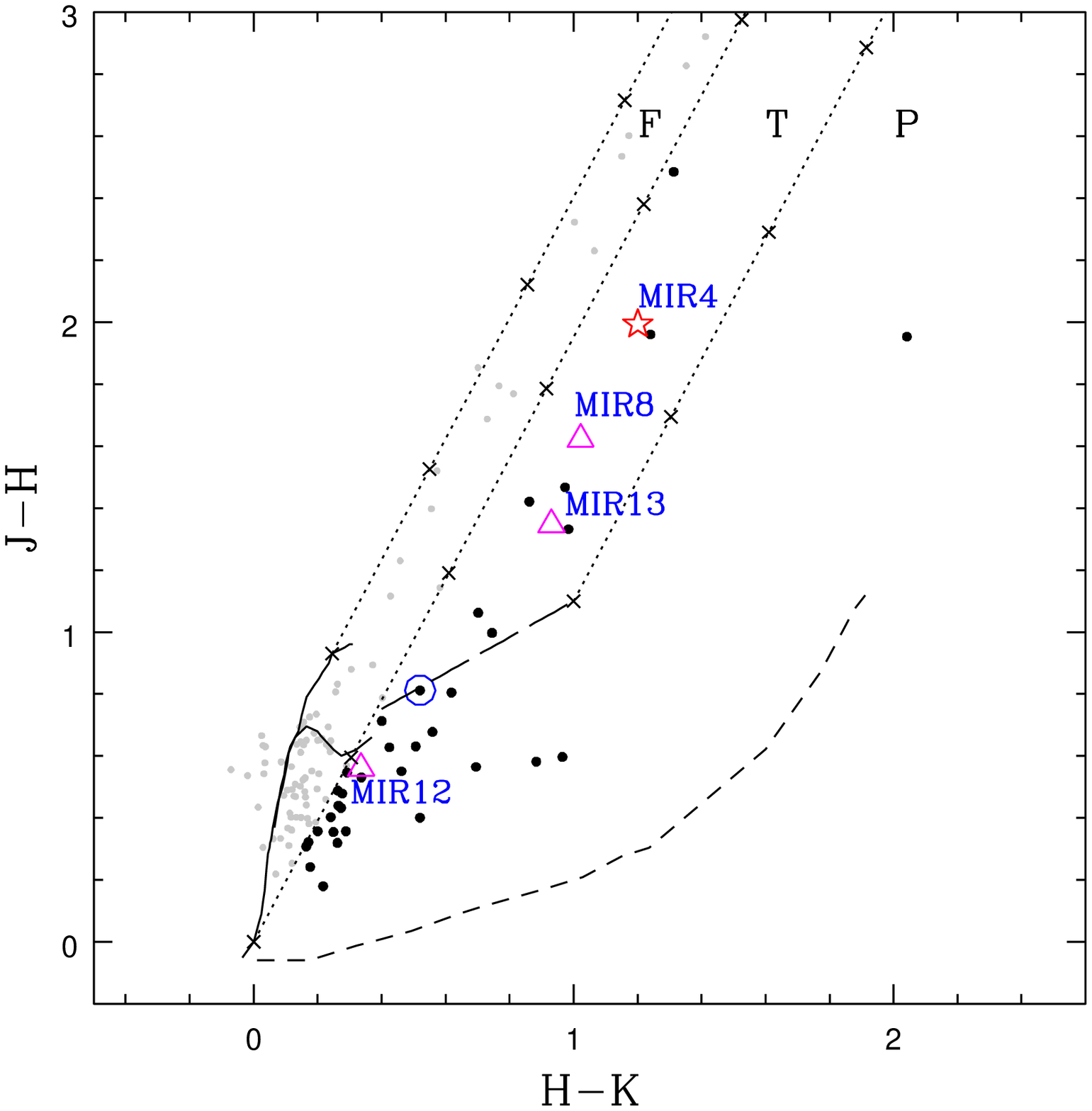} 
\caption{Color-color diagram of the 2MASS sources in the G333.73 region. The two solid curves represent the loci of giant (thick line) and main sequence stars (thin line) derived from \citet{1988PASP..100.1134B}. Classical T Tauri locus from \citet{1997AJ....114..288M} is represented with a long-dashed line. The parallel dotted lines are the reddening vectors with crosses placed at intervals corresponding to 5 magnitudes of visual extinction. We have assumed the interstellar reddening law of \citet{1985ApJ...288..618R} (A$_\textrm{J}$/A$_\textrm{v}$=0.282, A$_\textrm{H}$/A$_\textrm{v}$=0.175 and A$_\textrm{K}$/A$_\textrm{v}$=0.112). Short-dashed line represents the locus of the Herbig AeBe stars \citep{1992ApJ...393..278L}. The plot is classified into \textquotedouble{F}, \textquotedouble{T} and \textquotedouble{P} regions (see text for details). The colors and the curves shown in the figure are transformed  to \citet{1988PASP..100.1134B} system. Sources that are classified as main sequence or giants are represented by gray dots whereas pre-main sequence sources are denoted by black dots. IRAC YSOs identified from mid-infrared color-color diagram having 2MASS counterparts are also shown and labeled in the image. Circled dot denotes central object corresponding to source S2.}
\label{nircc}
\end{figure}

\par In order to study the YSO population within this IRDC, we searched the Glimpse I$'$07 Archive for mid-infrared point sources. For this, we have considered all sources lying within the 3$\sigma$ contour of the 1.2~mm emission. By proceeding in this way, we attempt to identify sources associated with the IRDC and eliminate other field objects that are not related to G333.73. However, as the mid-infrared emission from S1 extends beyond the  3$\sigma$ contour of the 1.2~mm emission, we have also considered sources within a circular region of radius 1.5$\arcmin$ around S1 (centred at $\alpha_\textrm{J2000}$=$16^h20^m09.4^s$, $\delta_\textrm{J2000}$=$-49^{\circ}36^{\arcmin}21^{\arcsec}$). Among these two groups of \textit{Spitzer} sources, we find that 73 are detected in all the four IRAC bands. We have also carried out a visual inspection of this region in all the IRAC images by scaling the images to identify sources embedded in mid-infrared nebulosity and find that 16 sources were not identified in the GLIMPSE catalog. Hence we have performed aperture photometry on these sources using the task {\tt qphot} in IRAF software. For this, we have selected a 5$\arcsec$ aperture. The inner and outer radii of the sky annulus are 6$\arcsec$ and 8.5$\arcsec$ respectively. In order to identify the YSO candidates, we used the methods prescribed by \citet{2004ApJS..154..367M} and \citet{2004ApJS..154..363A}. For the sources detected in all IRAC bands, we employed the [5.8]-[8.0] versus [3.6]-[4.5] color-color diagram to locate the YSO sources. The color-color diagram is shown in Fig.~\ref{mircc}(a). The regions occupied by Class I, Class II and reddened Class II sources, based on the predictions of existing models for disks and envelopes \citep{2004ApJS..154..367M}, are also shown in the image. A total of 22 YSO candidates are detected in the color-color diagram. Of these, 5 are Class I sources, 9 are Class II sources and 8 are reddened Class II sources. 
\begin{table*}
\scriptsize
\caption{Coordinates and magnitudes of IRAC YSO candidates.}
\begin{center}
\setlength{\tabcolsep}{3pt}
\begin{tabular}{l c c c c c c c c c c c} \hline \hline
 &  \\
YSO     &$\alpha_{2000}$&$\delta_{2000}$&J &H&K  &3.6~$\mu$m&4.5$\mu$m&5.8$\mu$m&8.0$\mu$m&24.0$\mu$m  &Classification$^\star$\\
&($^{h~m~s}$) &($^{\circ~\arcmin~\arcsec}$) &(mag)&(mag)&(mag)&(mag)&(mag)&(mag)&(mag)&(mag)&\\
\hline\\
MIR1&16:19:50.894&$-$49:38:21.36&- &- &- &$7.0 \pm 0.03$& $5.05 \pm 0.07$&$3.84 \pm 0.03$&$3.01 \pm 0.02$&$1.09 \pm 0.02$&Class I\\
MIR2&16:19:51.091&$-$49:37:49.88&-&-&-&$11.99 \pm 0.07$ & $10.42 \pm 0.06$ & $9.53 \pm 0.05$ & $9.53 \pm 0.08$&-&Reddened Class II\\
MIR3&16:19:52.317&$-$49:37:55.90&-&-&$13.46 \pm 0.04$&$12.30 \pm 0.06$ & $11.90 \pm 0.10$ & $11.39 \pm 0.10$ & $10.20 \pm 0.27$&-&Class I\\
MIR4&16:19:52.879&$-$49:37:37.00&$15.71 \pm 0.07$&$13.72 \pm 0.05$&$12.51 \pm 0.04$&$11.15 \pm 0.07$&$10.51 \pm 0.09$&$10.37 \pm 0.11$&$10.09 \pm 0.12$&$4.74 \pm 0.04^\dagger$&Reddened Class II\\
MIR5&16:19:53.038&$-$49:37:34.59&-&-&-&$13.31 \pm 0.11$&$11.51 \pm 0.10$&$10.75 \pm 0.11$&$10.08 \pm 0.13$&$4.74 \pm 0.04^\dagger$&Class I\\
MIR6&16:19:55.084&$-$49:37:42.55&-&-&-&$12.76 \pm 0.10$  &$12.17 \pm 0.11$  &$11.09 \pm 0.10$  &$10.43 \pm 0.31$&-&Class II\\
MIR7&16:19:58.049&$-$49:37:04.62&- &$14.90 \pm 0.09$&$10.75 \pm 0.03$&$7.45 \pm 0.04$&$6.58 \pm 0.07$&$5.87 \pm 0.03$&$5.54 \pm 0.03$&$4.20 \pm 0.03$&Reddened Class II\\
MIR8&16:19:58.291&$-$49:37:47.05&$15.10 \pm 0.04$&$13.47 \pm 0.02$&$12.45 \pm 0.04$&$10.76 \pm 0.05$&$10.10 \pm 0.05$&$9.69 \pm 0.05$&$8.88 \pm 0.06$&$5.57 \pm 0.07$&Class II\\
MIR9&16:20:00.555&$-$49:36:06.85&-&$14.59 \pm 0.06$&$13.25 \pm 0.04$&$11.88 \pm 0.07$&$11.46 \pm 0.08$&$10.78 \pm 0.06$&$10.24 \pm 0.06$&$7.09 \pm 0.69$&Class II\\
MIR10&16:20:00.701&$-$49:37:14.31&-&-&-&$13.71 \pm 0.12$  &$12.23 \pm 0.10$  &$11.37 \pm 0.09$  &$10.92 \pm 0.15$&-&Class I\\
MIR11&16:20:01.848&$-$49:37:28.12&-&-&-&$12.20 \pm 0.06$  &$10.85 \pm 0.06$  &$10.13 \pm 0.05$  &$10.26  \pm 0.05$&-&Reddened Class II\\
MIR12&16:20:02.657&$-$49:36:30.42&$13.03 \pm 0.03$  &$12.47 \pm  0.06$  &$12.14 \pm  0.04$ &$11.78  \pm 0.07$  &$11.75 \pm  0.09$&$11.58 \pm 0.11$&$10.98 \pm 0.20$&-&Class II\\
MIR13&16:20:03.535&$-$49:36:46.71&$16.30  \pm 0.10$  &$14.95  \pm 0.10$  &$14.02  \pm 0.09$&$12.90  \pm 0.08$  &$12.68 \pm  0.12$  &$12.25 \pm  0.27$  &$11.36 \pm 0.13$ &-&Class II\\
MIR14&16:20:05.035&$-$49:36:35.31&-&$14.53 \pm 0.11$ & $13.39 \pm 0.05$&$12.06 \pm 0.06$ & $11.46 \pm 0.07$ & $10.98 \pm 0.08$&  $10.17 \pm 0.07$&-&Class II\\
MIR15&16:20:08.602&$-$49:37:16.20&-&-&-&$13.52 \pm 0.14$&$12.86 \pm 0.14$&$11.62 \pm 0.26$&$10.98 \pm 0.14$ &$4.60  \pm 0.25$&Class II\\
MIR16&16:20:09.786&$-$49:36:17.02&-&-&$12.33 \pm 0.06$&$8.86 \pm 0.19$  &$7.44 \pm 0.14$  &$6.37 \pm 0.06$  &$5.55 \pm 0.03$&-&Class I\\
MIR17&16:20:14.246&$-$49:36:25.56&-&$14.01 \pm  0.05$&$12.54 \pm  0.04$&$11.62 \pm 0.08$  &$11.34 \pm 0.07$  &$11.11 \pm 0.13$  &$10.55  \pm 0.20$&-&Class II\\
MIR18&16:20:16.272&$-$49:34:41.82&-&-&$14.01 \pm 0.07$&$12.40 \pm 0.07$  &$11.58 \pm 0.09$  &$10.92 \pm 0.09$  &$10.49  \pm 0.07$&-&Reddened Class II\\
MIR19&16:20:18.677&$-$49:36:23.27&-&-&- &$12.59 \pm 0.07$&$11.84 \pm 0.11$&$11.04 \pm 0.09$&$10.48 \pm 0.15$&$3.24 \pm 0.27$&Reddened Class II\\
MIR20&16:20:21.415&$-49$:34:36.64&-&-&-&$12.74 \pm 0.07$&  $11.96 \pm 0.08$&  $11.85 \pm 0.27$&  $11.31 \pm 0.18$&-&Reddened Class II\\
MIR21&16:20:24.991&$-$49:34:44.93&-  &$14.31 \pm 0.05$&$11.82 \pm 0.02$&$9.63 \pm 0.05$&$8.91 \pm 0.05$&$8.38 \pm 0.04$&$7.60 \pm 0.03$ &$3.94 \pm 0.06$&Class II\\
MIR22&16:20:27.422&$-$49:35:32.74&-&-&-&$12.05 \pm 0.06$ & $10.83 \pm 0.06$ & $10.35 \pm 0.08$ & $10.36 \pm 0.07$&-&Reddened Class II\\
\hline
\multicolumn{9}{l}{%
\begin{minipage}{7.5cm}
\scriptsize{$^\star$ : Classification based on IRAC color-color diagram
\\
$^\dagger$ : Upper limit
}  
\end{minipage}%
}\\
\end{tabular}
\label{ysomag}
\end{center}
\end{table*}


\par The 24~$\mu$m data, if available, serves as an additional tool to discriminate between Class I and Class II objects. The spectral energy distribution (SED) of Class II objects are flat or declining near 24~$\mu$m whereas it is rising for  Class I YSOs \citep{2013AJ....145...78K}. We have, therefore, employed the flux densities from the 24~$\mu$m point source catalog of \citet{2015AJ....149...64G} to plot the IRAC-MIPS color-color diagram. From the catalog, we have identified 7 sources within both our regions of interest. We again carry out a visual inspection of the MIPS 24~$\mu$m image and find that there are 5 additional sources that are not listed in the catalog. We performed aperture photometry on these sources using the task {\tt qphot} in IRAF. The parameters for the photometry are taken from \citet{2015AJ....149...64G}. This leads to a total of 12 sources at 24~$\mu$m. Of these, 9 sources have IRAC magnitudes in all the 4 bands. Among the 12, two sources whose photometry is carried out by us are detected only in the 24~$\mu$m band. These sources are likely to be Class 0 protostellar objects although we cannot exclude the possibility of these being background objects. 

\par We have constructed a color-color diagram, [3.6]-[5.8] versus [8.0]-[24.0], based on IRAC and MIPS colors which is used to identify the Class II and Class 0/I sources. This color-color diagram is shown in Fig.~\ref{mircc}(b). The class I YSO MIR1, lies in the \textquotedouble{hot excess} region  and is of interest as this region is occupied mostly by Herbig AeBe stars. Class II YSOs with large extinction (A$_v$ $>$25) or Class 0/I objects with an extra hot component from unusually active accretion, can also fall in the \textquotedouble{hot excess} region \citep{2006ApJ...643..965R}. The reddened Class II YSO, MIR7, also lies just outside the boundary of \textquotedouble{hot excess} objects. The object MIR4 lies in the region to the right of Class II objects. This source has a large [8.0]-[24.0] excess compared to [3.6]-[5.8] color. A visual inspection of MIR4 reveals that another YSO, designated MIR5 (Table~\ref{ysomag}) lies in close vicinity of this source (angular separation of 3$\arcsec$). Hence the flux attributed to MIR4 in that catalog apparently has contribution from MIR5 due to resolution effects.

\par
\begin{figure*}
\hspace*{-0.6cm}
\centering
\includegraphics[scale=0.55]{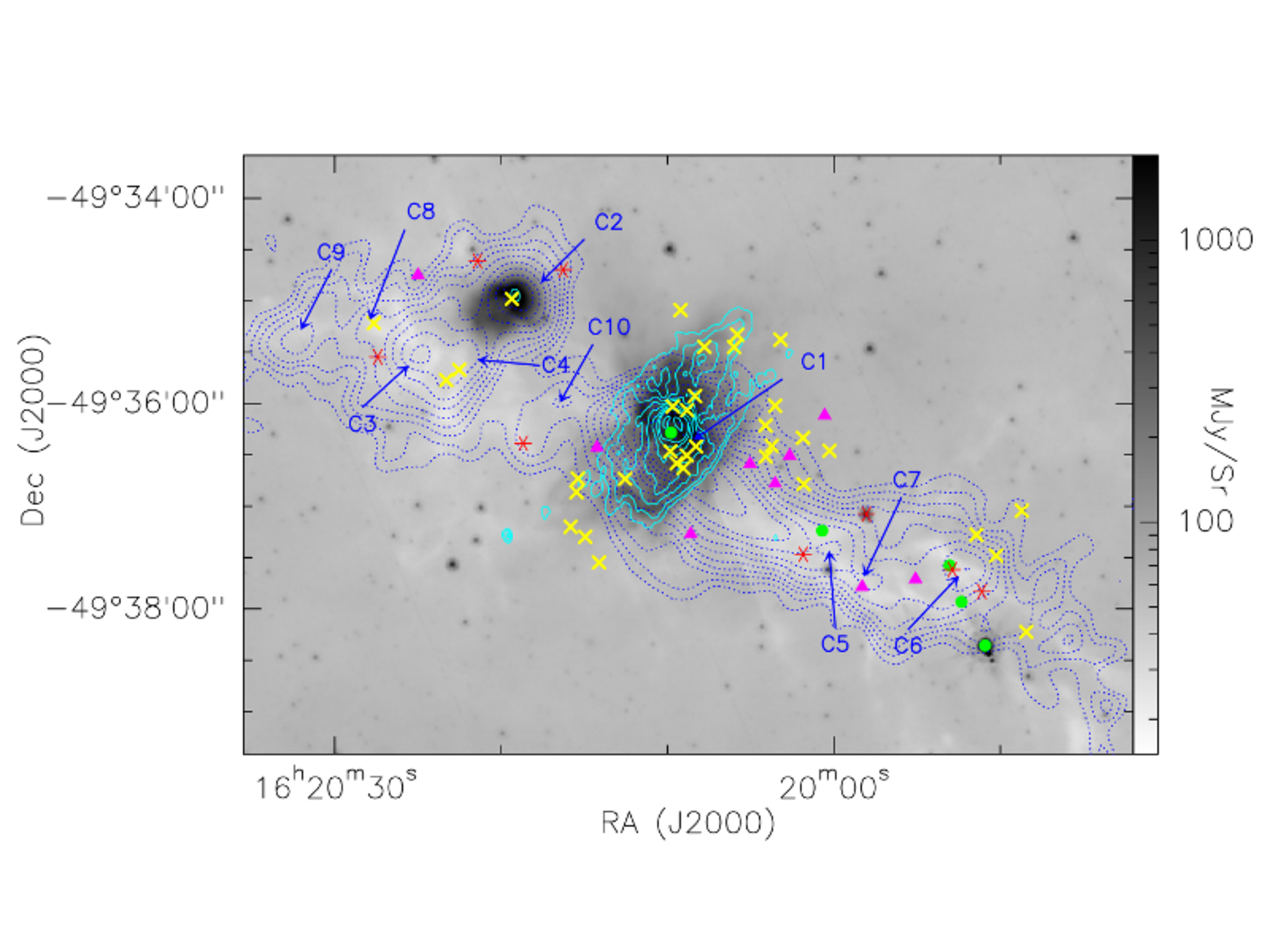}
\caption{IRAC 8~$\mu$m image of the IRDC G333.73 overlaid with 1.2~mm cold dust contours (dotted) and low resolution 1300~MHz contours (solid). Also marked in the image are the different classes of YSOs identified from near and mid-infrared color-color diagrams. IRAC Class I YSOs are denoted by filled circles (green) whereas Class II YSOs are marked as triangles (magenta). Asterisk (red) represents reddened Class II objects. Locations of the pre-main sequence objects identified from the 2MASS color-color diagram are indicated by crosses (yellow). The 1.2~mm contour levels are from 40~mJy/beam to 1500~mJy/beam in steps of 70~mJy/beam. 1300~MHz contours levels are the same as those shown in Fig.~\ref{1300}(b). Also marked are the locations of the 10 millimeter clumps.}
\label{yso8m}
\end{figure*}

 Even though we have identified the YSO population associated with this region using the mid-infrared color-color diagrams, not all sources are detected in all the four IRAC bands. This is due to the fact that the [3.6] and [4.5] bands are more sensitive compared to others \citep{2004ApJS..154...10F}. Moreover,  [5.8] and [8.0] bands traces emission from PAHs that can confuse the detection and photometry of point sources. Hence, we have resorted to the near-infrared (NIR) 2MASS ($H-K_s$) versus ($J-H$) color-color diagram to identify the young population of sources. We have selected sources with good photometric magnitudes (read flag = 2) which are detected in all the three bands. A total of 127 sources are detected within both the regions described earlier. This color-color diagram, shown in Fig.~\ref{yso8m},  is classified into three distinct regions \citep{{2002ApJ...565L..25S},{2006A&A...452..203T}}. The \textquotedouble{F} sources are located within the reddening bands of main sequence and giant stars and are believed to be either field stars, Class III objects or Class II objects with small NIR excess. \textquotedouble{T} sources occupy the region towards the left of \textquotedouble{F} region, and right of the reddening vector corresponding to T Tauri locus. The sources in this region are mostly classical T Tauri stars (Class II) with large NIR excess although there could be a few Herbig AeBe stars with small NIR excess. Towards the right of \textquotedouble{T} is the \textquotedouble{P} region that is occupied by sources which are at relatively younger (Class I or Herbig AeBe stars). From our sample, we find 38 sources that show NIR color excess (i.e. populating the \textquotedouble{P} and \textquotedouble{T} regions). The details of these YSOs are listed in Appendix~B with labels NIR1, NIR2,..., NIR 38. Eleven sources fall in the \textquotedouble{T} region and 27 in \textquotedouble{P} region. Among the YSOs identified in NIR, 4 objects are already classified as Class II sources based on IRAC colors.

\let\cleardoublepage\clearpage

\par Fig.~\ref{yso8m} shows the distribution of all the 56 YSO candidates identified in this region: 22 IRAC YSOs and 34 NIR YSO candidates. Nearly 80$\%$ of the YSOs detected solely using NIR colors are located in the proximity of S1.
This is explicable as the sources away from S1 along the length of the IRDC have low probability of detection due to higher extinction, also evident from the column density map. This YSO sample is limited by sensitivity as well as nebulosity. Hence we would like to bring attention to the fact that this is a representative sub-sample of the total YSO population in this IRDC. 
\begin{table*}[hbt!]
\caption{Parameters of the models shown in Fig.~\ref{robit}. Col. 4 -- 9 give  Mass, Effective Temperature, Luminosity, Inclination angle, Envelope accretion rate, Disk Mass, Extinction and Age, respectively. The parameters listed are for the best fit model and the range of all the ten best fits.}
\scriptsize
\begin{center}
\vskip 0.4cm
\label{spec}
\hspace*{-2cm}
\setlength{\tabcolsep}{3.5pt}
\begin{tabular}{c c c c c c c c c c c} \hline \hline \\

Source &&$\chi^2$ &Mass  &T$_{eff}$  &Luminosity  &Inc. angle & Env. accretion rate  &Disk mass  & A$_V$   & Age\\
 & &&(M$_{\odot}$) &(K) &(L$_{\odot}$) & (Deg.) &(M$_{\odot}$/yr) &(M$_{\odot}$)  & (mag) &(Myr)\\
\hline \hline \\
\multirow{ 2}{*}{MIR1 (IRAS 16161--4931)} & Best fit &373.8 & 10.0 & 25790 & 6873.0 & 56.6 &0  & 0.2 & 55.5 &1.00\\ 
&Range &$373.8-5122.0$& $7.7-10.3$ &$21850-25790$&$2581.0-6873.0$ &$31.8-75.5$ &  0 & $9.6\times10^{-3}-0.3$ & $0-55.5$ &$1.0-3.3$\\ 
\hline  \\

\multirow{ 2}{*}{MIR4} & Best fit &8.5 & 3.7 & 4847& 30.3 & 81.4 &$1.1\times 10^{-5}$  & $3.4\times 10^{-3}$ & 7.2 &0.38\\ 
&Range &$8.5-14.7$& $3.7-5.3$ &$4153-5945$&$14.9-31.8$ &$56.6-81.4$ &  $0-2.4\times 10^{-5}$ & $4.3\times10^{-4}-0.05$ & $3.3-13.1$ &$0.08-2.2$\\ 
\hline  \\

\multirow{ 2}{*}{MIR5} & Best fit &3.6 & 4.0 & 6580& 100.1 & 81.4 &$3.2\times 10^{-8}$  & $1.3\times 10^{-4}$ & 83.3 &0.97\\ 
&Range &$3.6-3.9$& $2.7-4.0$ &$6580-12950$&$55.8-118.0$ &$49.5-81.4$ &  $0-3.2\times 10^{-8}$ & $1.3\times10^{-4}-0.05$ & $78.6-84.3$ &$0.97-9.51$\\ 
\hline  \\

\multirow{ 2}{*}{MIR7} & Best fit &398.1 & 5.8 & 18020& 1543.0 & 63.3 &0  & 0.01 & 34.0 &1.90\\ 
&Range &$398.1-1267.0$& $4.6-6.4$ &$15720-19240$&$84.2-1543.0$ &$31.8-75.5$ & 0 & $3.3\times10^{-3}-0.1$ & $28.2-34.0$ &$1.53-4.38$\\ 
\hline  \\

\multirow{ 2}{*}{MIR8} & Best fit &33.6 & 3.1 & 12160& 83.7 & 81.4 &0  & 0.02 & 7.8 &9.63\\ 
&Range &$33.6-66.4$& $1.7-3.2$ &$4335-12160$&$20.6-83.7$ &$18.2-81.4$ &  $0-1.9\times 10^{-6}$ & $2.5\times10^{-3}-0.02$ & $5.2-12.1$ &$0.18-9.80$\\ 
\hline  \\

\multirow{ 2}{*}{MIR9} & Best fit &8.8 & 1.7 & 4401& 15.4 & 18.2 &$2.7\times 10^{-6}$  &0.06 & 17.1 &0.23\\ 
&Range &$8.8-11.5$& $1.0-3.3$ &$4134-4975$&$8.6-19.1$ &$18.2-69.5$ &  $6.3\times10^{-8}-2.9\times 10^{-6}$ & $2.9\times10^{-5}-0.06$ & $14.5-17.8$ &$0.18-0.71$\\ 
\hline  \\

\multirow{ 2}{*}{MIR15} & Best fit &1.8 & 2.9 & 4385& 66.8 & 81.4 &$8.6\times 10^{-5}$  &0.01 & 0 &0.07\\ 
&Range &$1.8-3.7$& $0.6-6.4$ &$3793-4772$&$9.1-336.8$ &$41.4-87.1$ &  $3.5\times10^{-6}-1.5\times 10^{-3}$ & $4.9\times10^{-5}-0.17$ & $0-24.9$ &$0.01-0.20$\\ 
\hline  \\

\multirow{ 2}{*}{MIR19} & Best fit &0.8 &2.8 & 4542& 36.5 & 81.4 &$1.2\times 10^{-5}$  &$2.0\times10^{-4}$ & 5.6 &0.18\\ 
&Range &$0.8-2.4$& $2.7-5.6$ &$4367-5950$&$36.5-221.0$ &$49.5-87.1$ &  $8.3\times10^{-6}-1.5\times 10^{-4}$ & $2.0\times10^{-4}-0.07$ & $0-20.9$ &$0.05-0.31$\\ 
\hline  \\

\multirow{ 2}{*}{MIR21} & Best fit &50.8 & 5.6 & 4724& 158.8 & 31.8 &$7.1\times 10^{-5}$  &0.02 & 29.3 &0.14\\ 
&Range &$50.8-83.8$& $4.3-5.6$ &$4724-15130$&$137.8-408.4$ &$18.2-81.4$ &  $0-7.1\times 10^{-5}$ & $1.2\times10^{-5}-0.13$ & $20.1-29.6$ &$0.14-1.92$\\ 
\hline  \\

\multirow{ 2}{*}{NIR31 (S2)} & Best fit &25.9 & 5.6 & 6462& 267.0 & 18.2 &$1.3\times 10^{-6}$  &0.03 & 4.7 &0.30\\ 
&Range &$25.9-44.6$& $5.1-7.1$ &$4445-6462$&$166.7-351.6$ &$18.2-63.2$ &  $1.3\times10^{-6}-7.8\times 10^{-4}$ & $1.6\times10^{-5}-0.15$ & $1.5-4.7$ &$0.04-0.37$\\ 

 \hline \\
\end{tabular}
\label{robit_tb}
\end{center}
\end{table*}

\subsection{SED modelling of YSOs}
\begin{figure*}
\centering

\includegraphics[scale=0.62]{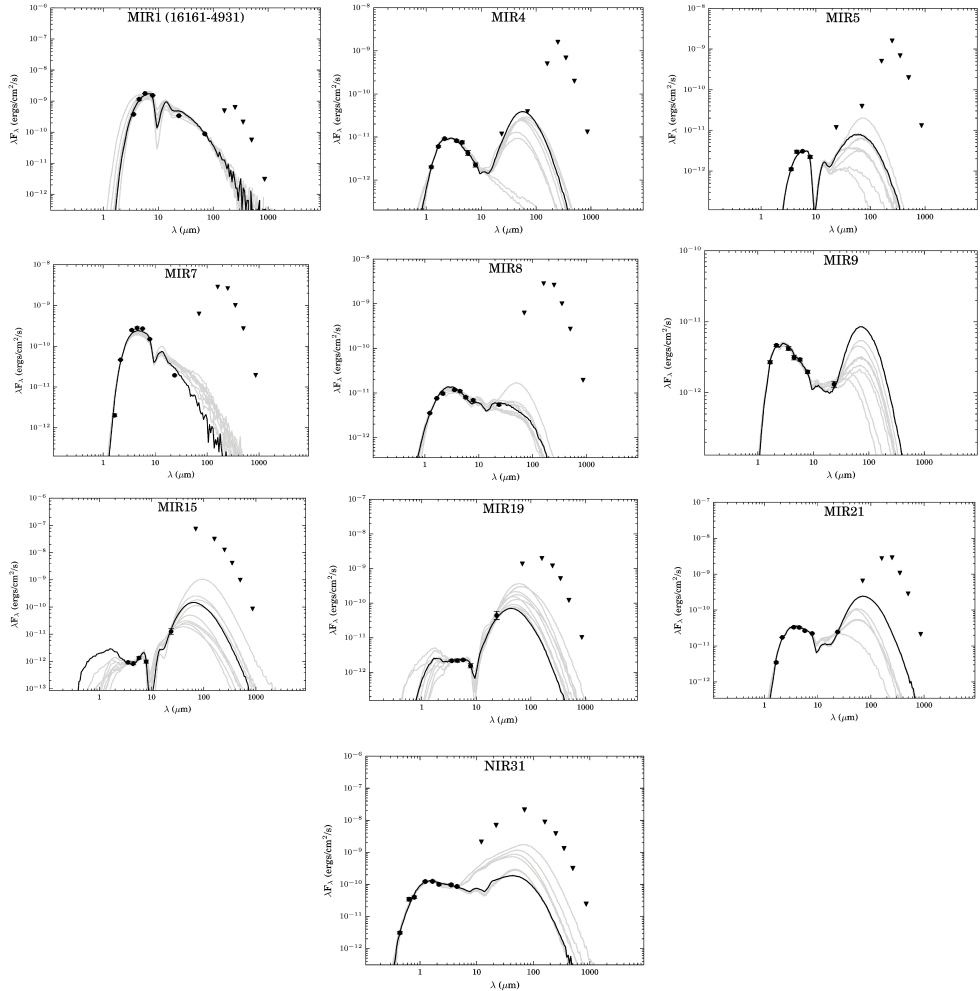} 
\vspace*{-2cm}
\caption{Infrared SEDs of YSO candidates MIR1, MIR4, MIR5, MIR7, MIR8, MIR9, MIR15, MIR19, MIR21 and near-infrared source NIR31 fitted with the SED models of \citet{2007ApJS..169..328R}. The filled circles represent input fluxes and triangles represent upper limits in the far-infrared and submillimeter wavelengths. For NIR31, we have also given the fluxes in the optical bands (B and R). Best fit model is shown as black line while the grey lines represent subsequent nine good fits.}
\label{robit}
\end{figure*}

Subsequent to the identification of YSOs, we are interested in gaining an insight into their characteristics such as mass, evolutionary stage, stellar temperature, envelope accretion rate, disk mass etc. To achieve this, we resort to the radiative transfer models of \citet{2007ApJS..169..328R} and use them to fit the spectral energy distributions of the YSOs. We have fitted SEDs of 9 YSO candidates: MIR1, MIR4, MIR5, MIR7, MIR8, MIR9, MIR15, MIR19 and MIR21, whose 24~$\mu$m fluxes are known. The reason for selecting these sources for SED modelling is that the data at longer wavelength, if available, serves as a better tool to constraint the models. For MIR4 and MIR5, only a single common flux density at 24~$\mu$m is at hand, and we use it as an upper limit for both these objects. In addition, we have selected a YSO identified based on NIR colors, NIR31, that is in close vicinity ($\sim3''$) of the radio peak towards S2. Besides, the source is detected at 3.6 and 4.5~$\mu$m IRAC bands. It is also detected in the optical bands and we have used the flux densities at B and R bands from USNO catalog \citep{1998usno.book.....M} and I band from the DENIS catalogue \citep{1998IAUS..179..106E} in the construction of the SED. As the 24~$\mu$m image is saturated, we have used the 12 and 22~$\mu$m fluxes from WISE catalog \citep{2012yCat.2311....0C} as upper limits (owing to the poor resolution of these images). Some of the sources appear point-like in the $\textit{Herschel}$ far-infrared images at 70 and 160~$\mu$m. For these sources, we have estimated the flux densities within circular apertures of radius 10$\arcsec$ and 20$\arcsec$, respectively. These apertures have been taken considering the radial profile of bright point-like sources in this region. For other YSOs as well as other wavelengths, we have considered flux densities of the corresponding clumps as upper limits. The YSO  MIR9 lies outside the $3\sigma$ threshold of the 1.2~mm map and hence for this object, we carried out the fitting for wavelengths upto 24~$\mu$m. The SED fitting of the 10 YSOs described above has been carried out using the command line version of the SED fitting tool.  The results are shown in Fig.~\ref{robit}. The fitted parameters along with ranges corresponding to first 10 best-fit models are given in Table~\ref{robit_tb}.

\begin{table}
\scriptsize
\caption{YSO evolutionary stages based on IRAC color-color diagram and classification scheme by \citet{2006ApJS..167..256R}.}
\begin{center}
\setlength{\tabcolsep}{1.8pt}
\hspace*{-0.2cm}
\begin{tabular}{l c c c c} \hline \hline
 &  \\
YSO     &$\dot{\textrm{M}}_\textrm{env}$/M$_\star$&M$_\textrm{disk}$/M$_\star$&Classification&Classification\\
&&&(IRAC CC diagram)&(SED modelling)\\
\hline\\
MIR1$^\star$&0&$9.3\times10^{-4}-0.04$&Class I&Stage II\\
MIR4&$0-6.5\times10^{-6}$&$8.1\times10^{-5}-0.01$&Red Class II&Stage I or II\\
MIR5&$0-1.2\times10^{-8}$&$3.3\times10^{-5}-0.02$&Class I&Stage II\\
MIR7$^\star$&0&$5.2\times10^{-4}-0.02$&Red Class II&Stage II\\
MIR8&0&$7.8\times10^{-4}-0.01$&Class II&Stage II\\
MIR9&$1.9\times10^{-8}-2.9\times10^{-6}$&$8.8\times10^{-6}-0.06$&Class II&Stage I or II\\
MIR15&$5.5\times10^{-7}-2.5\times10^{-3}$&$7.6\times10^{-6}-0.28$&Class II&Stage I or II\\
MIR19&$1.5\times10^{-6}-5.6\times10^{-5}$&$3.6\times10^{-5}-0.03$&Red Class II&Stage I\\
MIR21&$0-1.7\times10^{-5}$&$2.1\times10^{-6}-0.03$&Class II&Stage I or II\\
NIR31&$1.8\times10^{-7}-1.5\times10^{-4}$&$2.2\times10^{-6}-0.03$&-&Stage I or II\\
\hline\\
\multicolumn{4}{l}{%
\begin{minipage}{7.5cm}
\scriptsize{$^\star$ : Hot excess sources based on MIPS-IRAC CC diagram
}  
\end{minipage}%
}
\end{tabular}
\label{rob_class}
\end{center}
\end{table}

\par From the best-fit models, the masses of all sources fall in the range $1.7\leq\textrm{M}_\star\leq10.0$. This suggests that all these sources are intermediate to high mass objects. MIR1, the YSO that has been identified as IRAS 16161--4931 earlier, is the most massive one (10~M$_\odot$) according to the models. The ages of all the objects other than MIR8, are $\lesssim$2~Myr hinting at the youth of these sources. We have classified the YSO candidates based on their evolutionary stages using the method described by \citet{2006ApJS..167..256R}. This classification scheme divides the sources into three broad categories based on three physical properties: mass of the central object (M$_\star$), envelope accretion rate ($\dot{\textrm{M}}_\textrm{env}$) and disk mass (M$_\textrm{disk}$). Stage 0/I objects are those with $\dot{\textrm{M}}_\textrm{env}$/M$_\star>$10$^{-6}$~yr$^{-1}$ and are believed to be objects with significant infalling envelopes and possibly disks. If $\dot{\textrm{M}}_\textrm{env}$/M$_\star<$10$^{-6}$~yr$^{-1}$ and M$_\textrm{disk}$/M$_\star>$10$^{-6}$, the object is classified as a Stage II source that has an optically thick disk and possible remains of an infalling envelope. If $\dot{\textrm{M}}_\textrm{env}$/M$_\star<$10$^{-6}$~yr$^{-1}$ and M$_\textrm{disk}$/M$_\star<$10$^{-6}$, it is a Stage III object with an optically thin disk. The advantage of using this classification scheme along with the classification based on the slope of infrared SED is that it can avoid possible confusion between observable and physical properties and thereby provides a more physical basis for YSO classification. Table~\ref{rob_class} gives the values for $\dot{\textrm{M}}_\textrm{env}$/M$_\star$ and M$_\textrm{disk}$/M$_\star$ for the YSOs including the classification based on different methods. We notice that M$_\textrm{disk}$/M$_\star$ corresponding to all the 10 YSOs are above the threshold of 10$^{-6}$ for having optically thick disk, suggesting the fledgling nature of these objects.  

\par According to the classification  scheme of \citet{2006ApJS..167..256R}, there is a single Stage I source, there are 5 Stage I/II and 4 Stage II objects. We find that there is a broad corroboration between this classification scheme and those based on infrared colors. Two YSOs that show deviant behaviour between the schemes are MIR1 and MIR7.  From the SED modelling, we see that MIR1 and MIR7 fall into Stage II category with zero envelope accretion rate. According to IRAC color-color diagram, they are classified as Class I and Class II objects, respectively. But the IRAC-MIPS colors categorise them as \textquotedouble{hot excess} objects. As mentioned earlier, these are suspected to be Class I sources or Class II objects with large extinction (A$_v>$25~mag). The extinctions estimated from the models are relatively large: 56 and 34~mag for MIR1 and MIR7, respectively. This is consistent with that predicted for a Class II object falling in the \textquotedouble{hot excess} region. Hence MIR1 and MIR7 are speculated to be Class II objects  with large extinction. The age of these sources are also relatively higher ($1-4$~Myr) compared to Stage I and I/II sources ($<$1~Myr). The object NIR31 that is associated with S2 is classified as a Stage I/II object from SED modelling and has a mass of $\sim$7~M$_\odot$. If radio emission from S2 is due to NIR31, we may be probing the radio emission from an intermediate YSO object. We explore this possibility later in Sect 4.1.2. We would also like to point out that although we have applied these models for fitting the SEDs, the parameters are considered as representative at best as the models are based on assumptions that the SEDs of intermediate/massive YSOs are scaled-up versions of their lower mass counterparts.


\section{Discussion}
In this section, we first discuss about the active regions: S1 and S2, to gain an insight into their morphologies and identify the likely sources of excitation. Thereafter, we probe the properties of the IRDC using various tracers to fathom the star formation potential of the cloud. Finally, we surmise about the evolutionary stage of the cloud itself based on various lifetime indicators.  

\subsection{Morphology of radio sources}

\subsubsection{S1}
 From the radio emission that traces the ionised gas (Fig.~\ref{1300}), we perceive that S1 can be categorised as a HII region with a shell-like morphology. According to \citet{1989ApJS...69..831W} and \citet{1994ApJS...91..659K}, shell-like regions are the rarest (less than 5$\%$) of all the \hii~regions detected. Recent studies of Sgr~B2 and W49 star forming regions \citep[][and references therein]{2005ApJ...624L.101D} show a higher percentage of detection (28\%) of shell-like \hii~regions. Strong stellar-winds from the central OB stars as well as the pressure from the ionizing radiation are believed to induce the formation of shell-like/bubble structures in their vicinity \citep{{1977ApJ...218..377W},{1980ApJ...238..860S}}. While radiation pressure on dust grains are capable of producing shell-like structures in ultracompact \hii~regions \citep{1974A&A....37..149K}, they are generally less important compared to the stellar winds \citep{1984ApJ...277..164T}. An important probe of stellar winds is the presence or absence of extended 24~$\mu$m emission near the center of bubbles. The centrally peaked 24~$\mu$m favours the explanation that the grains are mostly heated by the absorption of Lyman continuum photons, that are abundant near the exciting star \citep{2010A&A...523A...6D}. In few cases, a void is found to exist in the 24~$\mu$m emission towards the centre. This orifice could be produced either by stellar winds or due to the radiation pressure of the central star \citep{{2008ApJ...681.1341W},{2009ApJ...694..546W}}. Towards S1, we anticipate that the overall distribution of 24~$\mu$m emission is similar to that of ionised gas although the former is saturated towards the central region. This leads us to believe that the stellar winds have not yet succeeded in clearing out dust from the central region. The emission at 24~$\mu$m and radio, is surrounded by 8~$\mu$m shell of enhanced emission. Towards the centre, several high extinction filamentary structures are perceived at 8~$\mu$m, evident from Fig.~\ref{s1colcom}(a). 

\par We first explore the possibility of stellar winds from the central star being responsible for the observed shell-like morphology of the ionised gas. Note that the 8~$\mu$m shell envelopes the ionised gas shell. The interaction of massive stellar winds with the ambient medium will sweep up dense shells of gas that expand away from the central source and the swept-up shell(s) are exposed to the ionizing radiation from the newly formed star. This shell material may be partially or completely ionised and the radius of the shell increases as a function of time. The radius ($R_{shell}$) and expansion velocity ($V_{shell}$) of the shell can be estimated using the following expressions \citep{{1975ApJ...200L.107C},{1999PASP..111.1049G}}:

\begin{figure*}
\centering
\hspace*{0cm}
\includegraphics[scale=0.4]{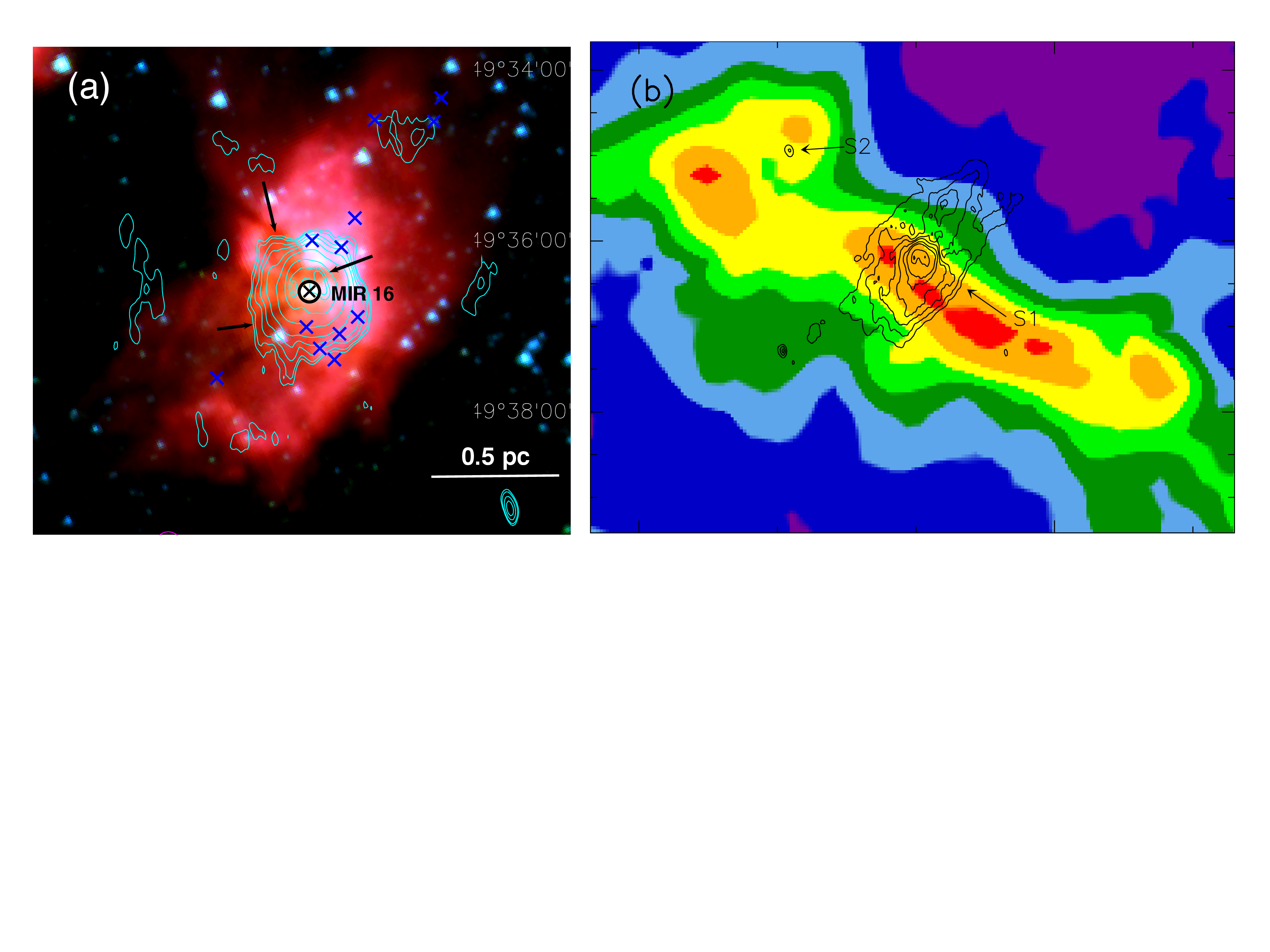} 
\vspace*{-4cm}
\caption{(a) $\textit{Spitzer}$-IRAC color composite image of S1. Blue represents 3.6~$\mu$m, green 4.5~$\mu$m and red 8~$\mu$m. Radio emission at 1300~MHz is overlaid as cyan contours. All YSO candidates within the radio contours are marked as $\times$. The location of the mid-infared Class I object MIR16 which is at the centre of the radio emission is marked and labeled in the image. Arrows point toward high extinction filamentary structures. (b) Smoothed column density map of G333.73 overlaid with 1300~MHz radio contours (black).}
\label{s1colcom}
\end{figure*}

\begin{equation}
R_{shell}=0.042\left[\frac{L_w}{10^{36}~\textrm{erg\,s}^{-1}}\right]^{1/5}\left[\frac{n_0}{10^5~\textrm{cm}^{-3}}\right]^{-1/5}\left[\frac{t}{10^3~\textrm{yr}}\right]^{3/5}~\textrm{pc}
\end{equation}

\begin{equation}
V_{shell}=24.7\left[\frac{L_w}{10^{36}~\textrm{erg\,s}^{-1}}\right]^{1/5}\left[\frac{n_0}{10^5~\textrm{cm}^{-3}}\right]^{-1/5}\left[\frac{t}{10^3~\textrm{yr}}\right]^{-2/5}~\textrm{km\,s}^{-1}
\end{equation}

 \noindent where $L_w$ is the mechanical luminosity of the stellar wind, $n_0$ is the density of the molecular cloud and $t$ is the shell expansion time. Based on the 1300~MHz radio image, the radius of the shell is taken as 0.05~pc. We adopt an archetypal shell expansion velocity of 10~km\,s$^{-1}$ \citep[e.g.,][]{{1986ApJ...309..553G},{1998ApJ...506..727B},{2009ApJ...693.1696H}} and consider the cloud density as $1.6\times10^4$~cm$^{-3}$. Using these values, we estimate the mechanical luminosity and shell expansion time to be $1.5\times10^{34}$~erg\,s$^{-1}$ and $2.9\times10^3$~yr, respectively.  

\par The expansion of the \hii~region could be due to the following: (i) the pressure difference between the ionised gas and ambient medium, and (ii) the stellar wind from the exciting star. A comparison of the expansion rates of both these mechanisms can aid in the determination of the stage of expansion, i.e. whether the stellar wind dominates the classical expansion or vice-versa. From previous studies \citep{{1980ApJ...238..860S},{1999PASP..111.1049G}}, it is seen that the stellar wind is more important when the following condition is satisfied.

\begin{equation}
\left[\frac{L_w}{10^{36}~\textrm{erg\,s}^{-1}}\right]>0.33\left[\frac{\textrm{N}_\textrm{Lyc}}{10^{49}~\textrm{s}^{-1}}\right]^{2/3}\left[\frac{n_0}{10^5~\textrm{cm}^{-3}}\right]^{-1/3}
\end{equation}

\noindent Considering the $\textrm{N}_\textrm{Lyc}$ estimated from radio flux, we estimate right hand side as $4.8\times10^{34}$~erg\,s$^{-1}$, that is nearly three times larger compared to the mechanical luminosity of stellar wind $\sim1.5\times10^{34}$~erg\,s$^{-1}$. This suggests that the effect of stellar wind is lower that that surmised from classical expansion. However, we would like to allude to the fact that the derived stellar wind luminosity is based on a typical expansion velocity of 10~km\,s$^{-1}$. If we increase $V_{shell}$ to 20~km\,s$^{-1}$, $L_w$ increases by an order of magnitude, to $1.2\times10^{35}$~erg\,s$^{-1}$. On the other hand, if we decrease $V_{shell}$ to 5~km\,s$^{-1}$, $L_w$ changes to $1.9\times10^{33}$~erg\,s$^{-1}$.  Thus, the expansion velocity is a crucial parameter that decides the stage of expansion of the \hii~region.

We also detect a large scale diffuse emission in the radio waveband associated with S1, which conforms to the morphology of emission from warm dust. The large scale morphology of radio emission can be attributed to the density gradient where the \hii~region expands out towards regions of lower density \citep{{1978A&A....70..769I},{1979A&A....71...59T}}. Fig.~\ref{s1colcom}(b) shows the distribution of ionised gas with respect to the column density map. The large scale radio emission is distributed nearly perpendicular to the long-axis of the cloud where the column density is high. This explains the expansion of ionised gas towards regions of lower density i.e. towards north-west and south-east, whereas towards the east and west of the radio peak, it is constricted by the high density gas, consistent with the champagne flow model.

\par To unravel the stars responsible for the ionised gas emission, we probe the region for YSO population, particularly towards the geometric centre (RA: $16^h20^m09.89^s$, Dec: $-49^{\circ}36\arcmin15.32\arcsec$) of the bubble/shell. The Lyman continuum flux predicts O9.5-B0 as the single ZAMS star exciting the shell. In this estimate, the diffuse emission is not taken into consideration. While we have been unable to detect any object at the geometric centre (in near or mid-infrared), we have detected a Class I source, MIR16, close to the centre (angular separation of 2$\arcsec$). The lack of detection of any source at the geometric centre is probably due to the nebulosity and high extinction in this region,  reinforced by the high extinction filamentary structures observed in the near- and mid-infrared wavelength bands, visible in Fig.~\ref{s1colcom}(a). We cannot rule out the possibility of an ionizing source being deeply embedded in the filamentary structures. In addition, we detect ionised peaks around the shell with lower flux density at 1300 MHz, shown in Fig.~\ref{s1colcom}(a). This would suggest that the large scale radio emission could be the result of multiple objects rather than a single ionizing source, although it is possible that this is the fragmented emission from the nebulous gas. We also detect six YSOs around the radio shell. The distribution of these objects around the radio shell is explicable on the basis of lower extinction as well as nebulosity in these regions.

\subsubsection{S2}

\par
\begin{figure}
\centering
\includegraphics[scale=0.25]{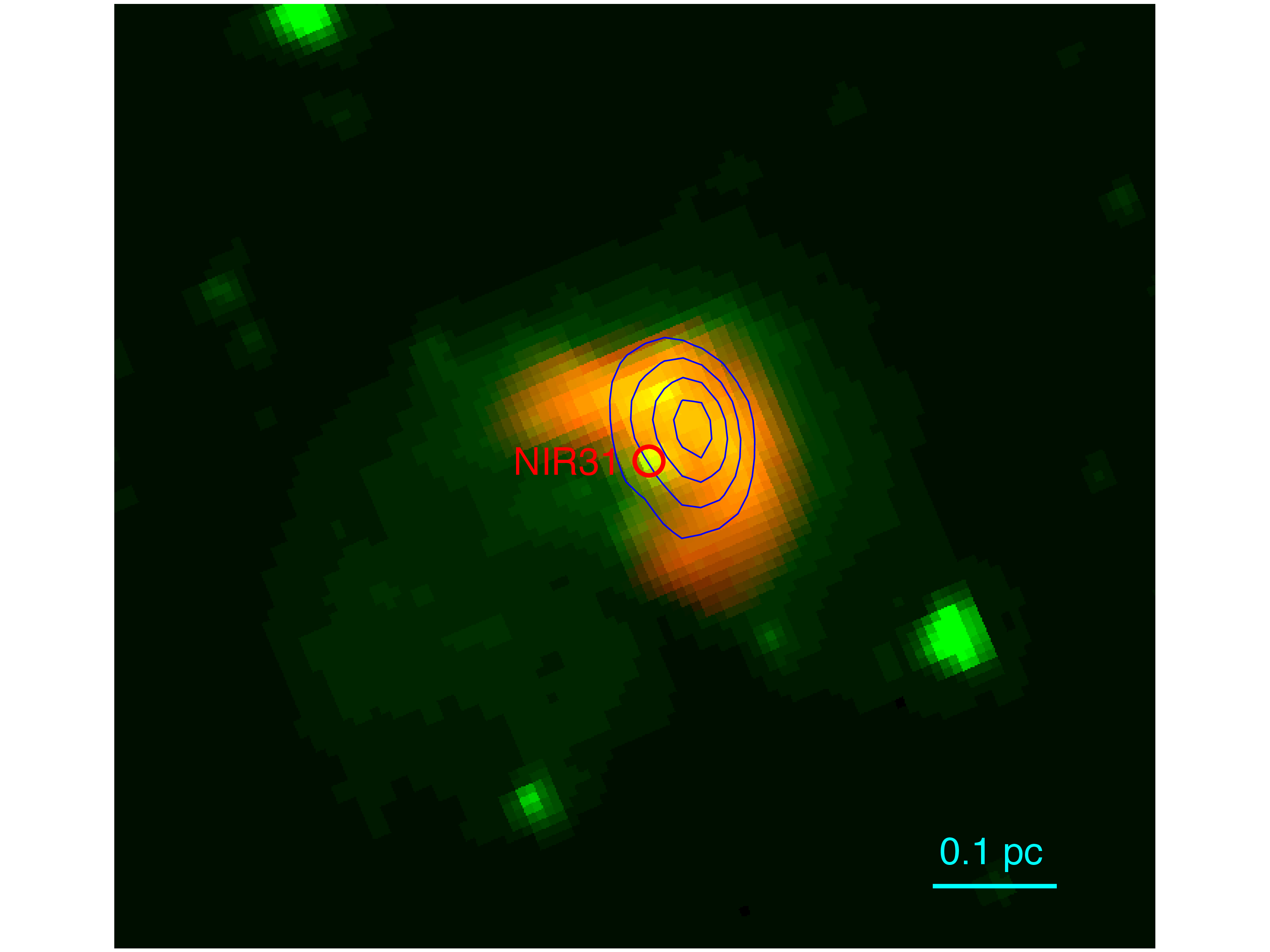}
\caption{$\textit{Spitzer}$-IRAC color composite image of S2. The 4.5~$\mu$m emission is shown in green and 8~$\mu$m is in red. Circle marks the YSO source NIR31. High resolution radio map at 1300~MHz is overlaid as blue contours. Contour levels are same as that of Fig.~\ref{1300}(b).}
\label{s2colcom}
\end{figure}

 In this subsection, we examine the morphology of the compact \hii~region S2 (see Fig.~\ref{s2colcom}), that is located towards the north-east of S1 and possesses an arc-like structure when viewed in the mid-infrared warm dust emission. Diffuse nebulosity at lower flux levels in mid-infrared, spanning a region $0.4\times0.3$~pc$^2$, is directed away from the concave edge of the arc. This cometary shaped object is oriented along the NW-SE direction (see Fig.~\ref{s2colcom}). At 24~$\mu$m, the emission is saturated and hence the distribution of emission is indecipherable. Such mid-infrared arc-shaped features have been observed towards other star forming regions \citep[e.g.,][]{{2008ApJ...689..242P},{2016AJ....152..146N}} and could be attributed to (i) expansion of an \hii~region, (ii) bow-shocks due to high velocity stars \citep{2008ApJ...689..242P} or (iii) dust or bow-wave \citep{2014A&A...563A..65O}. The radio emission towards S2 is compact but shows hints of extension towards the diffuse emission affirming the cometary outlook of the warm dust emission. The radio emission peaks approximately mid-way on the arc. The YSO, NIR31, is located $\sim3\arcsec$ away from the radio peak. This source is classified as a pre-main sequence star based on the NIR color-color diagram as well as from the SED modelling. The mass of this object is estimated to be $5-7$~M$_\odot$, an intermediate mass YSO. Radio emission has been detected from several low and intermediate mass YSOs and are ascribed to stellar winds \citep{{1975A&A....39....1P},{1996ApJ...473.1051M}} or collimated ionised jets \citep{1986ApJ...304..713R}. We look into the possibility of NIR31 being the ionizing source of S2.

\par  We first consider the bow-shock model as the origin of the mid and far-infrared arc-shaped emission \citep[e.g.,][]{{2007ApJ...655..920F},{2016ApJS..227...18K}}. Massive stellar objects with energetic winds generate strong shocks in the surrounding medium. If the relative motion between the star and ambient medium is large, the shock will be bent back around the star. For supersonic velocities, the ambient gas is swept up into an arc-shaped bow-shock and has been observed around several massive objects \citep[e.g.,][]{{2012A&A...538A.108P},{2015A&A...578A..45P}}. We use simple analytic expressions to calculate the shock parameters. For a star moving supersonically in the plane of the sky, the bow-shock is expected to trace a parabola. The shock occurs at a stand-off distance $R_0$ from the star where the stellar wind momentum flux equals the ram pressure of the ambient medium. The stand-off distance in the thin shell limit can be calculated using the expression \citep{1996ApJ...459L..31W}:

\begin{equation}
R_0=\sqrt{\frac{\dot{m}_{*}v_{w}}{4\pi \rho_a v_{*}^2}}
\end{equation}
\\

Here $\dot{m}_{*}$ is the stellar wind mass-loss rate, $v_{w}$ is the wind's terminal velocity, $\rho_a$ is the mass density of the ambient gas and $v_{*}$ is the relative velocity of the star through the medium. The stellar wind mass-loss rate ($\dot{m}_{*}$) and wind terminal velocity ($v_{w}$) are calculated using the expressions \citep{1991ApJ...369..395M}

\begin{equation}
\left[\frac{\dot{m}_*}{\rm{10^{-6}M_{\odot}yr^{-1}}}\right]=2\times10^{-7}\left[\frac{L}{\rm{L_{\odot}}}\right]^{1.25}
\end{equation}

\begin{equation}
{\rm{log}}\left[\frac{v_w}{\rm{10^8cm\,s^{-1}}}\right]=-38.2+16.23\ \rm{log}\left[\frac{T_{eff}}{\rm{K}}\right]-1.70\ \left(\rm{log}\left[\frac{T_{eff}}{\rm{K}}\right]\right)^2 
\end{equation} 
  \\

\par From the radio continuum emission we estimate the spectral type of the ionizing source as B0-B0.5. Considering a B0.5 type star, we have adopted luminosity $L=1.1\times10^4$~$L_\odot$ and effective temperature $T_{eff}=26,200$~K, \citep{1973AJ.....78..929P}. Using these values, we get $\dot{m}_{*}=0.23\times10^{-7}$~M$_{\odot}$yr$^{-1}$ and $v_{w}=1060$~km\,s$^{-1}$. We can estimate mass density using the electron number density derived in Sect. 3.5, using the expression $\rho_a = \mu m_{\textrm{H}}n_e$ where $n_e = 1.3\times 10^3$~cm$^{-3}$. Here $\mu$ is mean nucleus number per hydrogen atom taken as 1.4 and $m_{\textrm{H}}$ is the mass of hydrogen atom. The distance between NIR31 and radio peak is  0.04~pc that corresponds to 8251~AU. Considering this as $R_0$, we obtain the stellar velocity $v_{*}$ as 6.2~km\,s$^{-1}$. Typical stellar velocities observed in bow shock regions are $\sim$10~km\,s$^{-1}$ \citep{{1992ApJ...394..534V},{2016ApJ...825...16N}} which is similar to our estimate. The arc-shaped emission is also symmetric around MIR31. Therefore, we suggest that the observed mid-infrared arc could be a result of bow-shock due to MIR31. The arc-shaped features resulting from the stellar wind bow shocks are also detected in far-infrared wavelengths \citep[e.g.,][]{{2012A&A...537A..35C},{2012A&A...548A.113D}}. Density gradients present in the cloud can also affect the bow shock symmetries \citep{2000ApJ...532..400W}.  

\par The expansion of the \hii~region towards a low density medium could result in a cometary or arc-shaped morphology. A close examination of the morphology of S2 in 70 and 160~$\mu$m bands reveals that the far-infrared dust emission also follows an arc-like morphology similar to that seen at mid-infrared wavebands. At longer wavebands, the resolution prohibits us from distinguishing the morphology of cold dust emission in detail. If the nebulosity (seen in mid- and far-infrared with an inkling in radio) is due to a local density gradient in this direction, we would expect the constriction of ionised flow by a high density medium on the far side of the head or arc. By comparing the radio emission and column density distribution in Fig.~\ref{s1colcom}(b), we observe a local maxima in the column density towards the north of S2. As mentioned earlier, the radio emission is slightly extended towards south of column density peak. Hence density gradients might be responsible for the observed morphology of S2. Apart from pure bow-shocks and density gradients, there are various hybrid models that incorporate the effects of stellar winds into existing models \citep[e.g.,][]{{1994ApJ...432..648G},{2006ApJS..165..283A}}. Hence it is also possible that the arc-like morphology of S2 is a combination of bow shock and density gradient.

\par An alternate possibility that has been considered to justify the arc-shaped emission is the dust wave model \citep{2014A&A...563A..65O}. Here, the cometary morphology is a consequence of the interaction of radiation pressure of the star with the dust carried along by the photo-evaporative flow. In our case, the cometary structure in S2 is unlikely to be the outcome of a bow-wave as we do not perceive any bubble structure around S2.

\subsection{Evolutionary stages of clumps}

We next examine the star forming properties of the clumps in the IRDC using our multiwavelength approach. We estimate the relative evolutionary stages of clumps based on the evolutionary sequence proposed by \citet{2009ApJS..181..360C} and \citet{2010ApJ...721..222B}. According to \citet{2009ApJS..181..360C}, in an IRDC, the star formation begins with a quiescent clump which evolves later into an active clump (containing enhanced 4.5~$\mu$m emission called ``green fuzzy'' and a 24 $\mu$m point source) and finally becomes a red (enhanced 8 $\mu$m emission) clump. \citet{2010ApJ...721..222B} further modified this classification scheme by incorporating radio emission and suggesting that the red clumps are diffuse ones without associated millimeter peaks. They discuss four important star formation tracers: (1) Quiescent clump (no signs of active star formation), (2) Intermediate clumps that exhibit one or two signs of active star formation (such as shock/outflow signatures or 24 $\mu$m point source), (3) Active clumps that exhibit three or four signs of active star formation (\textquotedouble{green fuzzies}, 24 $\mu$m point source, UC\hii~region or maser emission), and (4) Evolved red clumps with diffuse 8 $\mu$m emission. \citet{2013A&A...550A..21S} also proposed an evolutionary sequence where clumps are classified as either Type 1 or Type 2 owing to their detection in infrared/millimeter images. A clump is classified as Type 2 if it has associated mid-infrared emission and Type 1 in the absence of mid-infrared emission. Thus, the quiescent clumps are Type 2 whereas intermediate/active clumps are Type 1. 

\par We have searched for MIPS 24 $\mu$m point sources associated with the clumps within a search radius of 10$\arcsec$ from the peak position. We also sought the locations of masers in this region in literature.  We find that while there has been a search for 22~GHz water maser and 6.7~GHz methanol maser (half power beam widths of 7$\arcmin$ and 3.3$\arcmin$) towards this star forming region, the masers were not detected \citep{{1989A&AS...77..465B},{1993MNRAS.261..783S},{1997MNRAS.291..261W}}. Hence, we have used solely the radio and mid-infrared data to classify the clumps. The results are presented in Table~\ref{clump_ev}. The columns in the table list the clump name, association with radio and 8~$\mu$m peaks, and 24~$\mu$m point source. Column 5 testifies to the clump activity (Q-quiescent, A-active, I-intermediate, E-evolved) and Column 6 shows whether a clump is in Type 1 or Type 2 evolutionary stage. Among the ten clumps considered, two (C1 and C2) are active/evolved clumps, two (C6 and C7) are intermediate clumps and the rest eight are quiescent clumps. The two active/evolved (A/E) clumps, C1 and C2, have higher dust temperatures (T$_{\rm{d}}$~$>$~20~K) compared to other clumps. The molecular gas associated with C1 also shows evidence of protostellar infall. The star formation activity in C1 and C2 have already been discussed in previous sections and these clumps corresponds to the star forming regions S1 and S2. The intermediate clumps: Clump C6 harbours two mid-infrared YSOs MIR4 and MIR5 whereas Clump C7 is associated with the YSO MIR8.

\begin{table}[hbt!]
\footnotesize
\caption{The classification of clumps based on multi-wavelength signposts of star formation.}
\begin{center}
\hspace*{-0.5cm}
\begin{tabular}{c c c c c c} \hline \hline
Clump &Radio &IRAC &MIPS & Clump&Evolutionary  \\
No.&source& 8~$\mu$m peak&24~$\mu$m source&activity&Stage\\
\hline
C1 &\checkmark &\checkmark &Saturated &A/E&Type 2\\
C2 &\checkmark &\checkmark &Saturated &A/E&Type 2\\
C3 &\text{\sffamily X} &\text{\sffamily X}&\text{\sffamily X}&Q&Type 1\\
C4 &\text{\sffamily X} &\text{\sffamily X}&\text{\sffamily X} &Q&Type 1\\
C5 &\text{\sffamily X} &\text{\sffamily X}&\text{\sffamily X} &Q&Type 1\\
C6 &\text{\sffamily X} &\text{\sffamily X} &\checkmark &I&Type 2\\
C7 &\text{\sffamily X} &\text{\sffamily X} &\checkmark  &I&Type 2\\
C8 &\text{\sffamily X} &\text{\sffamily X} &\text{\sffamily X} &Q&Type 1\\
C9 &\text{\sffamily X} &\text{\sffamily X}&\text{\sffamily X} &Q&Type 1\\
C10 &\text{\sffamily X} &\text{\sffamily X} &\text{\sffamily X} &Q&Type 1\\
\hline 
\multicolumn{5}{l}{\textsuperscript{a}\footnotesize{Q-quiescent, I-intermediate, A-active, E-evolved based on }}\\
\multicolumn{5}{l}{\footnotesize{\citet{2010ApJ...721..222B}}}\\
\multicolumn{5}{l}{\footnotesize{\checkmark - Detection; \text{\sffamily X} - Non detection}}\\
\end{tabular}
\label{clump_ev}
\\
\end{center}
\end{table}

\begin{table}[hbt!]
\footnotesize
\caption{Properties of the 3 quiescent clumps.}
\begin{center}
\hspace*{-0.5cm}
\begin{tabular}{c c c c c} \hline \hline
Clump &M$_\textrm{max}$&$\tau_{ff}$&$\tau_{dyn}$& $\tau_{ff}$/$\tau_{dyn}$\\
No.&(M$_\odot$)&(Myr)&(Myr)&\\
\hline
C3 &10.4&0.2&0.4&0.5 \\
C4 &9.7&0.2&0.4&0.5 \\
C5 &8.5&0.2&0.4&0.5 \\
C8 &2.9&0.2&0.3&0.7\\
C9 &10.3&0.2&0.4&0.5 \\
C10 &8.6&0.1&0.3&0.3 \\
\hline 
\end{tabular}
\label{quiescent}
\\
\end{center}
\end{table} 

\subsubsection{Star forming potential of quiescent clumps}

\par We probe the dynamical state of the quiescent clumps by comparing their free-fall and dynamical timescales. The parameter free-fall time ($\tau_{ff}$) is defined as the timescale on which an object will collapse into a point under its own gravity. We estimate the free-fall time of the quiescent clumps using the expression

\begin{equation}
\tau_{ff}=\left[\frac{3\,\pi}{32\,G\,\rho}\right]^{1/2}=3.4\,\left[\frac{100}{n_\textrm{H2}}\right]^{1/2} \textrm{Myr}
\end{equation}

\noindent where $\rho$ is the density of the molecular clump and $n_\textrm{H2}$ is the number density of H$_2$ molecules \citep{2014prpl.conf....3D}. The free-fall time of the 6 quiescent clumps are listed in Table~\ref{quiescent} and they lie in the range $0.1-0.2$~Myr. It is to be noted that the estimates of $\tau_{ff}$ are based on the current density of the cloud and if the clump is undergoing collapse, then the initial free-fall times may not be the same as the present day estimates. 

\par The dynamical or crossing time $\tau_{dyn}$, defined as the time required for the sound waves to cross the clump and can be estimated using the expression \citep{2006ApJ...641L.121T}

\begin{equation}
\tau_{dyn}=\left[\frac{\rm{R}}{\sigma_{obs}}\right]
\end{equation} 

\noindent Here R is the radius of the clump and $\sigma_{obs}$ is the velocity dispersion of the cloud. We estimate $\sigma_{obs}\sim0.9$~km~s$^{-1}$ using the FWHM of the optically thin H$^{13}$CO$^+$ line. The dynamical times of the quiescent clumps are listed in Column 4 of Table~\ref{quiescent} and they lie between $0.3-0.4$~Myr. The free-fall and dynamical time estimates can be used to predict whether a clump is gravitationally stable or not. A clump becomes gravitationally unstable if $\tau_{ff}<\tau_{dyn}$ \citep[e.g.,][]{{2017MNRAS.466..340C},{2017arXiv170408264S}}. To analyse the potential of collapse for these quiescent clumps, we specify the ratio $\tau_{ff}$/$\tau_{dyn}$, tabulated in Column 5 of Table~\ref{quiescent}. All the six quiescent clumps have $\tau_{ff}$/$\tau_{dyn}$~$<$~1 and these are prone to gravitational collapse. We note that these are representative values and a more rigorous analysis would entail high resolution molecular line observations to gauge the dispersion velocity in each clump. 

 Under the assumption that the quiescent clumps undergo collapse, we analyse their potential to form high mass stars. For this, we use the formulation of  \citet{2016ApJ...822...59S} to find the mass of the most massive star ($\rm{M}_{max}$), likely to form in the clump, using a stellar Initial Mass Function (IMF) of \cite{2001MNRAS.322..231K}. $\rm{M}_{max}$ is estimated using the expression

\begin{equation}
 \rm{M}_{max} = 20\,\left[\frac{\rm{\sigma}_{sf}\,\, \rm{M}_{clump}}{0.3\times1064\rm{M}_\odot} \right]^{1/1.3}~\rm{M}_\odot
 \end{equation}

\noindent Here, $\rm{\sigma}_{sf}$ is the star forming efficiency in the clump, taken as 30\% in the present work \citep{2003ARA&A..41...57L}. The estimated $ \rm{M}_{max}$ for the 6 quiescent clumps are listed in Table~\ref{quiescent}. The mass estimates range from $2.9-10.4$~M$_\odot$. This suggests that all the quiescent clumps in this region have the potential to form intermediate to massive stars.

\subsection{Fragmentation in the filamentary cloud}

From clumps, we move on to the expanse of the cloud in order to assimilate a larger picture of the IRDC. G333.73 is located in the fourth Galactic quadrant where multiple Giant Molecular Filaments (GMFs) are located. GMFs are tremendously long filamentary clouds (aspect ratio $\sim 50$ or larger) with lengths exceeding 100 pc. A recent study by \citet{2014ApJ...797...53G} suggests that these GMFs, also designated as `bones', can be used to constrain the spiral structure of the Milky Way. A number of bone-like filaments have been identified in our Galaxy \citep[e.g.,][]{{2015ApJ...815...23Z},{2016ApJS..226....9W}}. Large scale PV diagrams (i.e Galactic longitude versus velocity) reveal that many of these GMFs have velocity structures consistent with or close to that of the Scutum-Centaurus arm \citep[see][]{{2015ApJ...815...23Z},{2016A&A...591A...5L}}. For G333.73, based on its Galactic longitude (333.73) and LSR velocity of $-33.2$~km\,s$^{-1}$, we believe that the cloud is located in the inter-arm region, closer to the Scutum-Centaurus arm in comparison to the Sagittarius-Carina arm \citep{2016A&A...590A.131A}. The smaller size ($\sim$7 pc) and aspect ratio ($\sim$5) of G333.73 suggests that this filament is unlikely to be a GMF itself. This is corroborated by studies where star formation in nearby molecular clouds have demonstrated that the distribution of gas and dust is often filamentary comprising of either a single filament or a network of filaments. These are believed to trace the densest regions of GMFs \citep{{2005A&A...431..149H},{2009ApJ...700.1609M},{2016MNRAS.456.2041C}}. Recent surveys have found hundreds of filaments using \textit{Herschel} and ATLASGAL \citep{{2014ApJ...791...27S},{2016A&A...591A...5L}}. These filaments have lengths in the range $1-30$~pc, typical aspect ratios $\sim2-30$, and are dense ($N(\textrm{H}_2)\sim10^{21}-10^{22}$~cm$^{-2}$) and massive ($100-10^{5}$~M$_\odot$). The size, aspect ratio, mass (4700~M$_\odot$) and average column density ($\sim2.4\times10^{22}$~cm$^{-2}$) of G333.73 fall well within the range of values observed for other filamentary clouds.

We next compare the clump properties with those expected from the theoretical predictions of fragmentation of a filamentary cloud. Assuming that the clumps are governed by Jeans instability, we estimate the Jeans length of the homogeneous gas using the expression \citep{{1996ima..book.....C},{2014MNRAS.439.3275W}}

\begin{equation}
\lambda_\textrm{J} = c_s~\left(\frac{\pi}{G\rho}\right)^{1/2} = 0.066~\left(\frac{T}{10~\textrm{K}}\right)^{1/2}\left(\frac{n}{10^5~\textrm{cm}^{-3}}\right)^{-1/2}~\textrm{pc}
\end{equation} 

\noindent where $G$ is the gravitational constant and $c_s$ is the sound speed, $T$ is the temperature of the clump and $n$ is the number density. The expression for Jeans mass is given as

\begin{equation}
M_\textrm{J} =\frac{\pi^{5/2}\,c_s^3}{6\,\sqrt{G^3}\rho} = 0.877~\left(\frac{T}{10~\textrm{K}}\right)^{3/2}\left(\frac{n}{10^5~\textrm{cm}^{-3}}\right)^{-1/2} \textrm{M}_\odot
\end{equation}

\noindent Employing average values of (i) temperature $\sim16.8$~K and (ii) number density $\sim3.6\times10^4$~cm$^{-3}$, of the clumps in this region (Sect. 3.2), we determine $\lambda_\textrm{J}\sim0.2$~pc and M$_\textrm{J}\sim3.3$~M$_\odot$. Observationally, the radii of the clumps are found to lie between $0.3-0.9$~pc while the clump masses range from 87 to 1530~M$_\odot$. The radii of these clumps are consistent with the calculated Jeans length whereas the masses are larger by a factor of $\gtrsim25$ compared to the Jeans mass. The thermal pressure by itself predicts smaller Jeans mass than the observed clump masses, suggesting the dominance of turbulence in this region. This supports the idea that turbulence achieves greater significance in high mass star forming regions \citep{2008ApJ...672..410L}. As the molecular line width could have contributions from both thermal and non-thermal components, it is possible to estimate the magnitude of each of these effects. The mean thermal broadening can be calculated using the equation $V_\textrm{therm}$ = $\sqrt{k\,T_\textrm{ex}/\mu\,m_\textrm{H2}}$ and is found to be $\sim$0.2~km\,s$^{-1}$ for the average clump temperature. This is nearly a factor of four lower than the mean velocity dispersion of 0.9~km\,s$^{-1}$ implying the supersonic nature of the filament.

\par The clumps in G333.73 appear to be spaced regularly along the filament. The fragmentation along the filaments can explained using the `sausage instability' model \citep{1953ApJ...118..116C} that can approximate a filamentary cloud to an isothermal cylinder. The self-gravitating cylinder in equilibrium has a critical linear mass density above which it will undergo gravitational collapse. If the turbulent pressure dominates over thermal pressure, the critical linear mass density is given by the expression \citep{2010ApJ...719L.185J}:

\begin{equation}
(M/l)_{crit} = 84\, \Delta V^{2}~\textrm{M}_\odot~\textrm{pc}^{-1}
\end{equation}

\noindent For $\Delta V$ = 2.2~km\,s$^{-1}$, $(M/l)_{crit}$ is found to be $\sim$407~M$_\odot$~pc$^{-1}$, whereas the linear mass density of G333.73 is $(4700/7.2)\sim653$~M$_\odot$~pc$^{-1}$. We observe that this exceeds the critical limit and hence the cloud may be on the verge of global collapse. This is consistent with the infall motion seen in Clump C1 (Sect. 3.3.3) as well as accretion signatures observed on a larger scale (next section). The presence of YSOs \citep{2010A&A...518L.102A} and massive star formation \citep{2016A&A...591A...5L} substantiates the supercritical nature of the filamentary cloud. 

\subsection{Velocity gradient in G333.73: Rotation or accretion?}
As discussed in Sect 3.3.5, we have detected a fairly regular velocity gradient across the length of the filament. We first investigate whether this gradient could be a result of cloud rotation and estimate the ratio of rotational-to-gravitational energies, $\beta$, for a cylindrical cloud using the expression \citep[][and references therein]{2014MNRAS.439.1996J}

\begin{equation}
\beta = \frac{E_\textrm{rot}}{E_\textrm{grav}} \sim \frac{\Omega^2\,L^3}{36\,G\,M} \sim \frac{V^2_{\rm grad}\,L}{36\,G\,M}
\end{equation}

\noindent Here, $E_\textrm{grav}$ is the gravitational energy of the cylindrical cloud given by $E_\textrm{grav}$ = $\frac{3}{2}\frac{G\,M^2}{L}$, for a cloud of mass $M$ and length $L$ and $G$ is the gravitational constant. $E_\textrm{rot}$ is the kinetic energy due to the cloud's rotation defined as $E_\textrm{rot}$ = $\frac{1}{2}I\,\Omega^2$, where $\Omega$ is the angular speed of the cloud and $I$ is the moment of inertia given as $I \sim \frac{M\,L^2}{12}$ when the rotational axis is perpendicular to the long axis of the cylinder. $\Omega$ is estimated using the velocity gradient $V_{\rm grad}$ across the length, $ V_{\rm grad}/L$. For the case of G333.73, we consider $M\sim4700$~M$_\odot$, $L\sim7.2$~pc and $V_{\rm grad}\sim5/7.2\sim0.7$~km~s$^{-1}$~pc$^{-1}$. This gives $\Omega\sim2.3\times10^{-14}$~s$^{-1}$, and $\beta\sim0.3$. 
 \citet{2012ApJ...746..174R} cite a value of $\beta\sim\frac{1}{3}$ as the breakup speed for rotating clouds, with larger values of $\beta$ preventing cloud fragmentation and star formation. Our value of $\beta$ is close to the break-up speed. However this value is estimated using gradients from relatively low density tracers ($^{13}$CO) that can be dominated by turbulent motions \citep{{1986ApJ...303..356A},{1993ApJ...406..528G}}. This is also in accordance with what is observed by us. Therefore, our estimation of $\beta\sim0.3$ using $^{13}$CO is plausibly an overestimate. The presence of clumps as evidences of fragmentation suggests that any contribution of rotation to the velocity gradient is likely to be negligible. 

\par An alternate scenario to explain the velocity gradient is accretion flows along the filament \citep{{2013ApJ...771...24S},{2012ApJ...748...16T}}. Under this presumption, we estimate the gas accretion rate along the filament ($\dot{M}$) considering a simple cylindrical model \citep{2013ApJ...766..115K} using the following expression:

\begin{equation}
\dot{M}=\frac{V_\textrm{grad}\,M}{\tan(\alpha)}
\end{equation}

\noindent Here, $\alpha$ represents the angle of inclination of the long-axis of the cylinder with respect to the plane of the sky.  A value of $\alpha\sim0^\circ$ would prohibit the detection of velocity gradient, whereas $\alpha\sim90^\circ$ would prevent the cylindrical cloud appearing elongated to us.  Taking an average value of $\alpha \sim 45^\circ$, we find the accretion rate to be $4.6\times10^{-3}$~M$_\odot$\,yr$^{-1}$. This accretion rate is similar to those found in other star forming regions \citep[e.g.,][]{{2009ApJ...706.1036G},{2014MNRAS.439.1996J}}. We, therefore, believe that the accretion flows can reasonably explain the large scale velocity gradient found in G333.73, rather than rotation. The findings support the infall motion observed in the high density gas tracers and supercritical linear mass density derived earlier. However, we cannot rule out the residual effect of velocity coherence that has been widely observed in GMFs \citep{{2014A&A...568A..73R},{2015MNRAS.450.4043W}}.

\subsection{Age limit of the IRDC}

After examining the global traits of G333.73, our motivation is to uncover a limit to the age of the IRDC based on our multiwavelength analysis. The fragmentation of the cloud to clumps and the level of star forming activity within
should permit us to ascertain the broad evolutionary state of the IRDC \citep{2010ApJ...719L.185J}. The dark quiescent  filaments prior to the hot-core phase are believed to be in the earliest stages \citep{{2000ApJ...543L.157C},{2014MNRAS.439.3275W}}. The IRDCs that harbour young \hii~regions are likely to be in the intermediate phase \citep{{2016ApJ...819..117X},{2017arXiv170809098X}}. Thereafter, the \hii~regions evolves with time and disperses the IRDC, leading to the emergence of the embedded cluster which reveal the active sites of massive star formation such as Orion and NGC 6334. As G333.73 is associated with clumps that harbour \hii~regions and YSOs, it is consistent with the proposition that IRDCs are the precursors of massive star clusters and can form multiple high mass stars simultaneously. The expanding \hii~region in G333.73 is in the process of dispersing the natal cloud in its immediate vicinity [Fig.~\ref{s1colcom}(b)]. Therefore, we believe that G333.73 is in an intermediate phase as compared to prestellar filamentary clouds and highly evolved regions that harbour massive star clusters in our Galaxy.

\par In this work, we presented different timescales such as the age of \hii~region, age of YSOs from SED modelling, free-fall and dynamical timescales of the clumps. We now compare these estimates and attempt to derive a limit for the current age of G333.73. The expansion time scales of the compact \hii~regions, S1 and S2, are found to be 0.2 and 0.01~Myr respectively. We also estimated the free-fall and dynamical times of the quiescent clumps, that lie within the range: $0.1-0.4$~Myr.  The ages of YSOs in this region, based on the best fit to the SED using radiative transfer models, fall in the range $0.1-9.6$~Myr. All the YSOs except one, MIR8, are younger than  2~Myr according to the best-fit models. For MIR8, we find that the lower limit to the age among ten best fit models is 0.2~Myr. If we consider the YSOs other than MIR8, we could assign a lower limit to the age of G333.73 as 2~Myr. The age of giant molecular clouds (GMCs) are usually estimated to be nearly two to three times the free-fall timescale \citep{2011ApJ...729..133M}. The free-fall times of GMCs are estimated using a mean cloud density, based on the CO emission density that is converted directly to a hydrogen column density assuming a constant CO to H$_2$ abundance ratio \citep[e.g.,][]{{1975ApJ...199L.105S},{1976ApJ...208..346G},{1991ApJ...366...95S}}. We estimated the mean CO density of the cloud from our CO maps and find $n({\rm H_2})\sim 2300$~cm$^{-3}$. This corresponds to $\tau_{ff}^{cloud}\sim 0.7$~Myr giving an age estimate of $\sim2$~Myr. This is in conformity with those obtained by other means. However, a caveat is that we are sampling the relatively higher density region of the cloud and it may not be prudent to use these densities to obtain a measure of the limit to the age of the cloud. 

\par We reform our goal to estimate the age of the clumps as that can provide a lower limit to the age of the IRDC.  In order to carry this out, we refer to the characteristic timescales occupied by the starless and star-forming phases in a cloud.
\citet{2017ApJ...835..263B} carried out a pixel-by-pixel analysis of the cloud complex within a $2^\circ\times2^\circ$ of the Galactic plane and noticed that the starless phase occupies nearly $60-70\%$ of the dense molecular region (DMR) lifetime while the star forming phase is approximately $30-40\%$ of the total DMR lifetime. The total lifetime of DMRs $\tau_{\rm tot}$ based on the association with ultracompact \hii~regions (UC\hii) is given by the following expression \citep{2017ApJ...835..263B}:

\begin{equation}
\tau_{\rm tot} = \frac{\tau_\textrm{UC\hii}}{f_{\rm star}\,f_\textrm{UC\hii}}
\end{equation}   

\noindent where $\tau_\textrm{UC\hii}$ is the age of the UC\hii~region, $f_{\rm star}$ is the fraction of starry pixels and $f_\textrm{UC\hii}$ is the fraction of pixels that are associated with UC\hii~regions. We apply the above expression to clumps themselves in order to secure a limit to the age. Based on our clump analysis, we demonstrated that 4 clumps are in star-forming phase (i.e. active, evolved and intermediate) and rest 6 are in quiescent phase. This gives us the fraction of starry clumps as $40\%$. Of the 4 star-forming clumps, 2 harbour compact \hii~regions that leads us to $f_\textrm{UC\hii}\sim50\%$. $\tau_\textrm{UC\hii}$ is taken as 0.2~Myr, considering the \hii~region expansion timescale of S1 estimated in Sect. 3.5. This is in agreement with the age of ultracompact \hii~regions from chemical clocks \citep{2014A&A...569A..19T}. We then estimate a lower limit to the lifetime of G333.73 as 1~Myr. This is consistent with the absolute lifetime estimate of \citet{2017ApJ...835..263B} for massive star forming clouds which is $1-3$~Myr. We also obtained similar estimate based on the age of YSOs. Hence, the age of G333.73 is likely to be 1-2~Myr. An upper limit could be the lifetimes of GMCs in our Galaxy, which is found to lie between $10-20$~Myr \citep{{2011ApJ...729..133M},{2014prpl.conf....3D}} although GMFs are expected to be older \citep{2017MNRAS.470.4261D}.

\section{Conclusion} 
In this work, we present the continuum and molecular line study of the filamentary IRDC G333.73. The results are summarized as follows. \\

(i) G333.73 is an IRDC with filamentary morphology at a distance  of 2.6~kpc with a spatial extent of $7.2\times1.5$~pc. Mid-infrared images reveal the presence of two bright sources, S1 and S2, connected by high extinction structures. S1 is a bubble and S2 has an arc-like structure.\\

(ii) Far-infrared and submillimeter emission is detected towards G333.73. We detect 10 cold dust clumps in this region. The temperature of the clumps range between $14.3-22.3$~K with the peak located towards S1. The column density map exhibits multiple peaks coinciding with the locations of high density clumps. The total mass of this IRDC is estimated to be 4700~M$_\odot$.\\

(iii) Eight molecular species (HCO$^+$, H$^{13}$CO$^+$, HCN, HNC, N$_2$H$^+$, C$_2$H, $^{12}$CO and $^{13}$CO) are detected in this IRDC. The profiles of HCN and HNC molecules towards S1 exhibit signatures of protostellar infall. We also detect a fairly steady velocity gradient along the major axis of the filament. The observed gradient is explained using accretion flows along the filament.\\

(iv) Low frequency radio emission is detected towards S1 and S2 at 1300 and 610~MHz. S1 is associated with a shell-like \hii~region whereas S2 is a compact source. Assuming a single ZAMS exciting star in each region, the spectral types of these sources are consistent with late O or early B type stars. The expansion timescales of \hii~regions associated with S1 and S2 are 0.2 and 0.01~Myr respectively.\\ 

(v) We detected a total of 56 YSOs in different evolutionary stages within the IRDC. The radiative transfer modelling of ten sources having 24~$\mu$m counterparts suggest that these are intermediate to high mass objects.\\

(vi) The G333.73 filament is supercritical and harbours multiple objects in different evolutionary stages such as millimeter cores, \hii~regions and YSOs. A lower limit to the age of the IRDC is estimated as $1-2$~Myr. \\

\begin{acknowledgments}
\par We are grateful to the referee for the valuable comments and insights that significantly improved the quality of this paper. We thank the staff of GMRT, who made the radio observations possible. GMRT is run by the National Centre for Radio Astrophysics of the Tata Institute of Fundamental Research. Thanks are also due to R. Cesaroni for providing the 1.2~mm map of this region. This research made use of NASA/IPAC Infrared Science Archive, which is operated by the Jet Propulsion Laboratory, Caltech under contract with NASA. This publication also made use of data products from $Herschel$ (ESA space observatory). ASM is partially supported by the Collaborative Research Centre SFB 956, sub-project A6, funded by the Deutsche Forschungsgemeinschaft (DFG).
\end{acknowledgments}

\bibliography{ref}

\newpage
\appendix
\section{A : Clump SED fits}

\begin{figure*} [h]
\centering
\includegraphics[scale=0.7]{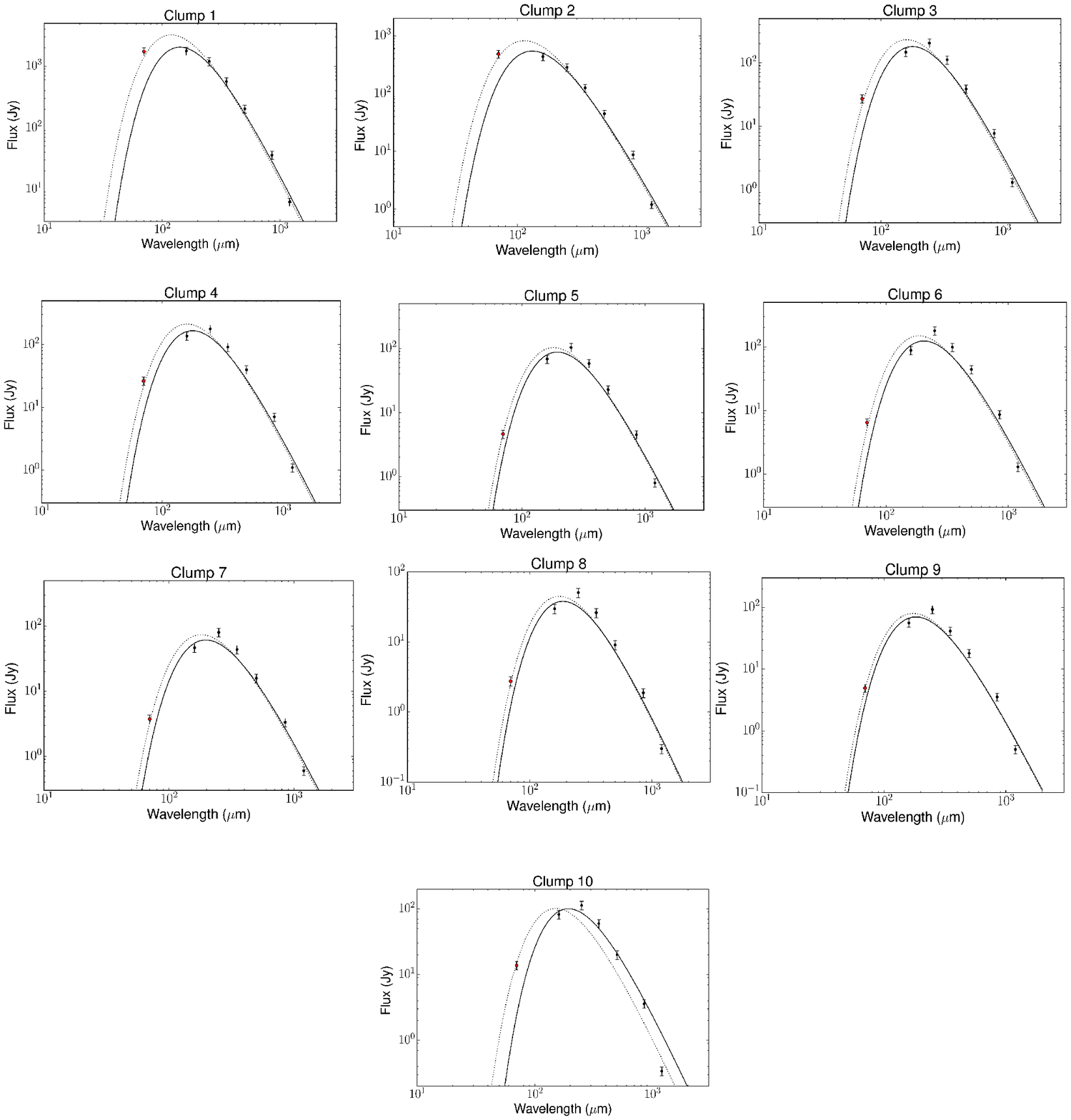}
\caption{Spectral energy distributions of the cold dust clumps in G333.73. Flux densities from $160~\mu\textrm{m}-1.2$~mm are represented as solid circles and 70~$\mu$m flux is represented as red dot. The best-fit modified blackbody function for each SED excluding 70~$\mu$m is shown as a solid line curve and the fit including 70~$\mu$m flux is shown as dotted line. The error bars correspond to 15\% uncertainties in flux densities. }
\label{clsed}
\end{figure*}


\section{B : NIR YSOs}
\begin{table}
\scriptsize
\caption{Coordinates and magnitudes of near-infrared pre-main sequence sources.}
\begin{center}
\setlength{\tabcolsep}{3pt}
\begin{tabular}{l c c c c c} \hline \hline
 &  \\
YSO     &$\alpha_{2000}$&$\delta_{2000}$&J &H&K  \\
&($^{h~m~s}$) &($^{\circ~\arcmin~\arcsec}$) &(mag)&(mag)&(mag)\\
\hline\\

NIR1 &16:19:48.421&$-$49:38:13.39 &$16.02 \pm 0.09$ & $13.54 \pm  0.04$ & $12.23 \pm  0.03$ \\
NIR2 &16:19:48.693 &$-$49:37:02.60 &$15.01 \pm  0.07$ & $14.46 \pm$  0.11 & $13.99 \pm  0.08$ \\
NIR3 &16:19:50.237 &$-$49:37:28.78 &$15.70 \pm  0.08$  &$14.64 \pm  0.09$  &$13.94  \pm 0.07$ \\
NIR4 &16:19:51.425 &$-$49:37:16.62 &$14.47 \pm  0.05$ & $13.84 \pm  0.05$  &$13.42 \pm  0.06$ \\
NIR5 &16:20:00.259 &$-$49:36:27.51 &$14.69 \pm  0.05$  &$13.98 \pm  0.05$  &$13.58 \pm  0.05$ \\
NIR6 &16:20:01.816 &$-$49:36:47.06 &$16.05 \pm  0.09$  &$14.58 \pm  0.13$  &$13.61 \pm  0.08$ \\
NIR7 &16:20:01.850 &$-$49:36:20.02 &$15.35 \pm  0.07$ & $13.93 \pm  0.07$ & $13.07 \pm  0.04$ \\
NIR8 &16:20:03.207 &$-$49:35:22.64 &$11.49 \pm  0.02$  &$11.31 \pm  0.05$ & $11.09 \pm  0.06$ \\
NIR9 &16:20:03.534 &$-$49:36:01.26 &$16.07 \pm  0.09$ & $15.08 \pm  0.13$ & $14.33 \pm  0.12$ \\
NIR10 &16:20:03.731 &$-$49:36:24.67 &$13.73 \pm  0.03$&  $13.24 \pm  0.04$ & $12.98 \pm  0.04$ \\
NIR11 &16:20:04.090 &$-$49:36:30.99 &$13.10 \pm  0.02$ & $12.78 \pm  0.04$ & $12.52 \pm  0.04$ \\
NIR12 &16:20:04.112 &$-$49:36:12.82 &$14.43 \pm  0.02$ & $13.99 \pm  0.02$ & $13.72 \pm  0.05$ \\
NIR13 &16:20:05.813 &$-$49:35:19.74 &$13.13 \pm  0.05$  &$12.80 \pm  0.04$  &$12.63 \pm  0.03$ \\
NIR14 &16:20:06.003 &$-$49:35:26.97 &$11.66 \pm  0.03$ & $11.35 \pm  0.04$  &$11.19 \pm  0.03$ \\
NIR15 &16:20:07.787 &$-$49:35:26.88 &$16.48 \pm  0.13$ & $14.52 \pm  0.06$ & $13.28 \pm  0.04$ \\
NIR16 &16:20:08.299 &$-$49:36:24.99 &$14.93 \pm  0.04$ & $14.33\pm   0.11$&  $13.37 \pm  0.14$ \\
NIR17 &16:20:08.346 &$-$49:35:55.60 &$12.98 \pm  0.03$ & $12.63 \pm  0.06$ & $12.38 \pm  0.07$ \\
NIR18 &16:20:08.824 &$-$49:36:03.85& $15.33 \pm  0.19$ & $13.38 \pm  0.20$ & $11.34 \pm  0.07$ \\
NIR19 &16:20:08.843 &$-$49:36:30.29 &$14.09 \pm  0.04$ & $13.41 \pm  0.07$  &$12.85 \pm  0.09$ \\
NIR20 &16:20:09.049 &$-$49:36:37.69 &$14.78 \pm  0.04$  &$14.15 \pm  0.08$ & $13.64 \pm  0.08$ \\
NIR21 &16:20:09.214 &$-$49:35:05.28 &$12.66 \pm  0.03$  &$12.26 \pm  0.05$  &$11.74 \pm  0.03$ \\
NIR22 &16:20:09.456 &$-$49:36:34.86 &$13.06 \pm  0.03$ & $12.53 \pm  0.04$ & $12.19 \pm  0.05$ \\
NIR23 &16:20:09.683 &$-$49:36:01.66 &$12.24 \pm  0.03$  &$11.44 \pm  0.06$ & $10.82 \pm  0.05$ \\
NIR24 &16:20:09.836 &$-$49:36:27.91 &$14.82 \pm  0.13$ & $14.24 \pm  0.22$ & $13.36 \pm  0.12$ \\
NIR25 &16:20:12.553 &$-$49:36:44.04 &$14.83 \pm  0.02$ & $14.26 \pm  0.07$ & $13.57 \pm  0.06$ \\
NIR26 &16:20:14.118 &$-$49:37:32.91& $13.99 \pm  0.02$ & $13.63 \pm  0.04$ & $13.43 \pm  0.05$ \\
NIR27 &16:20:14.949 &$-$49:37:17.82 &$14.04 \pm  0.05$ & $13.68 \pm  0.05$ & $13.39 \pm  0.07$ \\
NIR28 &16:20:15.384 &$-$49:36:43.86 &$14.36 \pm  0.09$ & $13.88 \pm  0.08$ & $13.60 \pm  0.07$ \\
NIR29 &16:20:15.470 &$-$49:36:51.72 &$13.82 \pm  0.03$ & $13.58 \pm  0.06$ & $13.40 \pm  0.06$ \\
NIR30 &16:20:15.827 &$-$49:37:11.98& $15.67 \pm  0.08$ & $14.34 \pm  0.05$ & $13.36 \pm  0.06$ \\
NIR31 &16:20:19.361 &$-$49:34:58.80 &$11.25 \pm  0.04$ & $10.43 \pm  0.04$ & $9.91 \pm  0.05$ \\
NIR32 &16:20:22.523 &$-$49:35:40.19 &$13.96 \pm  0.02$  &$13.55 \pm  0.03$ & $13.31 \pm  0.03$ \\
NIR33 &16:20:23.306 &$-$49:35:46.32 &$14.72 \pm  0.03$ & $14.28 \pm  0.05$  & $14.01 \pm  0.06$ \\
NIR34 &16:20:27.642&$-$49:35:13.01 &$14.78 \pm  0.04$  &$14.24 \pm  0.05$ & $13.94 \pm  0.05$ \\
\hline \\
\end{tabular}
\label{ysomag}
\end{center}
\end{table}


\end{document}